\documentclass[aps,preprint,superscriptaddress,nofootinbib,floatfix]{revtex4}
\usepackage{graphicx}
\usepackage{color}
\begin{document}
\preprint{ANL-HEP-PR-09-113, EFI-09-37, NUHEP-TH/10-01}
\global\long\def\shat{\hat{s}}
\newcommand{\gsim}{\mathrel{\hbox{\rlap{\lower.55ex \hbox{$\sim$}} \kern-.3em \raise.4ex \hbox{$>$}}}}
\newcommand{\lsim}{\mathrel{\hbox{\rlap{\lower.55ex \hbox{$\sim$}} \kern-.3em \raise.4ex \hbox{$<$}}}}

\title{Forward-Backward Asymmetry of Top Quark Pair Production}
\author{Qing-Hong Cao}
\email{caoq@hep.anl.gov}
\affiliation{HEP Division, Argonne National Laboratory, Argonne, IL 60439}
\affiliation{Enrico Fermi Institute, University of Chicago, Chicago, IL 60637}

\author{David McKeen}
\email{mckeen@theory.uchicago.edu}
\affiliation{Enrico Fermi Institute, University of Chicago, Chicago, IL 60637}

\author{Jonathan L. Rosner}
\email{rosner@hep.uchicago.edu}
\affiliation{Enrico Fermi Institute, University of Chicago, Chicago, IL 60637}

\author{Gabe Shaughnessy}
\email{g-shaughnessy@northwestern.edu}
\affiliation{HEP Division, Argonne National Laboratory, Argonne, IL 60439}
\affiliation{Department of Physics and Astronomy, 
Northwestern University, Evanston, IL 60208}

\author{Carlos E. M. Wagner}
\email{cwagner@hep.anl.gov}
\affiliation{HEP Division, Argonne National Laboratory, Argonne, IL 60439}
\affiliation{Enrico Fermi Institute, University of Chicago, Chicago, IL 60637}
\affiliation{Kavli Institute for Cosmological Physics, University of Chicago, 5640 S. Ellis Avenue, Chicago, Illinois 60637}

%\date{\today}

\begin{abstract}
We adopt a Markov Chain Monte Carlo method to examine various new physics
models which can generate the forward-backward asymmetry in top quark pair
production observed at the Tevatron by the CDF Collaboration.
We study the following new physics models: (1) exotic gluon $G^\prime$, 
(2) extra $Z^\prime$ boson with flavor-conserving interaction, 
(3) extra $Z^\prime$ with flavor-violating $u$-$t$-$Z^\prime$ interaction,
(4) extra $W^\prime$ with flavor-violating $d$-$t$-$W^\prime$ interaction, 
and (5) extra scalars $S$ and $S^\pm$ with flavor-violating $u$-$t$-$S$ and
$d$-$t$-$S^\pm$ interactions. After combining the forward-backward asymmetry with
the measurement of the top pair production cross section and the $t\bar{t}$ invariant 
mass distribution at the Tevatron, we find that an axial vector exotic gluon $G^\prime$ of mass about $1~{\rm TeV}$ or $2~{\rm TeV}$ or a $W^\prime$ of mass about $2~{\rm TeV}$ offer an improvement over the Standard Model.  The other models considered do not fit the data significantly better than the Standard Model.  
We also emphasize a few points which have been long ignored in the literature 
for new physics searches: (1) heavy resonance width effects, (2)
renormalization scale dependence, and (3) NLO corrections to the $t\bar{t}$ invariant mass spectrum.  We argue that these three effects are crucial
to test or exclude new physics effects in the top quark pair asymmetry. 

\end{abstract}

\maketitle

%--------------------------------------------------------------------%
%  Section I :          Introduction                                 %
%--------------------------------------------------------------------%

\section{Introduction}

The CDF Collaboration has observed a $2.3\sigma$ deviation in the
forward-backward (F-B) asymmetry of top quark pair production
at the Tevatron, using a data sample with $3.2\,{\rm fb}^{-1}$ integrated
luminosity~\cite{CDF:public}: 
\begin{equation}
A_{FB}^{p\bar{p}}(\cos\theta) = 0.193\pm0.065({\rm stat})\pm 0.024({\rm syst}).
\label{eq:cdf}
\end{equation}
This measurement improves the previous CDF result based on 
$1.9\,{\rm fb}^{-1}$~\cite{Aaltonen:2008hc},
\[
A_{FB}^{p\bar{p}}(\cos\theta)=0.17\pm0.08,\qquad{\rm and}
\qquad 
A_{FB}^{t\bar{t}}(\Delta\eta)=0.24\pm0.14,\]
where the results given in the lab ($p\bar{p}$) and the center-of-mass
(c.m.) frame of the top quark pair ($t\bar{t}$) are consistent with
the theoretically expected dilution of $30\%$ in passing from $t\bar{t}$
to $p\bar{p}$~\cite{Antunano:2007da}. It is also consistent with
the D0 result based on $0.9\,{\rm fb}^{-1}$~\cite{Abazov:2007qb}:
\[
A_{FB}^{obs}=0.19\pm 0.09({\rm stat})\pm 0.02({\rm syst}),
\qquad{\rm and}
\quad A_{FB}^{obs}=0.12\pm0.08({\rm stat})\pm0.01({\rm syst})
\]
for exclusive 4-jet events and inclusive 4-jet events, respectively.  Although
the value is
still consistent at a confidence level of $\sim1.5\%$ with the SM prediction, which is
\cite{Kuhn:1998jr,Kuhn:1998kw}
\begin{equation}
A_{FB}^{p\bar{p}}(\cos\theta) = 0.051\pm0.015,
\label{eq:cdf}
\end{equation}
it is interesting to ask whether or not the large central
value can be explained by new physics (NP) after one takes into account
other Tevatron experimental measurements of top quark pair production.
There has been recent excitement among theorists for this measurement at 
the Tevatron~\cite{Djouadi:2009nb,Jung:2009jz,Cheung:2009ch,Frampton:2009rk,
Shu:2009xf,Arhrib:2009hu,Ferrario:2009ee,Dorsner:2009mq,Jung:2009pi,
Cao:2009uz,Barger:2010mw}.

In this work we point out that a strong correlation exists between
$A_{FB}$ and $\sigma(t\bar{t})$ measurements and further derive
the bounds on NP from both measurements under the interpretation of a variety of models.  

One should also keep in mind that, thanks to $p\bar{p}$ collisions,
the Tevatron offers the best opportunity for measuring the asymmetry
of top quark pair production, because of the basic asymmetry of the
production process. At the Large Hadron Collider (LHC), the asymmetry
of top quark pair production is an odd function of the pseudorapidity
of the $t\bar{t}$ pair, due to the lack of definition of the forward
direction. Hence the LHC will improve the measurement of the total
cross section of top quark pairs, but has very limited reach for studying
the asymmetry. In this sense the Tevatron plays a unique role for
testing top quark interactions, and it would provide more accurate
measurements with future accumulated data. Projected bounds on both
$A_{FB}$ and $\sigma(t\bar{t})$ at the Tevatron with $10\,{\rm fb}^{-1}$
integrated luminosity are also presented.

The paper is organized as follows. In Sec.\ II we examine the correlation
between $A_{FB}$ and $\sigma(t\bar{t})$ based on the recent Tevatron
measurement, using the Markov Chain Monte Carlo method.  We then give examples
of a few interesting NP models generating the asymmetry, \textit{e.g.}, an
exotic gluon $G^{\prime}$ (Sec.\ \ref{sec:G-prime}), a model-independent
effective field theory approach (Sec.\ \ref{sec:eft}), a flavor-conserving
$Z'$ boson (Sec.\ \ref{sec:zprime-fd}), a flavor-violating $Z'$ or $W'$
(Sec.\ \ref{sec:Flavor-violating}), and a new scalar $S(S^{\pm})$ (Sec.\
\ref{sec:scalar}).  We then conclude in Sec.\ \ref{sec:conclusion}.

%--------------------------------------------------------------------%
%  Section II                                                        %
%                                                                    %
%   Correlation of AFB and top pair production cross section         %
%--------------------------------------------------------------------%
\section{Correlation of $A_{FB}$ and $\sigma(t\bar{t})$\label{sec:markov}}

The asymmetry $A_{FB}$ in the top quark pair production can be parameterized
as follows: 
\begin{eqnarray}
A_{FB}^{tot} & = &
 \frac{\sigma_{F}^{SM}-\sigma_{B}^{SM}+\sigma_{F}^{NP}-\sigma_{B}^{NP}}
      {\sigma_{F}^{SM}+\sigma_{B}^{SM}+\sigma_{F}^{NP}+\sigma_{B}^{NP}}
       \label{eq:AFB1}\\
 & = &
 \frac{\sigma_{F}^{NP}-\sigma_{B}^{NP}}{\sigma_{F}^{NP}+\sigma_{B}^{NP}}
 \times
 \left(1+
  \frac{\sigma_{F}^{SM}-\sigma_{B}^{SM}}{\sigma_{F}^{NP}-\sigma_{B}^{NP}}
 \right)
 \times
 \frac{\sigma_{tot}^{NP}}{\sigma_{tot}^{SM}+\sigma_{tot}^{NP}}\\
 & = & 
 A_{FB}^{NP}\times R+A_{FB}^{SM}\left(1-R\right)
 \label{eq:AFB}
\end{eqnarray}
where
\begin{equation}
A_{FB}^{NP}\equiv
\frac{\sigma_{F}^{NP}-\sigma_{B}^{NP}}{\sigma_{F}^{NP}+\sigma_{B}^{NP}}
\qquad{\rm ,}\qquad
A_{FB}^{SM}\equiv
\frac{\sigma_{F}^{SM}-\sigma_{B}^{SM}}{\sigma_{F}^{SM}+\sigma_{B}^{SM}}
\qquad{\rm and}\qquad
R=\frac{\sigma_{tot}^{NP}}{\sigma_{tot}^{SM}+\sigma_{tot}^{NP}}
\label{eq:def-afbnp-R}
\end{equation}
is the asymmetry induced by the NP, the asymmetry in the SM,
 and the fraction of the NP contribution
to the total cross section, respectively. In this work we consider
the case that the NP contribution to $A_{FB}$ occurs in the process
$q\bar{q}\to t\bar{t}$, for which the SM contributions do not generate
any asymmetry at all at LO.  However, at NLO a nonzero $A_{FB}^{SM}$ is generated.

It is worth while emphasizing the factorization of $A_{FB}^{NP}$ and $R$ in
Eq.~(\ref{eq:AFB}), as it clearly reveals the effects of NP on both the
asymmetry and the top quark pair production cross section.  For example, when
NP effects generate a negative forward-backward asymmetry, they still produce a
positive observed asymmetry as long as they give rise to a negative
contribution to $\sigma(t\bar{t})$.  This is important when the effects of
interference between the SM QCD channel and the NP channel dominate. 
Moreover, the possibility of negative contributions to $\sigma_F^{NP}$ or
$\sigma_B^{NP}$ means that $|A_{FB}^{NP}|$ can exceed 1.
 
Recently, the CDF collaboration~\cite{CDF:summer2009} has published 
new results on the $t\bar{t}$ cross section in the lepton plus jet 
channels using a neural network analysis, based on an integrated 
luminosity of $4.6~{\rm fb}^{-1}$,
\begin{eqnarray}
\sigma(m_t=171.0~\rm{GeV}) & = &
 [8.33 \pm 0.40({\rm stat}) \pm 0.39 ({\rm sys}) \pm 0.17 ({\rm theo})]~{\rm
 pb}~, \nonumber \\
\sigma(m_t=172.5~\rm{GeV}) & = & 
 [7.63 \pm 0.37({\rm stat}) \pm 0.35 ({\rm sys}) \pm 0.15 ({\rm theo})]~{\rm
 pb}~, \nonumber \\
\sigma(m_t=175.0~\rm{GeV}) & = & 
 [7.29 \pm 0.35({\rm stat}) \pm 0.34 ({\rm sys}) \pm 0.14 ({\rm theo})]~{\rm
 pb}~,
\label{eq:CDF2009summer}      
\end{eqnarray}
and also an analysis combining leptonic and hadronic channels with an integrated 
luminosity of up to $4.6~{\rm fb}^{-1}$~\cite{CDF:9913},
\begin{equation}
\sigma(m_t=172.5~\rm{GeV}) = 
[7.50 \pm 0.31({\rm stat}) \pm 0.34 ({\rm sys}) 
\pm 0.15 ({\rm theo})]~{\rm pb}~.
\label{eq:CDF2009combined}      
\end{equation} 
Note that the theory uncertainty is derived from the ratio with respect to the $Z$
cross section and the central value is quoted after reweighting to the central
values of the CTEQ6.6M PDF~\cite{Nadolsky:2008zw}.  
By means of the ratio with respect to the  $Z$
cross section, the luminosity-dependence of the theoretical $t\bar{t}$ cross
section is replaced with the uncertainty in the theoretical $Z$ boson
production cross section.  That reduces the total uncertainty to  $7\%$,
greatly surpassing the Tevatron Run II goal of $10\%$. 

In this work we fix the top quark mass to be 175~GeV as we also 
include the the CDF measurement of the invariant mass spectrum of top 
quark pairs in our study, which is based on $m_t=175~{\rm GeV}$. 
We rescale the combined CDF measurements at $m_t = 172.5~{\rm GeV}$ 
(cf. Eq.~\ref{eq:CDF2009combined}) to $m_t=175~{\rm GeV}$ which 
we estimate to be
\begin{equation}
\sigma(t\bar{t})=[7.0 \pm 0.5]~{\rm pb},
\label{eq:ttbar-we}
\end{equation}
on the basis of the approximate behavior of 
Eqs.~\ref{eq:CDF2009summer} and \ref{eq:CDF2009combined} and 
the theoretical calculation by Langenfeld, Moch, and Uwer~\cite{Langenfeld:2009wd}. 
It yields $\left|R\right|\leq 7\%$ at the $1\sigma$ level. Any asymmetry
induced by the NP ($A_{FB}^{NP}$) is highly suppressed by the SM
cross section due to the small fraction $R$; see Eq.~(\ref{eq:AFB}).

\subsection{Parameter estimation}

In this work we utilize a Markov Chain Monte Carlo (MCMC) to examine
the correlation of $A_{FB}$ and $R$. The MCMC approach is based
on Bayesian methods to scan over specified input parameters given
constraints on an output set. 
In Bayes' rule, the posterior probability of the model parameters, $\theta$, given the data, $d$, and model, $M$, is given by
\begin{equation}
p(\theta | d,M) = {\pi(\theta|M) p(d|\theta,M)\over p(d|M)},
\end{equation}
where $\pi(\theta|M)$ is known as the prior on the model parameters which contains information on the parameters before unveiling the data.  The $p(d|\theta,M)$ term is the likelihood and is given below in Eq.~\ref{eq:likelihood}.  The $p(d|M)$ term is called the evidence, but is often ignored as the probabilities are properly normalized to sum to unity.  In using the MCMC, we follow the Metropolis-Hastings algorithm, in which a random point, $\theta_{i}$,
is chosen in a model's parameter space and has an associated likelihood,
${\cal L}_{i}$, based on the applied constraints.  A collection of these points,
$\left\{ \theta_{i}\right\} $,
constructs the chain. The probability of choosing another point that
is different than the current one is given by the ratio of their respective
likelihoods: ${\rm min}(\frac{{\cal L}_{i+1}}{{\cal L}_{i}},1)$. Therefore,
the next proposed point is chosen if the likelihood of the next point
is higher than the current. Otherwise, the current point is repeated
in the chain. The advantage of a MCMC approach is that in the limit
of large chain length 
 the distribution of points, $\theta_{i}$, approaches the posterior
distribution of the modeling parameters given the constraining data.
In addition, the set formed by a function of the points in the chain,
$f(\theta_{i})$, also follows the posterior distribution of that function
of the parameters given the data. How well the chain matches the posterior
distribution may be determined via convergence criteria.  We follow the method
outlined in Ref.~\cite{Barger:2008qd} to verify convergence after generating
25000 unique points in the chain.

We adopt the likelihood
\begin{equation}
{\cal L}_{i}=e^{-\Sigma_{j}\chi_{j}^{2}/2}
      =e^{-\Sigma_{j}(y_{ij}-d_{j})^{2}/2\sigma_{j}^{2}},
\label{eq:likelihood}
\end{equation}
where $y_{ij}$ are the observables calculated from the input parameters
of the $i^{th}$ chain, $d_{j}$ are the values of the experimental
and theoretical constraints and $\sigma_{j}$ are the associated uncertainties.
In our case, the input parameter set is taken to be
$\theta_{i}=\{\sigma_{t\bar t}^{SM},\,\sigma_{t\bar t}^{NP},\, A_{FB}^{SM},\, A_{FB}^{NP}\}$.  We scan with flat priors for the unknown inputs with a range of
\begin{equation}
\begin{array}{ccccc}
-5 \text{ pb} &\le& \sigma_{t\bar t}^{NP} &\le& 5\text{ pb}\\
-5 &\le&A_{FB}^{NP}&\le&5
\end{array}
\end{equation}
(recall that as a result of its definition, $|A_{FB}^{NP}|$ may exceed 1),
while the known inputs are scanned with normal distributions about their
calculated central values,
\begin{equation}
\begin{array}{ccccc}
\sigma_{t\bar t}^{SM} &=& 6.38\text{ pb} &\pm& 0.50\text{ pb}\\
A_{FB}^{SM}&=& 0.051 &\pm& 0.015
\label{eq:ranges}
\end{array}~~~.
\end{equation}
The calculated total $t\bar{t}$ production cross section at NLO for
$m_{t}=175\,{\rm GeV}$ has been taken as~\cite{Kidonakis:2008mu,Nason:1987xz,Beenakker:1988bq}
\begin{equation}
\sigma_{th}(t\bar{t})=6.38_{-0.7}^{+0.3}{\rm (scale)}
_{-0.3}^{+0.4}\,{\rm (PDF)}\,{\rm pb},
\end{equation}
where the PDF uncertainty is evaluated using the CTEQ6.6M PDF~\cite{Nadolsky:2008zw}.  
The fully NNLO QCD correction to top pair production is highly desirable to make a more reliable prediction on the asymmetry. Since it is still not clear how the asymmetry will be affected by the complete NNLO QCD corrections, we consider the NLO QCD corrections to the top quark pair production throughout this work without including the partial NNLO QCD corrections computed in~\cite{Cacciari:2008zb,Moch:2008ai,Langenfeld:2009wd}.

\begin{figure}
\includegraphics[scale=0.35]{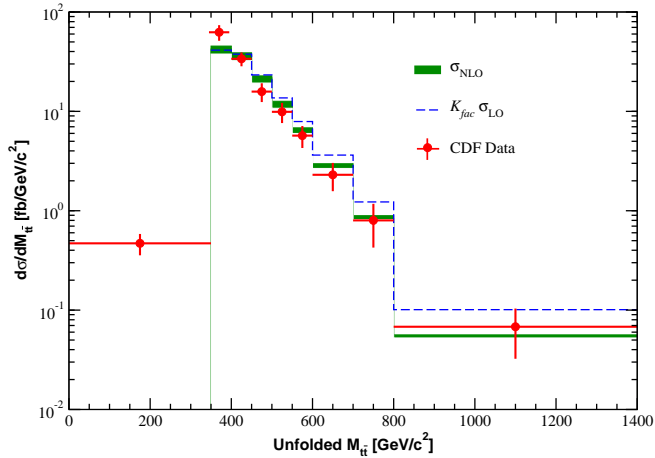}
\caption{The invariant mass spectrum measured by CDF assuming $m_t=175$ GeV.  The solid histogram is the CDF expectation taken from a LO calculation and Pythia.  The solid band indicates the full NLO SM prediction with a theoretical error due to scale uncertainty which we use in our scans.  The dashed line is $K\left(d\sigma_{\rm LO}/dM_{t\bar{t}}\right)$ with $K=\sigma_{\rm NLO}/\sigma_{\rm LO}$ which shows a large deviation from the data.  The data are taken from Ref.~\cite{Aaltonen:2009iz}.
\label{fig:mtt}}
\end{figure}

While both CDF and D0 have measurements of the invariant mass distribution~\cite{Aaltonen:2009iz,D0:winter2009}, only CDF presents an unfolded differential cross section. Therefore, we inspected the ${t\bar t}$ invariant mass spectra reported by CDF; see Fig.~\ref{fig:mtt}.  We take the 7 bins with $M_{t\bar t}> 400$ GeV in our fit and weight their $\chi^2$ by the number of included bins.  This assigns an equal weight between the $M_{t\bar t}$ measurement and the $\sigma_{t\bar t}$ and $A_{FB}$ measurements.  

% This is Figure 1

The observables are $d_{i}=\{\sigma(t\bar{t})^{exp},\, A_{FB}^{exp}\,\}$ 
in addition to the binned ${d\sigma\over d M_{t\bar t}}$ data and
define the output set.  
We use the combined cross section of Eq.~(\ref{eq:ttbar-we}).
We therefore assign $d_{i}=\{7.00,\,0.193\}$ and
$\sigma_{i}=\{0.50,\,0.069\}$ in our implementation of the likelihood defined 
above for the case we denote as ``Current" ($\int {\cal L} dt =4.6~{\rm fb}^{-1}$ for $\sigma_{t\bar{t}}$, $\int {\cal L} dt =3.2~{\rm fb}^{-1}$ for $A_{FB}$, and $\int {\cal L} dt =2.7~{\rm fb}^{-1}$ for ${d\sigma\over d M_{t\bar t}}$) while $d_{i}=\{7.00,\,0.193\}$ and
$\sigma_{i}=\{0.34,\,0.039\}$ for the case we denote as ``Projected," where $\int {\cal L} dt =10~{\rm fb}^{-1}$ of integrated luminosity is used for each measurement, in which we assume the central values are fixed and the uncertainties are scaled by a factor $1/\sqrt{\cal L}$.  We combine the chains to form iso-contours of $1\sigma$, $2\sigma$, 
and $3\sigma$ significance via their respective $p$-values.

% This is Figure 2

\begin{figure}
\includegraphics[scale=0.38]{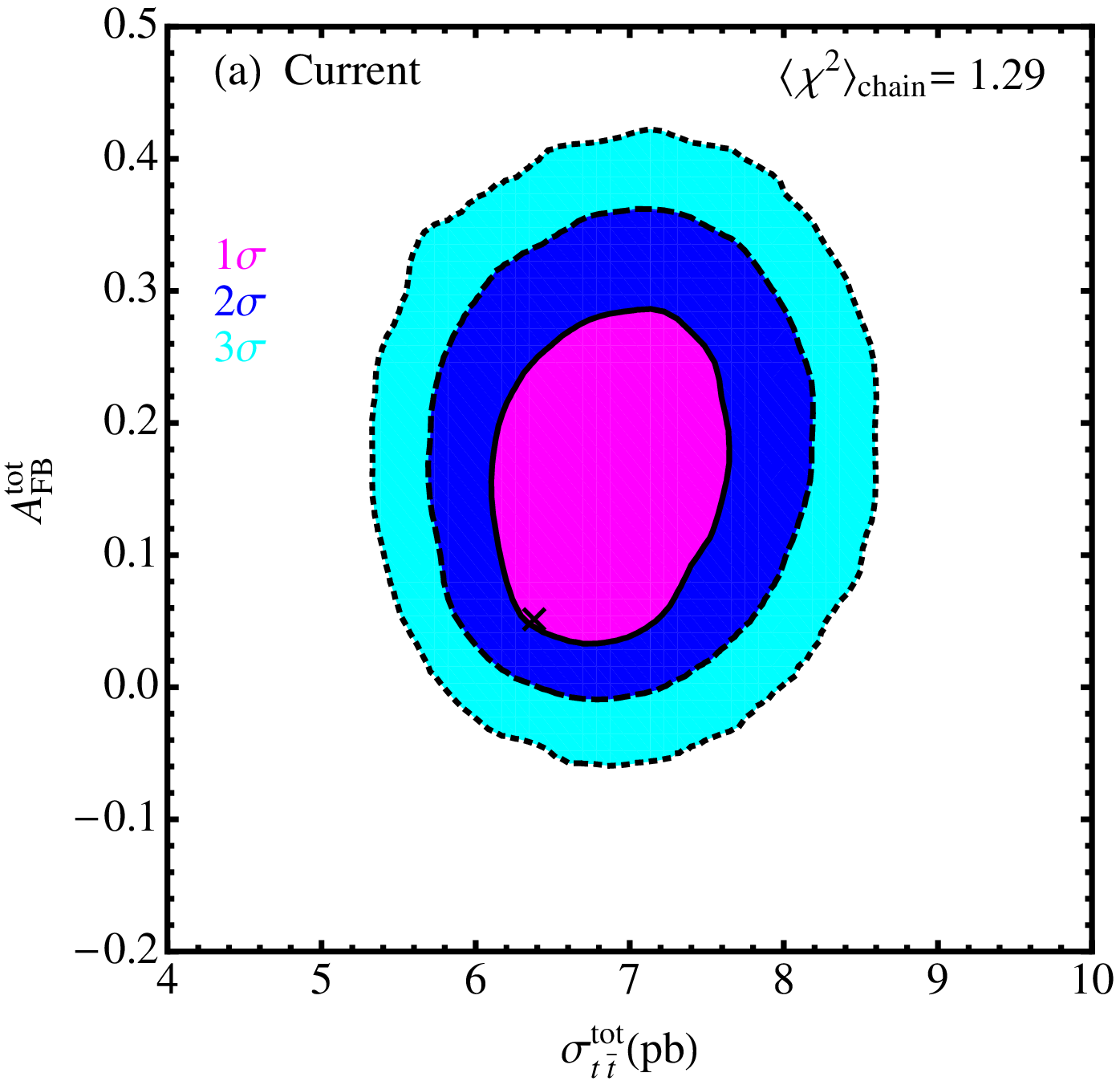}
\includegraphics[scale=0.38]{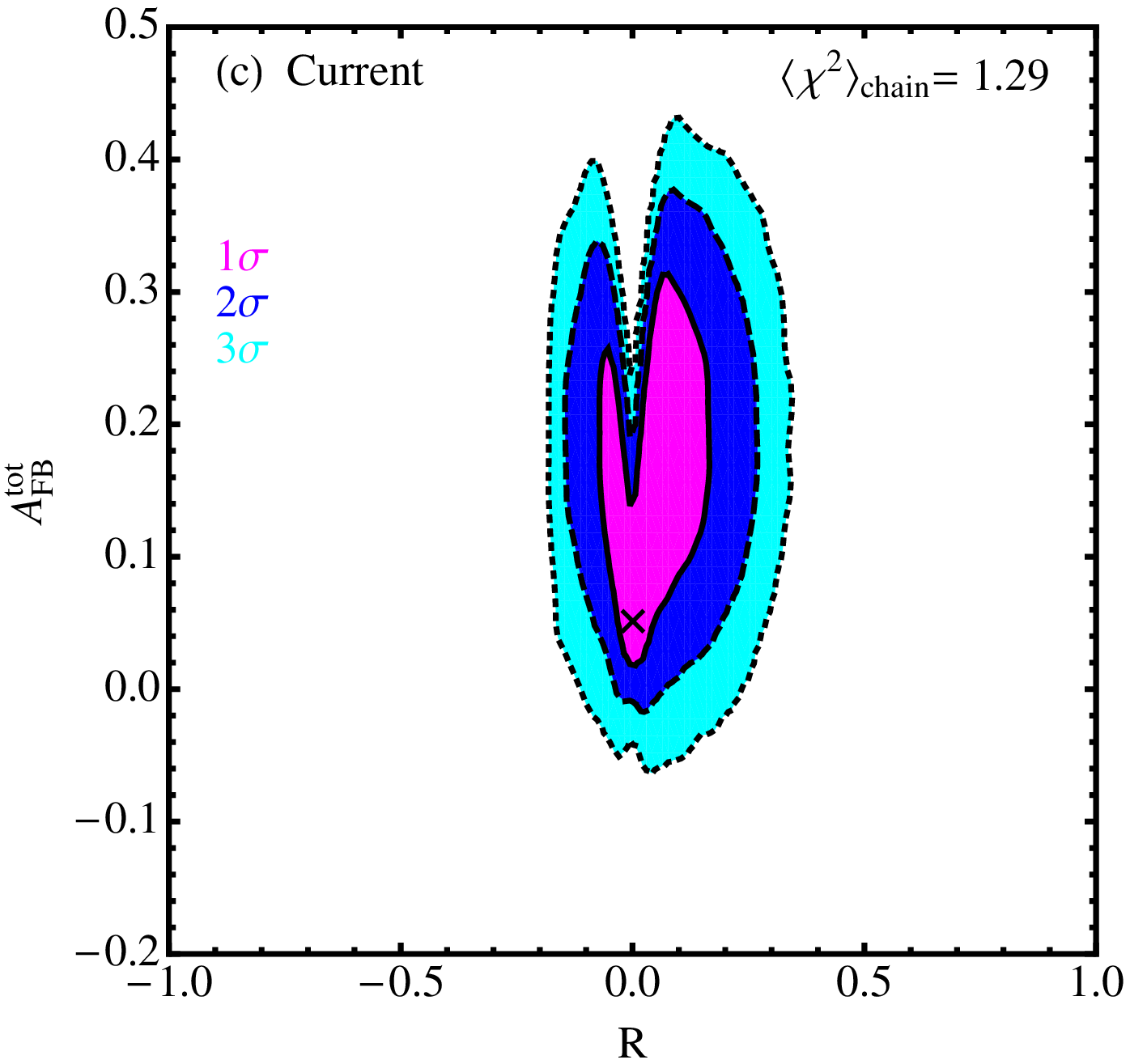} 
\includegraphics[scale=0.37]{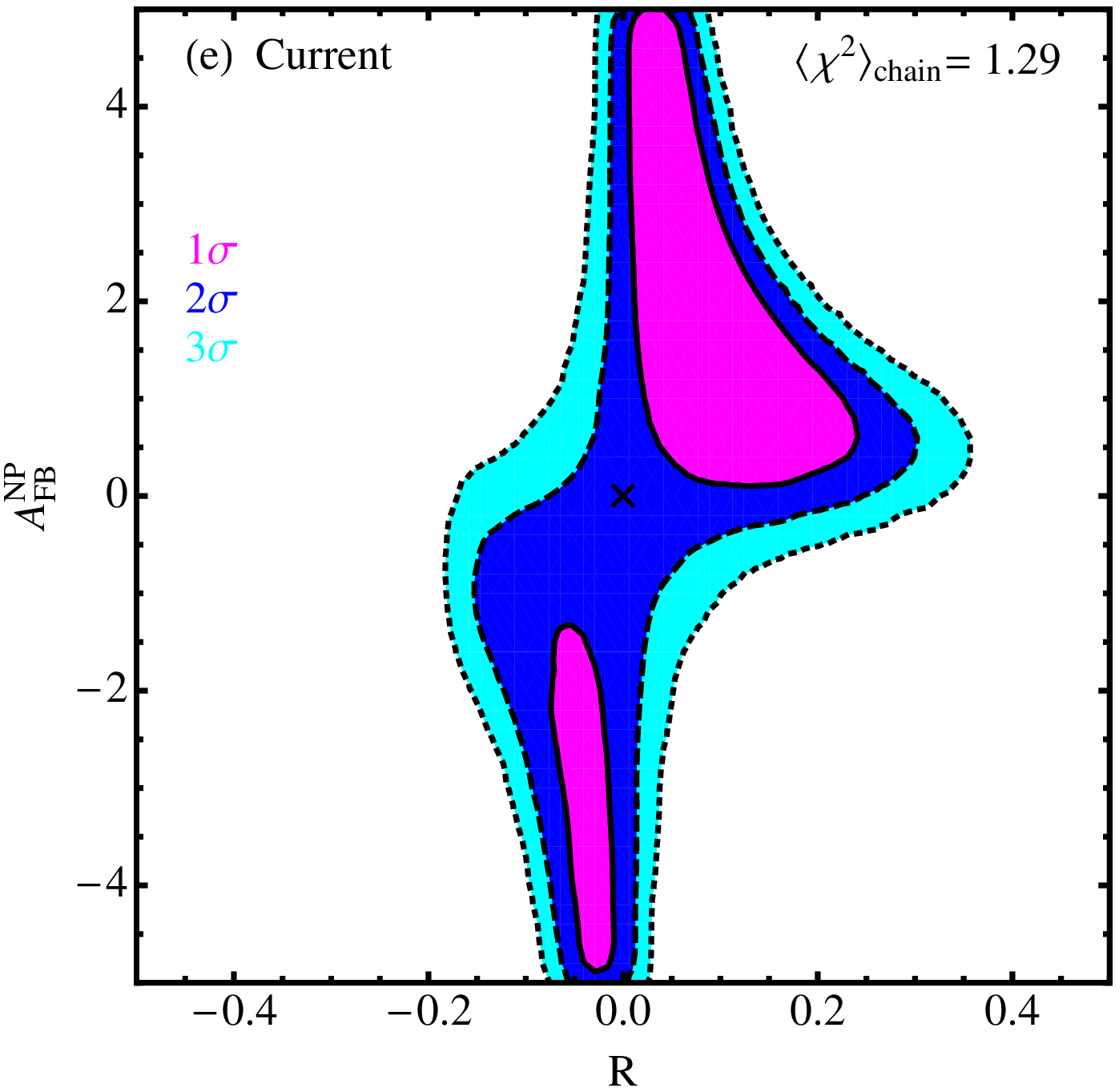}\\
\includegraphics[scale=0.38]{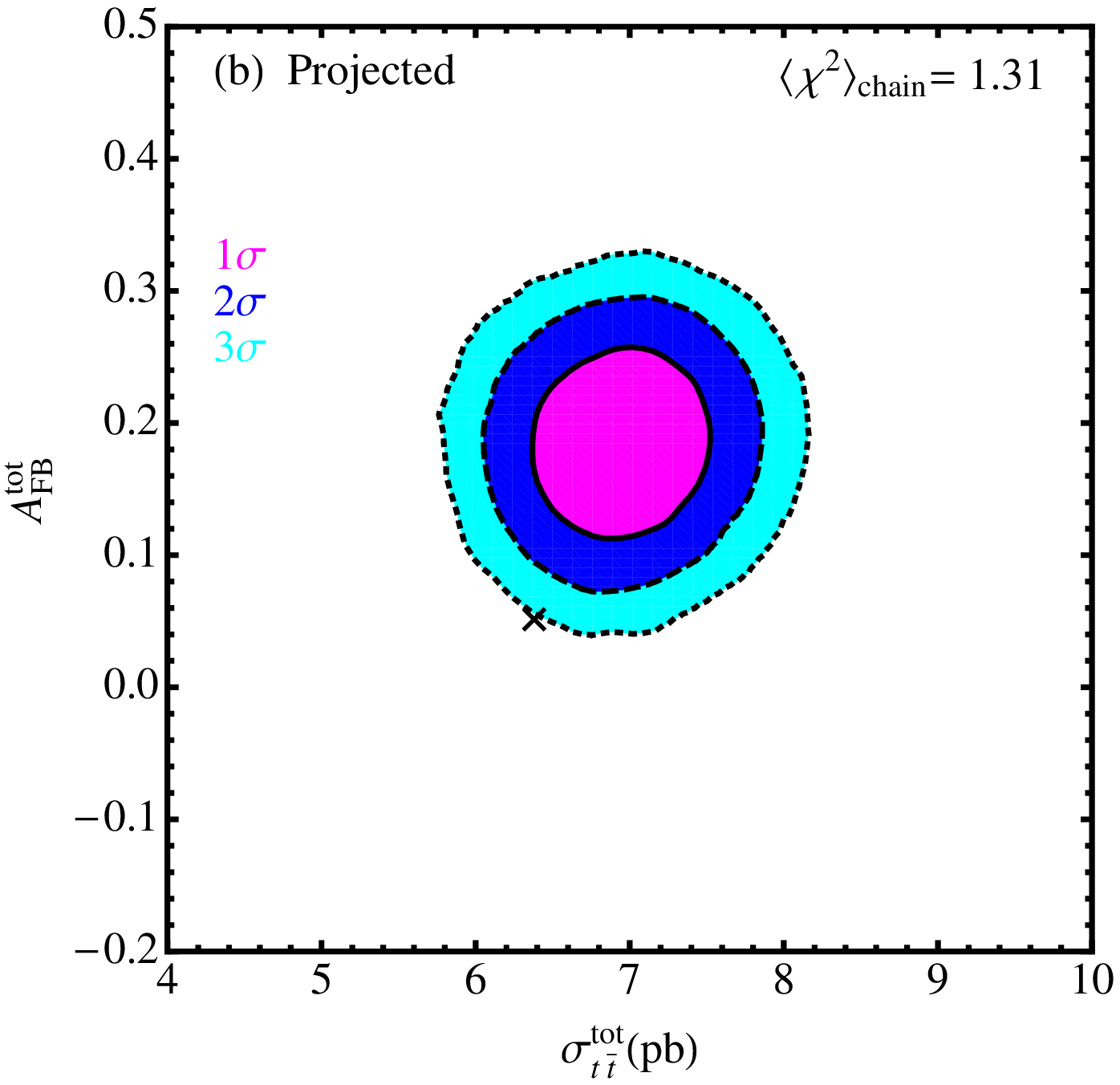}
\includegraphics[scale=0.38]{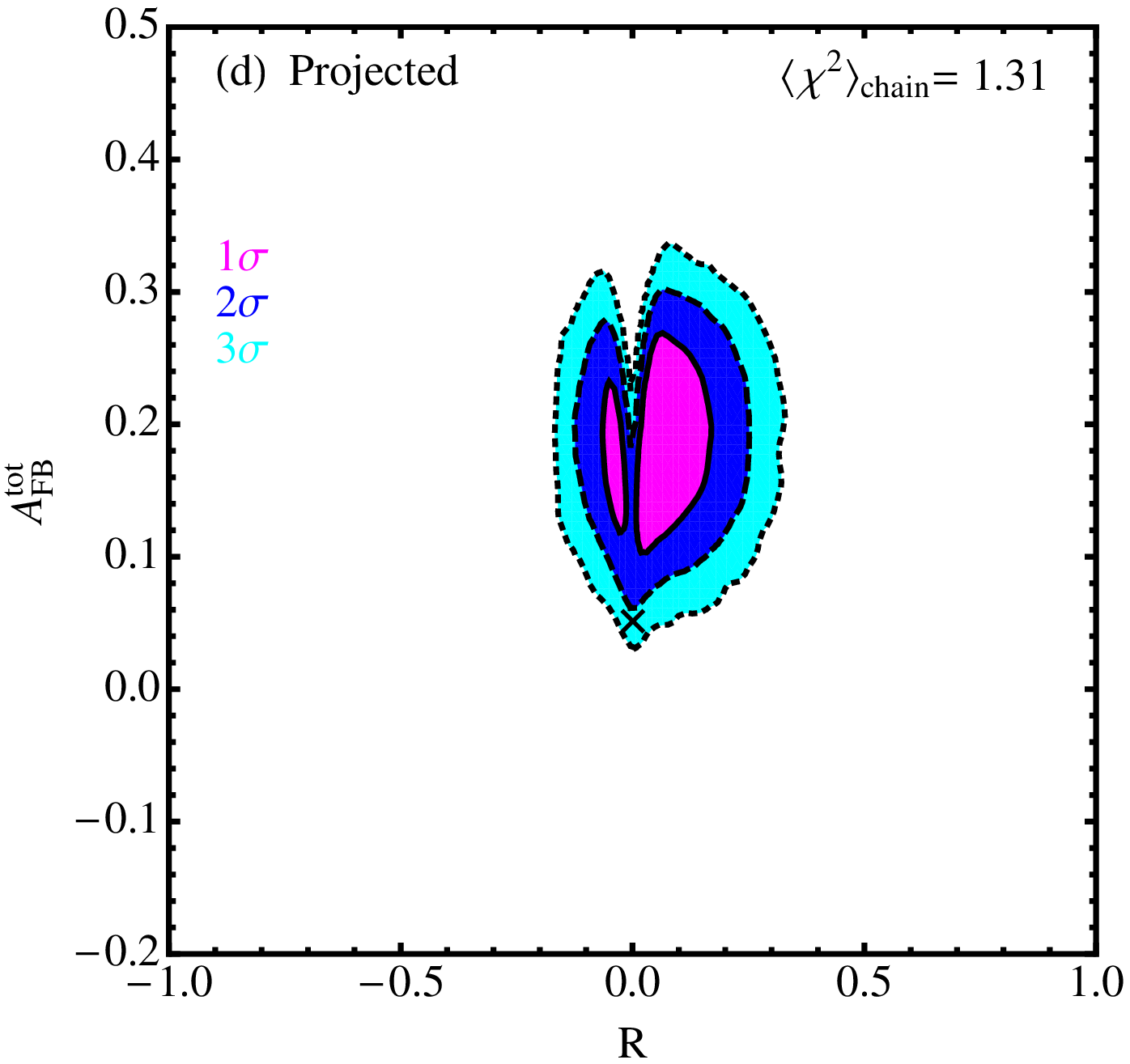}
\includegraphics[scale=0.37]{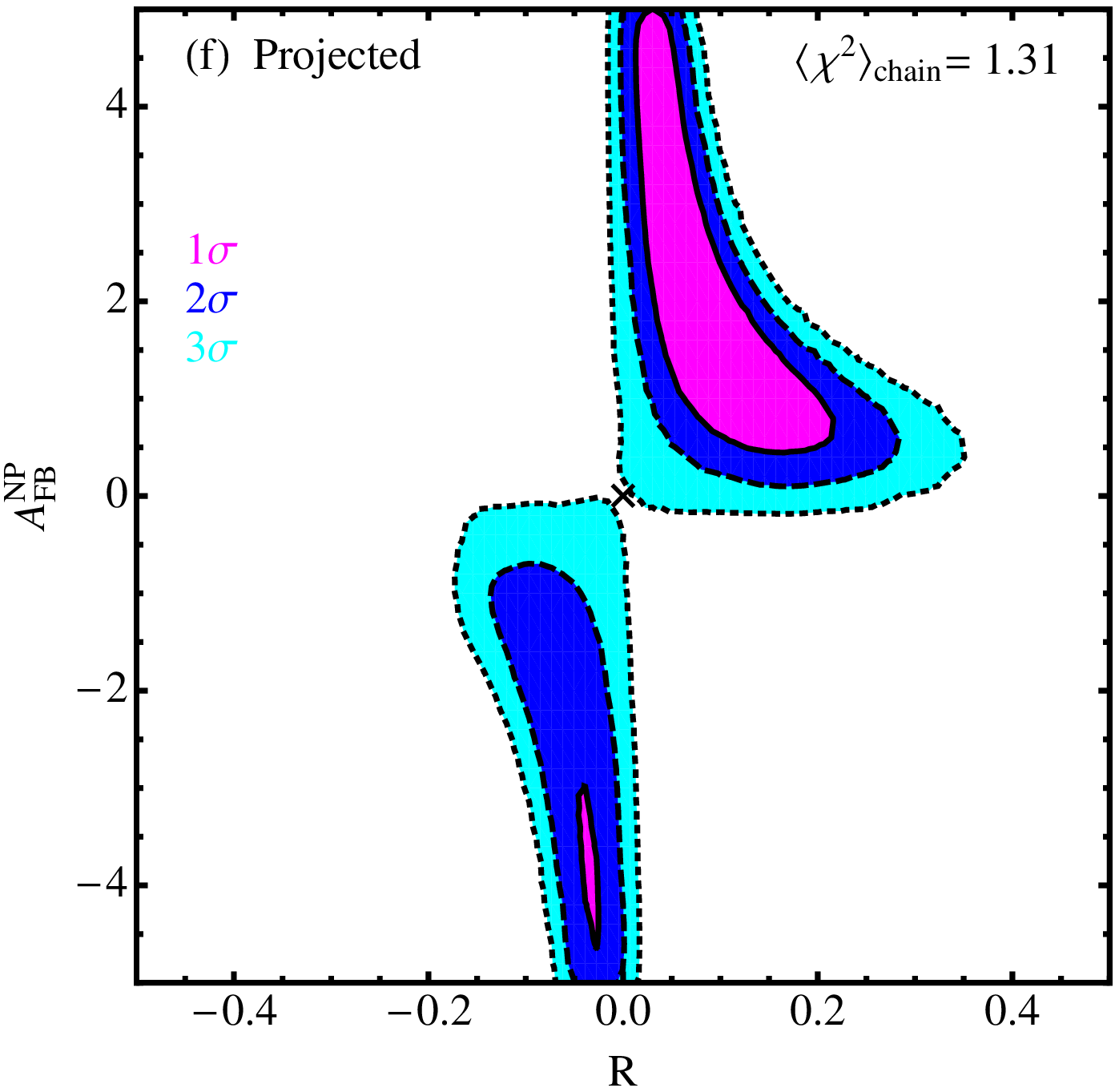}
\caption{(a) Correlation between observed $A_{FB}^{tot}$ and
$\sigma(t\bar{t})$ at the Tevatron with the current integrated luminosities; (b) same for $\int \mathcal{L} dt =
10\,{\rm fb}^{-1}$.  (c) Correlation between observed $A_{FB}^{tot}$ and
fraction of NP contribution to $\sigma(t\bar{t})$ for the current integrated luminosities; (d) same for $\int \mathcal{L} dt = 10\,{\rm fb}^{-1}$.
(e) Correlation between the NP-generated asymmetry $A_{FB}^{NP}$ and $R$
for the current integrated luminosities; (f) same for $\int \mathcal{L}
dt =10\,{\rm fb}^{-1}$.  Here, we do not include the $M_{t\bar t}$ spectrum constraint.  In each case the predicted correlations for
$\int \mathcal{L} dt = 10\,{\rm fb}^{-1}$ assume the same central values of
experimental data.  The crosses
correspond to
the Standard Model predictions of the $A_{FB}$ and $\sigma_{t\bar t}$.
Innermost contours correspond to $1 \sigma$ deviation from minimum-$\chi^2$
solutions; next-to-innermost correspond to $2 \sigma$; outermost correspond
to $3 \sigma$.  Note that these scans are performed by comparing only to the measurements of the total cross section and asymmetry and not the $M_{t\bar{t}}$ distribution.  These should be compared to a SM scan only subject to these two measurements which gives $\left<\chi^2\right>_{chain}=2.56$ for the current luminosity and
$\left<\chi^2\right>_{chain}=6.20$ for $10~{\rm fb}^{-1}$ if the central values are unchanged.
\label{fig:AFB.vs.xsectt}}
\end{figure}

For illustration, we plot these contours in the plane of $A_{FB}^{tot}$ and
$\sigma(t\bar{t})$ in Figs.~\ref{fig:AFB.vs.xsectt}(a).  We
note that the current average values of $A_{FB}$ and $\sigma(t\bar{t})$
are consistent with the SM within the $2\sigma$ level.  With an upgraded
integrated luminosity of $\int \mathcal{L} dt =10\,{\rm fb}^{-1}$ at the
Tevatron, the statistical uncertainty would be reduced significantly; see
Fig.~\ref{fig:AFB.vs.xsectt}(b). The deviation of $A_{FB}$ from
zero is then larger than $3\sigma$. Note that we also allow negative values of
$A_{FB}^{tot}$ in this work, though they are not preferred.
Taking the SM theory prediction, we translate $\sigma(t\bar{t})$
into $R$ defined in Eq.~(\ref{eq:def-afbnp-R}). The correlation of
$A_{FB}^{tot}$ and $R$ is shown in Figs.~\ref{fig:AFB.vs.xsectt}(c)
and (d). Finally, using Eq.~(\ref{eq:AFB}), we obtain the correlation
between $A_{FB}^{NP}$ and $R$ shown in Figs.~\ref{fig:AFB.vs.xsectt}(e)
and (f). Clearly, the smaller $R$ the larger $A_{FB}^{NP}$; see
the $1\sigma$ contour (solid black).

Note that since the MCMC is sensitive to the relative likelihood change in
going between two points, it is sensitive to only the $\Delta \chi^2$ values.
Therefore, the iso-contours of the $p$-values for 1, 2, and $3\sigma$ assume the given model.
To obtain an overall indication of how well the model in question fits
the data, we quote $\langle \chi^2 \rangle_{chain}$, the $\chi^2$ per degree of freedom values averaged
over the entire chain.  In cases where we include the ${d\sigma\over d M_{t\bar t}}|_{bin}$ constraint, $N_{dof} = 3$, otherwise $N_{dof} = 2$.  This quantity is an overall estimate of the model's
consistency with the data.  Generally, values of $\langle \chi^2\rangle_{chain}< 2$ are considered fairly good fits, while values much beyond that are not considered very good.

One might be tempted to search for the parameter set that yields the best fit  to the given data.  However, this is doing so without regard to the level of fine-tuning required to find such a point.  Explicitly, this can be seen as a set of points in parameter space by which the $\chi^2$ value is minimized, ideally to zero.  However, if a small deviation from these points provides a large increase in $\chi^2$, this particular set of points that provide a good fit can be seen as more fine-tuned compared with another solution set without such a steep increase in $\chi^2$.  Therefore, the MCMC approach does take into account the parameter space available that affords a good fit, preferentially solutions with low fine-tuning.

To compare the MCMC results of Fig.~\ref{fig:AFB.vs.xsectt} and subsequent Figures, we ran a MCMC with a pure SM explanation by explicitly setting $A_{FB}^{NP}$ and $\sigma_{t\bar t}^{NP}$ to zero and scanning over Eq.~\ref{eq:ranges} with gaussian priors.  We find that
\begin{equation}
\langle \chi^2\rangle_{chain}^{SM} = \Bigg\{ \begin{array}{ccc}  2.56 & &{\rm Current~Luminosity} \\
 6.20 & & {\rm Projected: }~\int\mathcal{L} dt = 10~{\rm fb }^{-1} \\
 \end{array}
  \Bigg. ,
  \label{eq:chisqsm1}
\end{equation}
where ``Current Luminosity" refers to the measurement of $\sigma_{t\bar t}$ with an integrated luminosity of 4.6 fb$^{-1}$ and the measurement of $A_{FB}$ with an integrated luminosity of 3.2 fb$^{-1}$.  For the projected integrated luminosity of 10 fb$^{-1}$, we assume the central values of $A_{FB}^{tot}$ and $\sigma_{t\bar t}^{tot}$ remain unchanged from the values taken in Eqs.~\ref{eq:cdf} and \ref{eq:ttbar-we} while the uncertainties reduce by a scale factor $\sqrt{{\cal L}}$.  When we examine specific models that could give rise to a larger $A_{FB}$ than the SM, we must also take into account the $M_{t\bar{t}}$ distribution measurement.  To compare these models against the SM, we again run a MCMC with a pure SM explanation scanning over $A_{FB}$ and $\sigma_{t\bar t}$ as above while also scanning over our NLO prediction (seen in Fig.~\ref{fig:mtt}) with gaussian priors for the last seven bins of the CDF $M_{t\bar{t}}$ distribution.  The bin nearest $t\bar{t}$ threshold accounts for the majority of the total cross section.  Since we already include the total cross section in our fit, we do not include this bin in our fit of the $M_{t\bar{t}}$ distribution so that we do not weight the total cross section too heavily.  If we include the measurement of the $M_{t\bar{t}}$ distribution and perform a MCMC scan over the SM, we find
\begin{equation}
\langle \chi^2\rangle_{chain}^{SM} = \Bigg\{ \begin{array}{ccc}  1.75 & &{\rm Current~Luminosity} \\
 4.22 & & {\rm Projected: }~\int\mathcal{L} dt = 10~{\rm fb }^{-1} \\
 \end{array}
  \Bigg. .
  \label{eq:chisqsm}
\end{equation}
Here, ``Current Luminosity" refers to the above values of integrated luminosity for the $\sigma_{t\bar t}$ and $A_{FB}$ measurements and 2.7 fb$^{-1}$ for the measurement of the $M_{t\bar{t}}$ distribution.  For the projected luminosity of 10~fb$^{-1}$, we again assume that the central values of all measurements remain the same while their errors scale as $1/\sqrt{{\cal L}}$.  We note that the values of $\langle \chi^2\rangle_{chain}$ in Eq.~(\ref{eq:chisqsm}) are less than those in Eq.~(\ref{eq:chisqsm1}).  This is because $\langle \chi^2\rangle_{chain}$ is a $\chi^2$ per degree of freedom.  There are two degrees of freedom in Eq.~(\ref{eq:chisqsm1}) and three in Eq.~(\ref{eq:chisqsm}) with the addition of the $M_{t\bar{t}}$ distribution.  The good agreement of the $M_{t\bar{t}}$ distribution in the SM with data (seen in Fig.~\ref{fig:mtt}) causes the $\chi^2$ per degree of freedom to decrease when it is included in the fit.  When comparing models, we can say that if the $\langle \chi^2\rangle_{chain}$ value for a given model is less than that for the SM with the appropriate data into account, the model will provide a better overall fit to the data than the SM.

The forward-backward asymmetry, defined in terms of a ratio of cross sections, 
is very sensitive to the renormalization and factorization scales, $\mu_R$ and $\mu_F$ respectively, at which the cross
sections are evaluated.  The uncertainties in the cross section associated with
those scales can be
considered as an estimate of the size of unknown higher order contributions. 
In this study, we set $\mu_{R}=\mu_{F}=\mu_{0}$ and vary it around the central
value of $\mu_{0}=m_{t}$, where $m_{t}$ is the mass of the top quark.
Typically, a factor of 2 is used as a rule of thumb. Large scale dependence
in the LO cross section can be significantly improved by including
the higher order QCD and EW corrections. In this work we calculate the SM top
pair production cross section with the NLO QCD corrections. 
Unfortunately, the QCD corrections to the $G^\prime$ induced top pair
production are not available yet. Therefore, we calculate the NP contributions
only at LO and rescale them by the $M_{t\bar{t}}$-dependent SM $q\bar{q}$ K-factors. Due to the mismatch between the SM and NP cross sections,
$A_{FB}$ calculated in this way depends on the choice of scale. 

% This is Table I
\begin{table}[b]
\caption{Predicted LO and NLO top pair production cross sections (pb) in the SM
at the Tevatron, with $\mu_0 = m_t=175~{\rm GeV}$. Note that the 
negative $\sigma(gq)$ and $\sigma(g\bar{q})$ cross sections are
due to the negative NLO PDFs~\cite{Nadolsky:2008zw}. } \label{tab:xsec_sm_175}
\begin{tabular}{c|ccc|ccc}
\hline
& \multicolumn{3}{c|}{LO} & \multicolumn{3}{c}{NLO}\tabularnewline
\cline{2-7}
& $\mu_{0}/2$ & $\mu_{0}$ & $2\mu_{0}$ & $\mu_{0}/2$ & $\mu_{0}$ & $2\mu_{0}$
\tabularnewline
\hline
\hline
$\sigma(q\bar{q})$ & 6.82 & 5.01 & 3.79 & 5.70 & 5.56 & 5.04\tabularnewline
\hline
$\sigma(gg)$       & 0.37 & 0.24 & 0.17 & 1.00 & 0.90 & 0.74\tabularnewline
\hline
$\sigma(gq)$       & 0.00 & 0.00 & 0.00 & 0.01 & -0.03 & -0.05\tabularnewline
\hline
$\sigma(g\bar{q})$ & 0.00 & 0.00 & 0.00 & 0.01 & -0.03 & -0.05\tabularnewline
\hline
$\sigma_{tot}$     & 7.19 & 5.26 & 3.96 & 6.72 & 6.39 & 5.69\tabularnewline
\hline
\hline
$\sigma(gg)/\sigma_{tot}$
                 & 0.05 & 0.05 & 0.04 & 0.15 & 0.14 & 0.13\tabularnewline
\hline
$K_{\rm fac}$      &      &      &      & 0.93 & 1.22 & 1.42\tabularnewline
\hline
\end{tabular}
\end{table} 
 
Table~\ref{tab:xsec_sm_175} shows the LO and NLO top quark pair production
cross sections in the SM at the Tevatron. We present the quark annihilation
and gluon fusion processes individually as well as their sum.  The 
CTEQ6.6M~\cite{Nadolsky:2008zw} and CTEQ6L~\cite{Pumplin:2002vw} 
PDF packages are used in the NLO and LO calculations, respectively. 
In the last row we also list the K-factor, defined as the ratio of NLO 
and LO cross sections, for three scales.
 
We argue that the higher order corrections cannot be estimated by a K-factor
(defined as the ratio of NLO and LO cross sections) because the K-factor is
very sensitive to the scale. Furthermore, the gluon fusion channel contributes
much more at the NLO (roughly about $13\sim 15\%$ of total cross section) than
at the LO (only about $5\%$). Hence, one also needs to take account 
of the gluon fusion channel contribution when calculating $A_{FB}$. 

Another uncertainty originates from the top quark mass.  In
Table~\ref{tab:xsec_sm} we show the top pair production cross section for
various top quark masses and three scales.  The central values of NLO theory
calculations for the three masses $m_t=171.0,~172.5,~175.0~\rm{GeV}$ are always ${\cal O}\left(1\sigma\right)$ below the recent CDF results given in Eqs.~(\ref{eq:CDF2009summer}) and~(\ref{eq:CDF2009combined}), suggesting
that the NP should contribute positively to $t\bar{t}$ production.  

\begin{table}
\caption{Predicted LO and NLO $t\bar{t}$ production cross sections (pb) for various top quark 
masses and three scales ($\mu_0/2$, $\mu_0$, $2\mu_0$ with $\mu_0=m_t$) at the
Tevatron.}\label{tab:xsec_sm}  
\begin{tabular}{c|ccc|ccc}
\hline
 & \multicolumn{3}{c|}{LO} & \multicolumn{3}{c}{NLO}     \tabularnewline
 \hline 
$m_{t}({\rm GeV})$ & $\mu_{0}/2$ &  $\mu_{0}$ & $2\mu_{0}$ 
& $\mu_{0}/2$ &  $\mu_{0}$ & $2\mu_{0}$\tabularnewline
\hline
171.0  & 8.08 & ~5.91 & ~4.45 & ~7.61 & ~7.23 & ~6.44\tabularnewline
172.0  & 7.84 & ~5.74 & ~4.32 & ~7.37 & ~7.01 & ~6.24\tabularnewline
172.5  & 7.72 & ~5.66 & ~4.26 & ~7.25 & ~6.90 & ~6.14\tabularnewline
173.0  & 7.61 & ~5.57 & ~4.19 & ~7.14 & ~6.79 & ~6.05\tabularnewline
174.0  & 7.40 & ~5.42 & ~4.07 & ~6.92 & ~6.58 & ~5.86\tabularnewline
175.0  & 7.19 & ~5.26 & ~3.96 & ~6.72 & ~6.39 & ~5.69\tabularnewline
176.0  & 6.98 & ~5.11 & ~3.84 & ~6.51 & ~6.19 & ~5.51\tabularnewline
177.0  & 6.78 & ~4.96 & ~3.73 & ~6.31 & ~6.01 & ~5.35\tabularnewline
\hline
\end{tabular}
\end{table}

In the following sections, we study a few interesting new physics models 
which can generate a significant deviation from the SM expectation for
$A_{FB}$ in the top quark pair production channel. We also comment
on the scale dependence in each new physics model. Without losing generality,
in the rest of this paper, we set $m_t=175~\rm{GeV}$.

%--------------------------------------------------------------------%
%  Section III                                                       %
%                                                                    %
%   S-channel diagram for g-prime boson                              %
%--------------------------------------------------------------------%
\section{Exotic gluon\label{sec:G-prime}}
We begin with an exotic gluon ($G^{\prime}$) model, as the other models can be easily
derived from the $G^\prime$ model result.  In Sec.~\ref{sec:xsec_asymm} we present analytic formulae for $\sigma_{t\bar{t}}$ and $A_{FB}$.  We calculate its width in Sec.~\ref{sec:width}.  In Secs.~\ref{sec:lefthanded}, \ref{sec:axigluon}, and \ref{sec:other_couplings} we perform MCMC scans over parameters in several $G^\prime$ scenarios subject to the experimental constraints.

\subsection{Differential cross section and asymmetry}\label{sec:xsec_asymm}
The $G^{\prime}$
boson couples to the SM quarks also via the QCD strong interaction,
\begin{eqnarray}
G^{\prime}q\bar{q} & : &
 ig_{s}t^{A}\gamma^{\mu}\left(f_{L}P_{L}+f_{R}P_{R}\right),\\
G^{\prime}t\bar{t} & : &
 ig_{s}t^{A}\gamma^{\mu}\left(g_{L}P_{L}+g_{R}P_{R}\right),
\end{eqnarray}
where we normalize the interaction to the QCD coupling, $g_s$, and use $q$ to denote light quarks of the first two generations.
Such an exotic gluon can originate from an extra-dimensional model
such as the Randall-Sundrum (RS) model~\cite{Djouadi:2009nb},
chiral color model~\cite{Pati:1975ze,Hall:1985wz,Frampton:1987dn,
Frampton:1987ut,Bagger:1987fz,Cuypers:1990hb,Cao:2005ba,Carone:2008rx,
Ferrario:2008wm,Martynov:2009en},
or top composite model~\cite{Contino:2006nn}. As discussed
below, the axial coupling of $G^{\prime}$ to the SM quarks is necessary
to create a forward-backward asymmetry. In the extra-dimensional model,
such non-vector coupling of Kaluza-Klein gluons to fermions arises
from localizing the left- and right-handed fermions at different locations
in the extra dimension.

The differential cross section
with respect to the cosine of the top quark polar angle $\theta$
in the $t\bar{t}$ center-of-mass (c.m.) frame is 
\begin{equation}
\frac{d\hat{\sigma}(G^{\prime})}{d\cos\theta}=
\mathcal{A}_{SM}+\mathcal{A}_{INT}^{G^{\prime}}+\mathcal{A}_{NPS}^{G^{\prime}} 
\label{eq:dsdz}
\end{equation}
where 
\begin{eqnarray}
\mathcal{A}_{SM} & = &
 \frac{\pi\beta\alpha_{s}^{2}}{9\hat{s}}\left(2-\beta^{2}
 +\left(\beta\cos\theta\right)^{2}\right), \label{eq:dsdz_sm}\\
\mathcal{A}_{INT}^{G^{\prime}} & = &
 \frac{\pi\beta\alpha_{s}^{2}}{18\hat{s}}
 \frac{\hat{s}\left(\hat{s}-m_{G^{\prime}}^{2}\right)}
{\left(\hat{s}-m_{G^{\prime}}^{2}\right)^{2}
+m_{G^{\prime}}^{2}\Gamma_{G^{\prime}}^{2}}
\left(f_{L}+f_{R}\right)\left(g_{L}+g_{R}\right)\nonumber \\
 & \times & 
 \left\{ \left(2-\beta^{2}\right)+
 2\frac{\left(f_{L}-f_{R}\right)\left(g_{L}-g_{R}\right)}{\left(f_{L}+f_{R}
 \right)\left(g_{L}+g_{R}\right)}\beta\cos\theta+
 \left(\beta\cos\theta\right)^{2}\right\} ,\label{eq:dsdz_int}\\
\mathcal{A}_{NPS}^{G^{\prime}} & = &
 \frac{\pi\beta\alpha_{s}^{2}}{36\hat{s}}
 \frac{\hat{s}^{2}}{\left(\hat{s}-m_{G^{\prime}}^{2}\right)^{2}
 +m_{G^{\prime}}^{2}\Gamma_{G^{\prime}}^{2}}\left(f_{L}^{2}+f_{R}^{2}\right)
 \left(g_{L}^{2}+g_{R}^{2}\right)\nonumber \\
 & \times & 
 \left\{ 1+\frac{2g_{L}g_{R}}{g_{L}^{2}+g_{R}^{2}}\left(1-\beta^{2}\right)
 +2\frac{\left(f_{L}^{2}-f_{R}^{2}\right)\left(g_{L}^{2}-g_{R}^{2}\right)}
 {\left(f_{L}^{2}+f_{R}^{2}\right)\left(g_{L}^{2}+g_{R}^{2}\right)}
 \beta\cos\theta+\left(\beta\cos\theta\right)^{2}\right\}.
 \label{eq:dsdz_np}
\end{eqnarray}
Here the angle $\theta$ is defined as the angle between the direction of
motion of the top quark and the direction of motion of the incoming quark
(e.g., the $u$-quark) in the $t \bar t$ c.m.\ system.  The
subscripts ``SM'', ``INT'' and  ``NPS'' denote the contribution from the SM,
the interference between the SM and NP, and the NP amplitude squared. 
For the $G^\prime$ model, the SM contribution is from the gluon-mediated
$s$-channel diagram, the NPS contribution from the exotic gluon $G^{\prime}$-%
mediated diagram, and the INT contribution from the interference between the
two.  The squared c.m.\ energy of the $t \bar t$ system is $\hat{s} =
\left(p_{q}+p_{\bar{q}}\right)^{2}$, and $\beta=\sqrt{1-4m_{t}^{2}/\hat{s}}$ is
the top quark velocity in the $t \bar t$ c.m.\ system. 

The forward-backward asymmetry of the top quark in the $t \bar t$ c.m.\ frame
is defined as 
\begin{equation}
A_{FB}^{t\bar{t}}=\frac{\sigma_{F}-\sigma_{B}}{\sigma_{F}+\sigma_{B}},
\label{eq:AFB_def}
\end{equation}
where 
\begin{equation}
\sigma_{F}\equiv\int_{0}^{1}\frac{d\sigma}{d\cos\theta}d\cos\theta,
\qquad\sigma_{B}\equiv\int_{-1}^{0}\frac{d\sigma}{d\cos\theta}d\cos\theta.
\label{eq:AFB_def2}
\end{equation}
We further parameterize the differential cross section
$d\sigma/d\cos\theta$ as follows: 
\begin{equation}
\frac{d\sigma_i}{d\cos\theta}=A_i+B_i\cos\theta+C_i\cos^{2}\theta,
\label{eq:dsdz_def}
\end{equation}
where the subindex $i$ denotes ``SM'', ``INT'' and ``NPS''. 
Hence, after integrating over the angle $\theta$, we obtain the asymmetry
and total cross section
\begin{equation}
A_{FB}=\frac{\sum_i B_i}{\sum_i (2A_i+\frac{2}{3} C_i)}~,
\quad{\rm and}\quad
\sigma_{tot}=\sum_i \left(2A_i+\frac{2C_i}{3}\right)~,
\label{eq:AFB_para}
\end{equation}
where the sums are over the SM, INT and NPS terms.  In reality the incoming
quark could originate from either a proton or an anti-proton, but it
predominantly comes from a proton due to large valence quark parton
distribution functions. Taking the quark from the anti-proton and the anti-quark from the proton contributes less than 1\% of the total $t\bar{t}$ cross section.  Therefore, in $\bar p p$ collisions at the Tevatron one
can choose the direction of the proton to define the forward direction. 

Now let us comment on a few interesting features of the asymmetry and cross section
generated by the INT and NPS effects individually, because both effects
are sensitive to different new physics scales: the former to a higher NP
scale and the latter to a lower scale.   First, we note that the asymmetry is
sensitive to the ratio of coupling (squared) differences and sums for the INT
(NPS) effects, e.g.,
\begin{eqnarray}
A_{FB}^{INT} & \propto & \frac{(f_L-f_R)(g_L-g_R)}{(f_L+f_R)(g_L+g_R)}
\times \frac{2\left< \beta \right>}
{2(2-\left<\beta^2\right>)+\frac{2}{3}\left<\beta^2\right>},
\label{eq:AFBINT}\\
A_{FB}^{NPS} & \propto & 
\frac{(f_L^2-f_R^2)(g_L^2-g_R^2)}{(f_L^2+f_R^2)(g_L^2+g_R^2)}
\times \frac{2\left< \beta \right>}
{2\left[1+\left(1-\left<\beta^2\right>\right)\left(2g_L g_R\right)/\left(g_L^2+g_R^2\right)\right]+\frac{2}{3}\left<\beta^2\right>},
\label{eq:AFBNPS}
\end{eqnarray}
where $\left<\beta\right>$ and $\left<\beta^2\right>$ denote the averaged
$\beta$ and $\beta^2$ after integration over the angle $\theta$ and convolution
of the partonic cross section with parton distribution functions.

\begin{figure}
\includegraphics[scale=0.6]{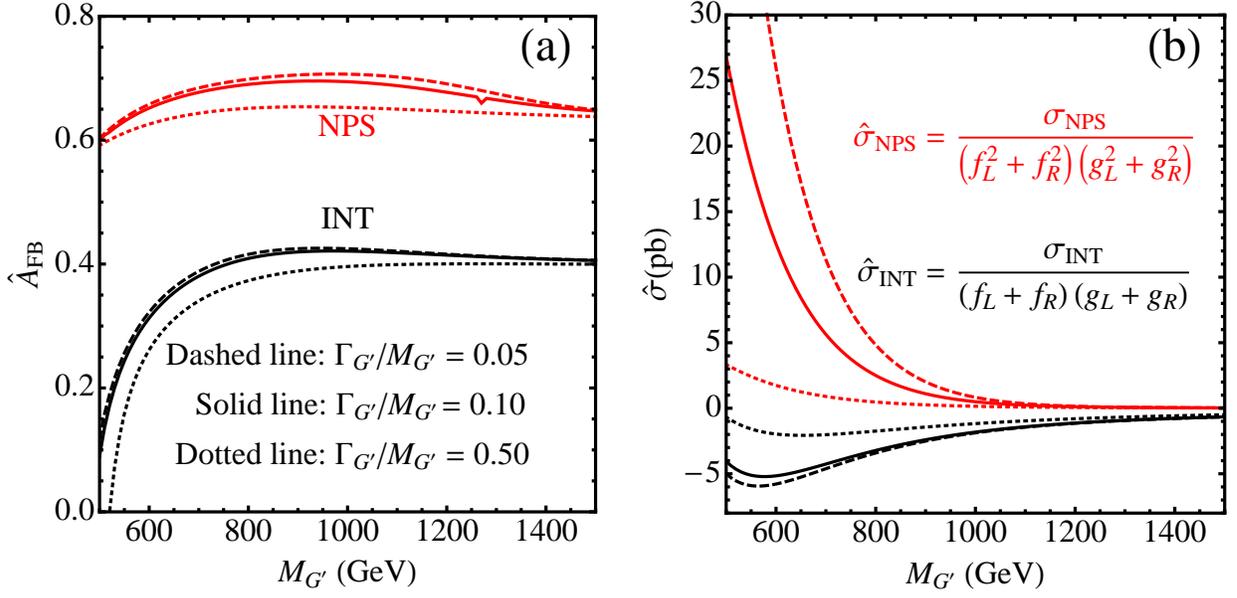}
\caption{(a) Reduced asymmetries defined in Eqs.~(\ref{eq:reduced_AFBINT}) and~(\ref{eq:reduced_AFB_NPS}): $\hat{A}_{FB}^{INT}$ (lower three curves) and $\hat{A}_{FB}^{NPS}$ (upper three curves); (b) Reduced cross sections defined in Eqs.~(\ref{eq:reduced_xse_INT}) and~(\ref{eq:reduced-xsec-NPS}): $\hat{\sigma}^{INT}$ (lower three curves) and $\hat{\sigma}^{NPS}$ (upper three curves).
\label{fig:theo-int-nps}}
\end{figure}

To make the physics source of the asymmetry more transparent, we define
the reduced asymmetry ($\hat{A}_{FB}$) and reduced cross section $\hat{\sigma}$
as follows:
\begin{eqnarray}
\hat{A}_{FB}^{INT} & = & A_{FB}^{INT}\Biggl/
\frac{(f_L-f_R)(g_L-g_R)}{(f_L+f_R)(g_L+g_R)} 
\label{eq:reduced_AFBINT}\\
\hat{A}_{FB}^{NPS} & = & A_{FB}^{NPS}\Biggl/
\frac{(f_L^2-f_R^2)(g_L^2-g_R^2)}{(f_L^2+f_R^2)(g_L^2+g_R^2)}
\label{eq:reduced_AFB_NPS}\\
\hat{\sigma}^{INT} & = & \frac{\sigma^{INT}}
{(f_L+f_R)(g_L+g_R)} 
\label{eq:reduced_xse_INT}\\
\hat{\sigma}^{NPS} & = & \frac{\sigma^{NPS}}
{(f_L^2+f_R^2)(g_L^2+g_R^2)} 
\label{eq:reduced-xsec-NPS}
\end{eqnarray}
The reduced asymmetries and cross sections are easily computed
universal functions that
allow us to focus on two separate limiting cases; the new physics contribution
to $\sigma_{t\bar{t}}$ and $A_{FB}$ is primarily from the INT term if it is 
produced by a heavy
resonance that interferes with the SM production process.  If the new
physics is due to a resonance that doesn't interfere with the SM production, then
the new contribution to $\sigma_{t\bar{t}}$ and $A_{FB}$ is given by the NPS term.
One simply has to multiply the reduced asymmetry or cross section by the
appropriate combination of couplings to obtain the full new physics
contribution to $\sigma_{t\bar{t}}$ and $A_{FB}$.
In Fig.~\ref{fig:theo-int-nps} we plot the reduced asymmetry (a) and the 
reduced cross section (b) as functions of $m_{G^\prime}$ for various
choice of $\Gamma_{G^\prime}/m_{G^\prime}$. 
The reduced asymmetry generated by the INT effects increases rapidly with
increasing $m_{G^\prime}$ and finally reaches its maximal value $\sim 0.4$.
The reduced asymmetry generated by the NPS effects is large, typically 
around 0.6-0.7. As expected, the reduced cross section of the NPS effects
is alway positive; cf.\ the upper three curves in Fig.~\ref{fig:theo-int-nps}(b). On the 
other hand, the reduced cross section of the INT effects is always negative
due to ($\shat-m_{G^\prime}^2$) in the numerator of Eq.~(\ref{eq:dsdz_int}).
Both reduced cross sections, especially the NPS effects, are sensitive to
the $G^\prime$ decay width.  They both go to zero when $G^\prime$ decouples.  

% This is Figure 3

% This is Figure 4
\begin{figure}
\includegraphics[scale=0.6]{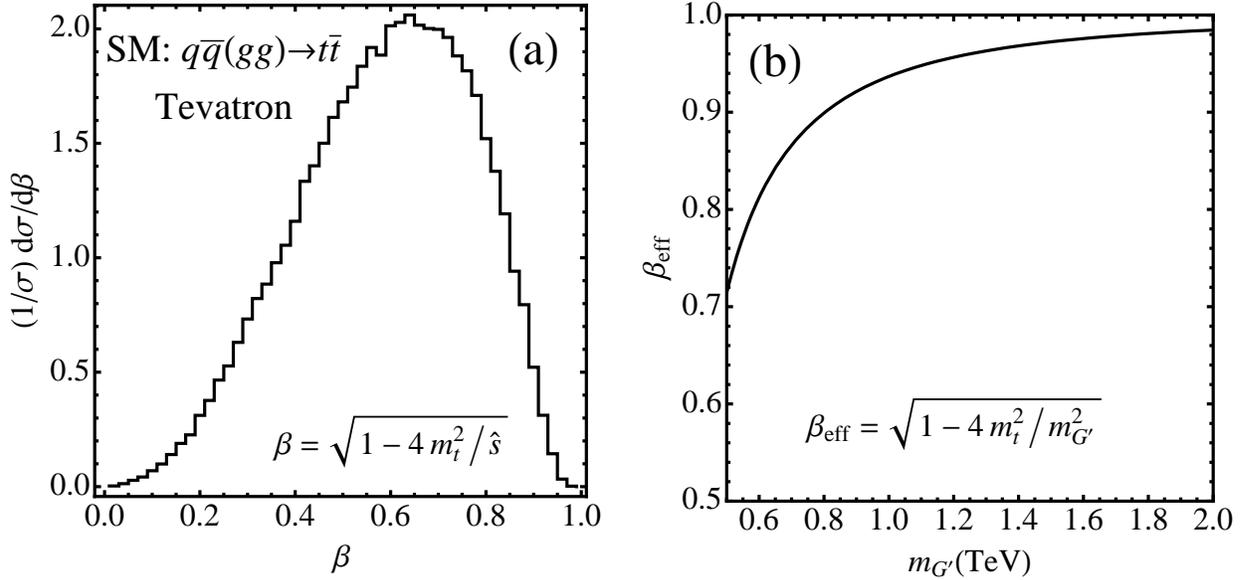}
\caption{
(a) Normalized differential cross section for $t\bar{t}$ production as a
function of $\beta=\sqrt{1-4 m_t^2/\shat}$ of a 175~GeV top quark 
at the Tevatron;
(b) $\beta_{\rm eff} =\sqrt{1-4 m_t^2/m_{G^\prime}^2}$ as a function of heavy
resonance $G^\prime$ mass.
\label{fig:theo-beta}}
\end{figure}    

The difference between the two reduced asymmetries can be easily
understood from the $\beta$ distribution shown in Fig.~\ref{fig:theo-beta}.
Figure~\ref{fig:theo-beta}(a) shows the normalized differential cross section
with respect to $\beta$ for 175~GeV top quark production in the SM at the 
Tevatron, which peaks around $\beta\sim 0.65$. The INT effects only slightly
shift the peak position. Substituting $\left<\beta\right>\sim 0.65$ into
Eqs.~(\ref{eq:AFBINT}) and (\ref{eq:reduced_AFBINT}), we obtain 
${A}_{FB}^{INT} \simeq 0.4$. On the contrary, the NPS effects prefer 
a much larger $\beta$ enforced by the heavy $G^\prime$ resonance. We plot
$\beta_{\rm eff}=\sqrt{1-4 m_t^2/m_{G^\prime}^2}$ in
Fig.~\ref{fig:theo-beta}(b), where $\beta_{\rm eff}\sim 0.98$ for a 2~TeV
$G^\prime$. Such a large $\beta_{\rm eff}$ leads to the large value of
$\hat{A}_{FB}^{NPS}$ in Fig.~\ref{fig:theo-int-nps}. For an extremely heavy
$G^\prime$, $\beta_{\rm eff}$ is equal to 1, yielding the well-known maximal 
$\hat{A}_{FB}^{NPS}=3/4$.
When both INT and NPS effects contribute, one cannot factorize out
the couplings as in Eqs.~(\ref{eq:AFBINT}-\ref{eq:reduced-xsec-NPS}) due to
the presence of both linear and quadratic coupling terms.

%  subsection : G-prime decay width                           

\subsection{$G^\prime$ decay width}\label{sec:width}
The $\mathcal{A}_{NPS}$ term contributes significantly in the vicinity
of $m_{G^{\prime}}$ where the decay width $\Gamma_{G^{\prime}}$
plays an important role. Hence, it is very important to use an accurate
decay width in the parameter scan. We consider the case that the $G^{\prime}$
boson decays entirely into SM quark pairs, yielding the following
partial decay width~\cite{Ferrario:2008wm}:
\begin{eqnarray}
 &  & \Gamma(G^{\prime}\to t\bar{t})=
 \frac{\alpha_{s}}{12}m_{G^{\prime}}\left[\left(g_{L}^{2}+g_{R}^{2}\right)
 \left(1-\frac{m_{t}^{2}}{m_{G^{\prime}}^{2}}\right)+6g_{L}g_{R}
 \frac{m_{t}^{2}}{m_{G^{\prime}}^{2}}\right]\\
 &  & \Gamma(G^{\prime}\to b\bar{b})=
 \frac{\alpha_{s}}{12}\left(g_{L}^{2}+g_{R}^{2}\right)m_{G^{\prime}},\\
 &  & \Gamma(G^{\prime}\to\sum q\bar{q})=
 N_{f}\frac{\alpha_{s}}{12}\left(f_{L}^{2}+f_{R}^{2}\right)m_{G^{\prime}},
\end{eqnarray}
where $N_{f}=4$ denotes the light quark flavors and we have assumed that $b_R$ couples to $G^\prime$ with the same strength as $t_R$.  In the limit of
$M_{G^{\prime}}\gg m_{t}$, the total decay width of $G^{\prime}$ is 
\begin{equation}
\Gamma_{G^{\prime}}=\frac{\alpha_{s}}{6}m_{G^{\prime}}
\left[\left(g_{L}^{2}+g_{R}^{2}\right)
    +2\left(f_{L}^{2}+f_{R}^{2}\right)
\right].
\end{equation}
When the couplings $g_{L/R}\simeq f_{L/R}$ are of order 1,
$\Gamma_{G^{\prime}}\simeq\alpha_{s}M_{G^{\prime}}\simeq0.1\, M_{G^{\prime}}$.
When $g_{L/R}\approx f_{L/R}\sim3$, $\Gamma_{G^{\prime}}\sim M_{G^{\prime}}$.
In the following parameter scan we vary the couplings of the $G^{\prime}$
boson in the range of $-3$ to $3$ when all couplings are present but $-5$ to $5$ when only two are non-zero. 

%  subsection : G-prime  Mass = 1000 GeV                           

\subsection{Left-handed $G^\prime$: $f_R=g_R=0$}\label{sec:lefthanded}

Since there are five independent parameters (four couplings and the
$G^{\prime}$ boson mass) in Eqs.~(\ref{eq:dsdz}-\ref{eq:dsdz_np}),
we turn off the right-handed
couplings $f_{R}$ and $g_{R}$ in order to make the physics origin of the
asymmetry more transparent.   We first consider $m_{G^{\prime}}= 1000\,{\rm GeV}$ in Sec.~\ref{sec:1TeV} and then $m_{G^{\prime}}= 2000\,{\rm GeV}$ in Sec.~\ref{sec:2TeV}.  We will comment on non-zero $f_R$ and $g_R$ in Secs.~\ref{sec:axigluon} and \ref{sec:other_couplings}.

\subsubsection{$m_{G^{\prime}}= 1000\,{\rm GeV}$}\label{sec:1TeV}

% This is Figure 5
\begin{figure}
\includegraphics[scale=0.5]{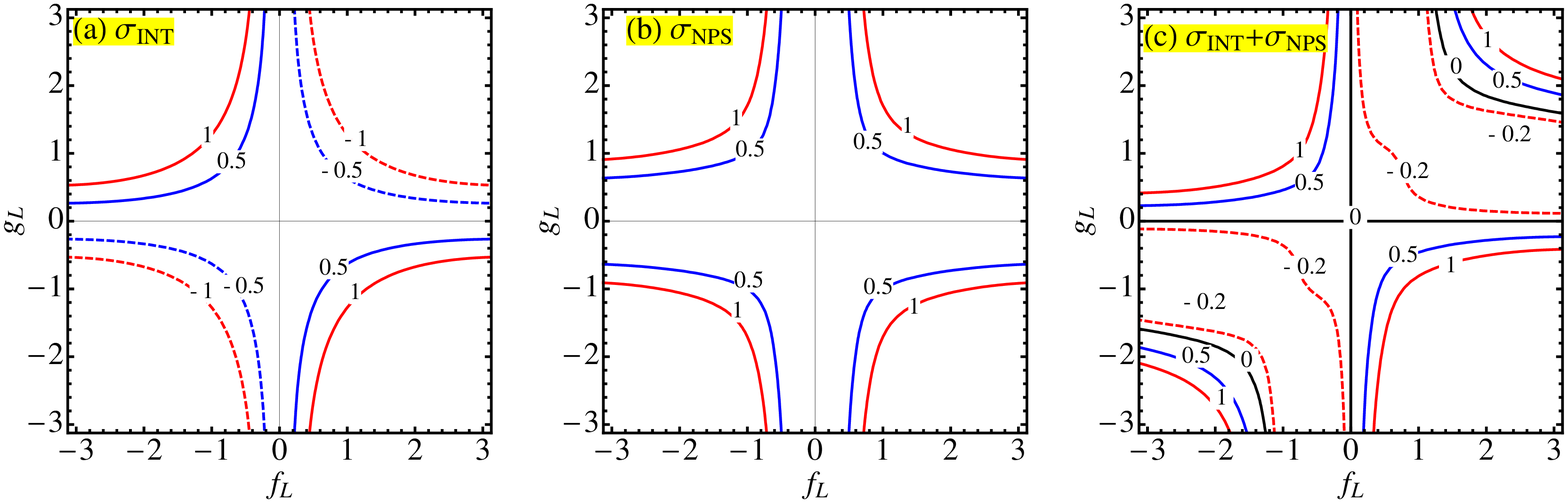}\\
\includegraphics[scale=0.5]{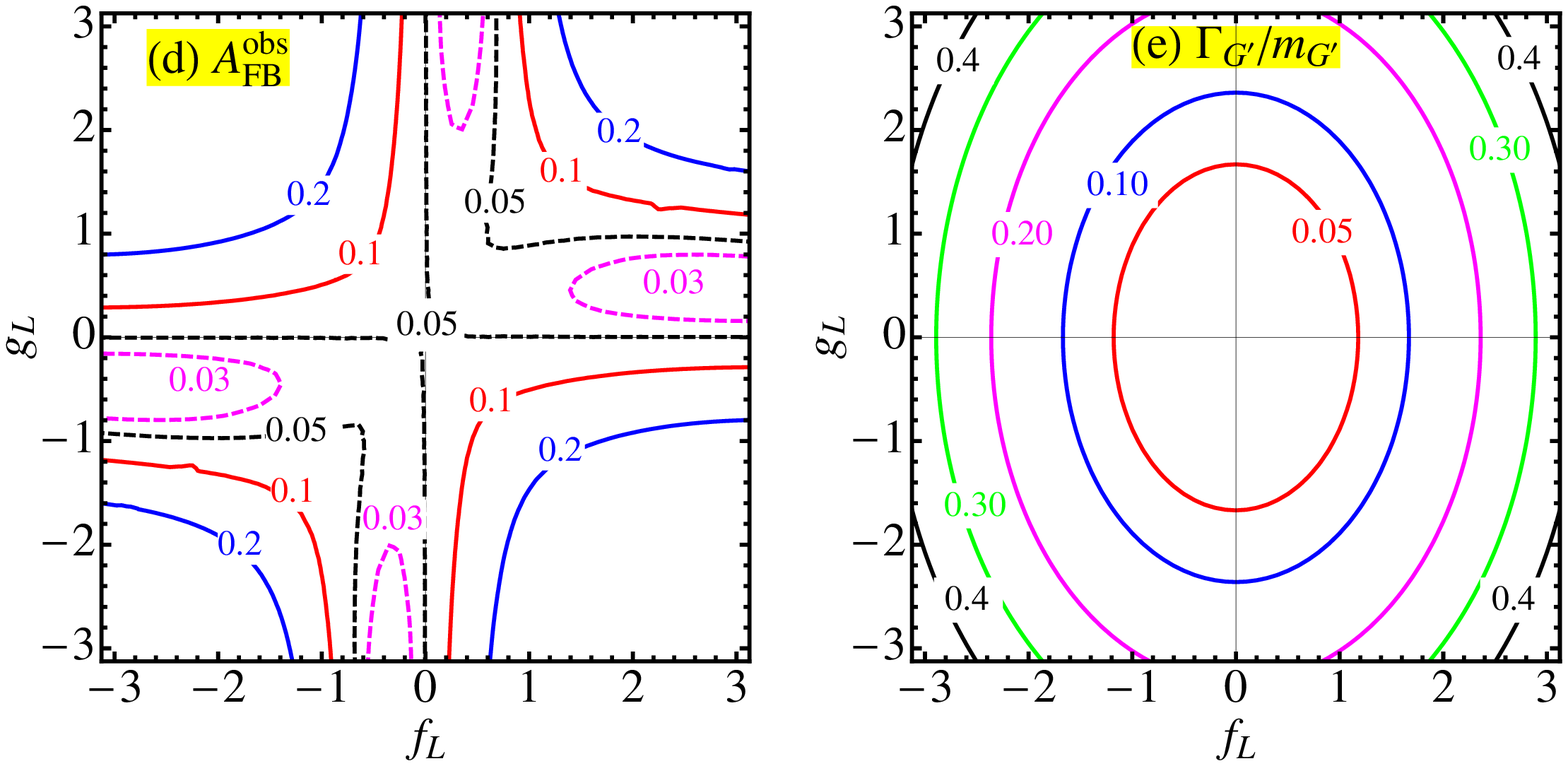}\\
\caption{Theoretical prediction contours in the plane of $f_{L}$ and $g_{L}$ with $f_{R}=g_{R}=0$ for a $1~{\rm TeV}$ $G^\prime$:
(a) contours of cross section (pb) induced by the INT effects,
(b) contours of cross section induced by the NPS effects,
(c) contours of net cross section of the INT and NPS effects,
(d) contours of the observed asymmetry,
(e) contours of $\Gamma_{G^{\prime}}/m_{G^{\prime}}$.
\label{fig:theo-gp1000}}
\end{figure} 

We first examine theoretical predictions of $A_{FB}^{obs}$ and $\sigma_{INT}$ 
and $\sigma_{NPS}$ before we perform a MCMC scan over the parameters. 
By ``theoretical'' we mean that the asymmetry and the top pair production cross
section are calculated independently, without regard to their correlation.
Fig.~\ref{fig:theo-gp1000}(a) displays the cross section contours
generated by the INT effects ($\sigma_{INT}$). The INT effects
could be either positive or negative, depending on the sign of the coupling
product $f_L g_L$. The INT effects dominate in the region of 
$\sqrt{\hat{s}} < m_{G^\prime}$, so their contribution to the top pair
production cross section can be written as
\begin{equation}
    \sigma_{INT} \propto -\left(m_{G^\prime}^2-\shat\right) f_L g_L.
\end{equation}
This expression thus yields a positive contribution to the cross section when
$f_L g_L < 0$ (i.e., the second and fourth quadrants in Fig.\
\ref{fig:theo-gp1000}) and a
negative contribution when $f_L g_L > 0$ (i.e., the first and third quadrants). 

On the contrary, the NPS contribution is always positive; 
see Fig.~\ref{fig:theo-gp1000}(b). Since the NPS effects contribute
mainly in the vicinity of $m_{G^\prime}$, i.e. $\shat\simeq m_{G^\prime}^2$,
their contribution to the top pair production cross section can be written
as follows:
\begin{equation}
\sigma_{NPS} \propto \frac{f_L^2 g_L^2}{\Gamma_{G^\prime}^2} 
      \sim \frac{f_L^2 g_L^2}{m_{G^\prime}^2 (2 f_L^2 + g_L^2)^2}.
\end{equation}
Hence, the contour pattern of $\sigma_{NPS}$ is determined by 
$f_L^2 g_L^2 / (2 f_L^2 + g_L^2)^2$. 
Fig.~\ref{fig:theo-gp1000}(c) shows the competition between the INT and NPS
effects.  For a $1~{\rm TeV}$ $G^\prime$, the INT generally dominates over the NPS.  Note that the contours of net cross section
are not symmetric between $f_L$ and $g_L$ due to the width effects.
In Fig.~\ref{fig:theo-gp1000}(d), we see that, except for small couplings in the upper-right and lower-left quadrants, a positive asymmetry can be generated in all four quadrants.  In the upper-right and lower-left quadrants, for $\left| f_L \right|,\left| g_L \right| \gsim 3$, the NPS term is large enough to generate a positive $A_{FB}$.
Note that negative values of $A_{FB}$, although not plotted in Fig.~\ref{fig:theo-gp1000}(d), are still
consistent with the Tevatron data within $3\sigma$ C.L. and are included
in the following analysis.

% This is Figure 6
\begin{figure}
\includegraphics[scale=0.45]{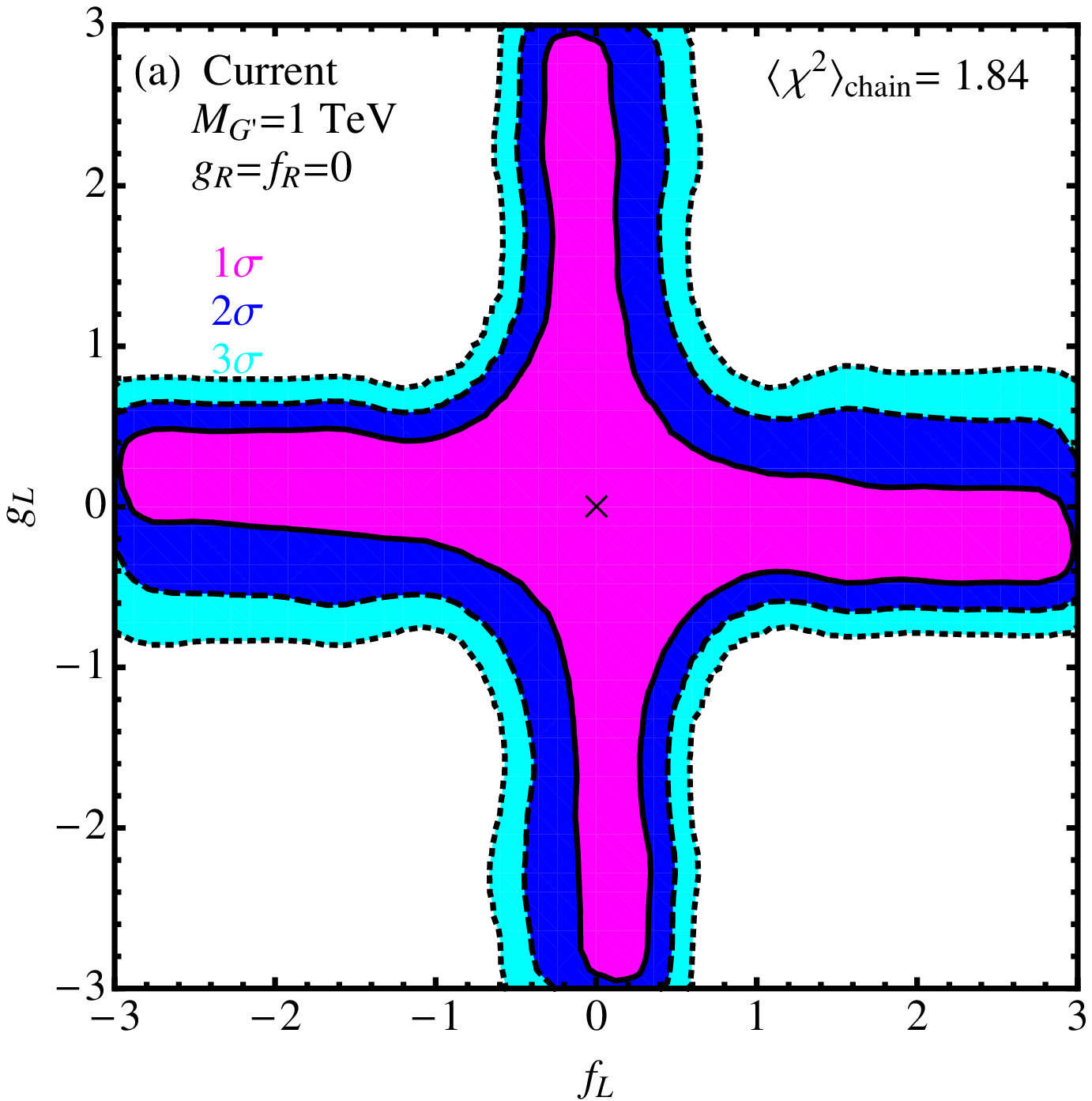}~~~~
\includegraphics[scale=0.45]{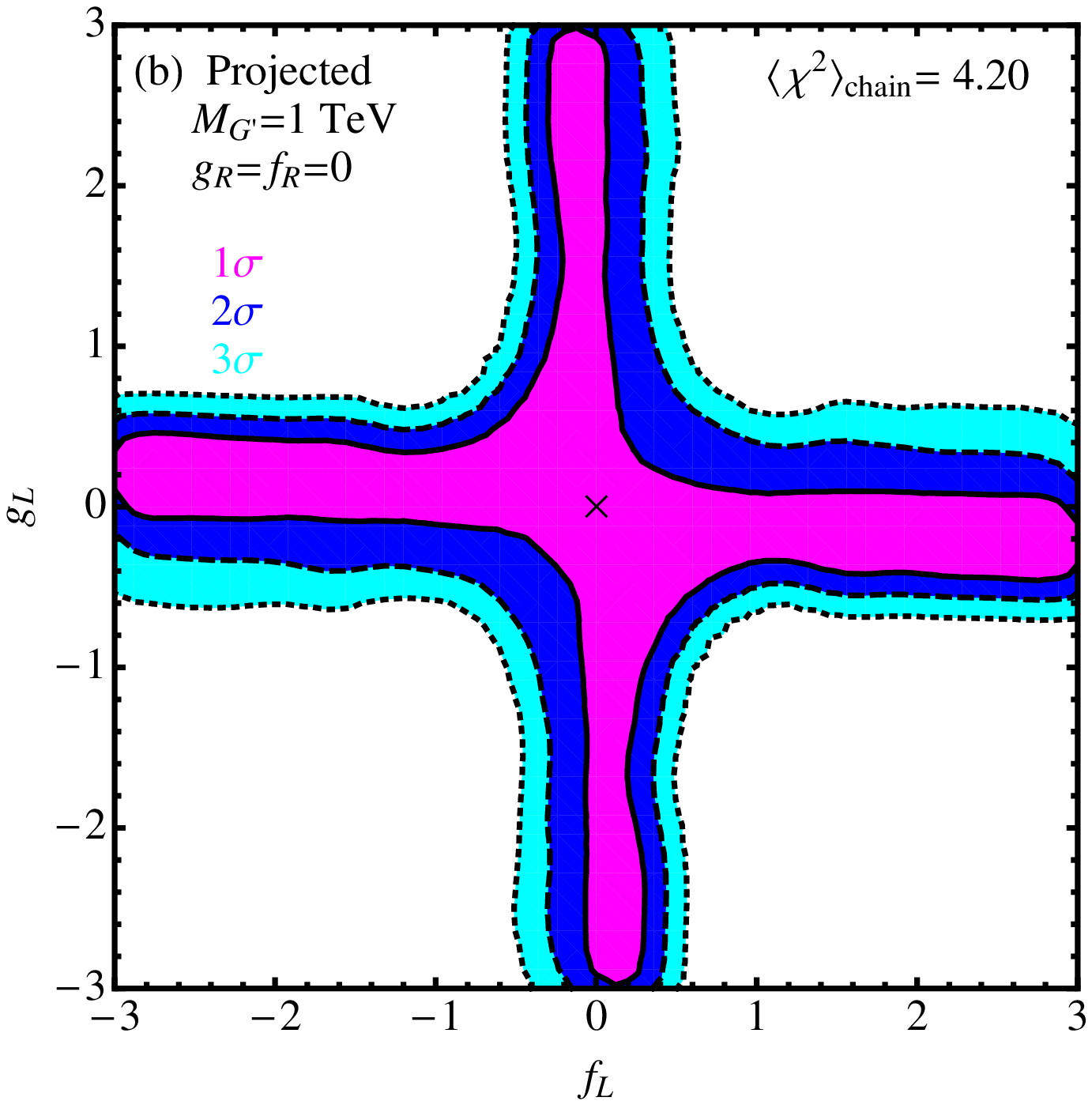}
\caption{Correlation of couplings for $m_{G^{\prime}}=1000\,{\rm GeV}$ with $f_{R}=g_{R}=0$.  Figures
on left are for the current integrated luminosity while those on right are for $\int \mathcal{L} dt
=10\,{\rm fb}^{-1}$.  The regions inside the contours are preferred,
corresponding to $1 \sigma$ (innermost), $2 \sigma$ (next-to-innermost), and
$3 \sigma$ (outermost).
The crosses correspond to $f_L = g_L = 0$.
\label{fig:gp1000}}
\end{figure}

Now we perform a MCMC scan over the parameter space after combining measurements of the $t\bar{t}$ asymmetry, the total cross section, and $\frac{d\sigma}{dM_{t\bar{t}}}$.  To obtain the $M_{t\bar{t}}$ distribution, we separate the contribution from the $q\bar{q}$ initial state (which includes the NP that we analyze) from that of the $gg$ initial state, noting that the $gq$ and $g\bar{q}$ contributions are negligible as seen in Table~\ref{tab:xsec_sm_175}.  We multiply these leading order results by the SM K-factors, $K_{q\bar{q}}$ and $K_{gg}$ respectively, which are obtained by using the Monte Carlo program MCFM~\cite{Campbell:2000bg} to calculate the full NLO SM differential cross section.  Each K-factor itself differs as a function of $M_{t\bar{t}}$ (as seen in \cite{Ahrens:2009uz}) and so we weight each bin in the $M_{t\bar{t}}$ distribution by the appropriate K-factors.  We vary the scale $\mu_0$ at which we evaluate the NLO differntial cross section between $m_t/2$ and $2m_t$ which gives a range of K-factors for each $M_{t\bar{t}}$ bin.  This is used in our fits as our estimate of the theoretical uncertainty.  This uncertainty is about 10\% in the first six bins and around 15\% in the last bin.  Observe that this procedure, when NP effects are decoupled, reproduces the exact NLO SM differential cross section seen in Fig.~\ref{fig:mtt}.  
In this and subsequent Figures of MCMC distributions, we adopt flat priors in
all variables scanned.  The priors for the SM-only contribution to the $t\bar t$ cross
section and $A_{FB}$ are given in Eq.~(\ref{eq:ranges}).
Fig.~\ref{fig:gp1000}(a) displays the parameter space consistent with both
measurements at the $1\sigma$ (innermost region), $2\sigma$ (next-to-innermost region) and
$3\sigma$ (outermost region) level, respectively, for $m_{G^\prime}= 1000 ~\rm{GeV}$.
Remember, the iso-contours of the $p$-values for 1, 2, and $3\sigma$ assume
the given model, while the $\langle \chi^2\rangle_{chain}$ value gives an
indication of the overall fit.  In this case, we get a somewhat worse fit to both
experimental results than in the SM, with
$\left<\chi^2\right>_{\rm chain}=1.84$.  Fig.~\ref{fig:gp1000}(b) shows the
estimated parameter space contours with an integrated luminosity of
$10~{\rm fb}^{-1}$, assuming the central values of both experimental
measurements are not changed.  
The quality of the fit is marginally better than in the SM, with 
$\langle \chi^2\rangle_{chain}=4.20$ vs.\ 4.22 in the SM.
We observe that the regions where $f_L$ or $g_L$ are small provide the best fit.  The boundaries of all three contours can be understood from the theoretical predictions in Figs.~\ref{fig:theo-gp1000}(c) and (d).  To explain the discrepancy in the total cross section and in the asymmetry, values of $f_L$ and $g_L$ in the top-left or bottom-right quadrants would be preferred.  However, couplings here inevitably worsen the $M_{t\bar{t}}$ distribution.  In the top-right and bottom-left quadrants the fit to the $M_{t\bar{t}}$ distribution is improved for small couplings ($\left| f_L \right|,\left| g_L \right| \lsim 1$) but the agreement with the total cross section is slightly worse and an asymmetry smaller than the SM value is generated.  For intermediate couplings in these two quadrants ($1\lsim \left| f_L \right|,\left| g_L \right| \lsim 2$), the fit to the $M_{t\bar{t}}$ distribution is improved but the total cross section is reduced too much and the asymmetry is not improved significantly.  Eventually, at large values of the couplings in these quadrants ($\left| f_L \right|,\left| g_L \right| \gsim 2.5$), a large asymmetry is generated and the fit to the total cross section is improved but the  $M_{t\bar{t}}$ distribution is greatly worsened.  Furthermore, we note that the bands along the $g_L=0$ axis are slightly wider than those along the $f_L=0$ axis due to the asymmetric contributions of $f_L$ and $g_L$ to $\Gamma_{G^\prime}$, cf. Fig.~\ref{fig:theo-gp1000}(e).

\subsubsection{$m_{G^{\prime}}= 2000\,{\rm GeV}$}\label{sec:2TeV}

% This is Figure 7
\begin{figure}[b]
\includegraphics[scale=0.55]{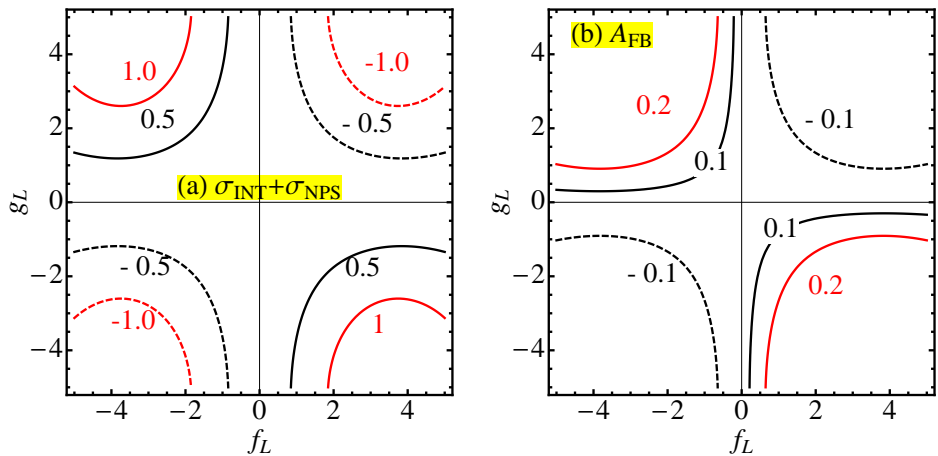}
\caption{Theoretical prediction contours for $m_{G^\prime} = 2000~\rm{GeV}$ with $f_{R}=g_{R}=0$
in the plane of $f_{L}$ and $g_{L}$  with $f_{R}=g_{R}=0$: (a) cross section (pb) 
and (b) $A_{FB}^{tot}$.
\label{fig:theo-gp2000}}
\end{figure} 

\begin{figure}
\includegraphics[scale=0.4]{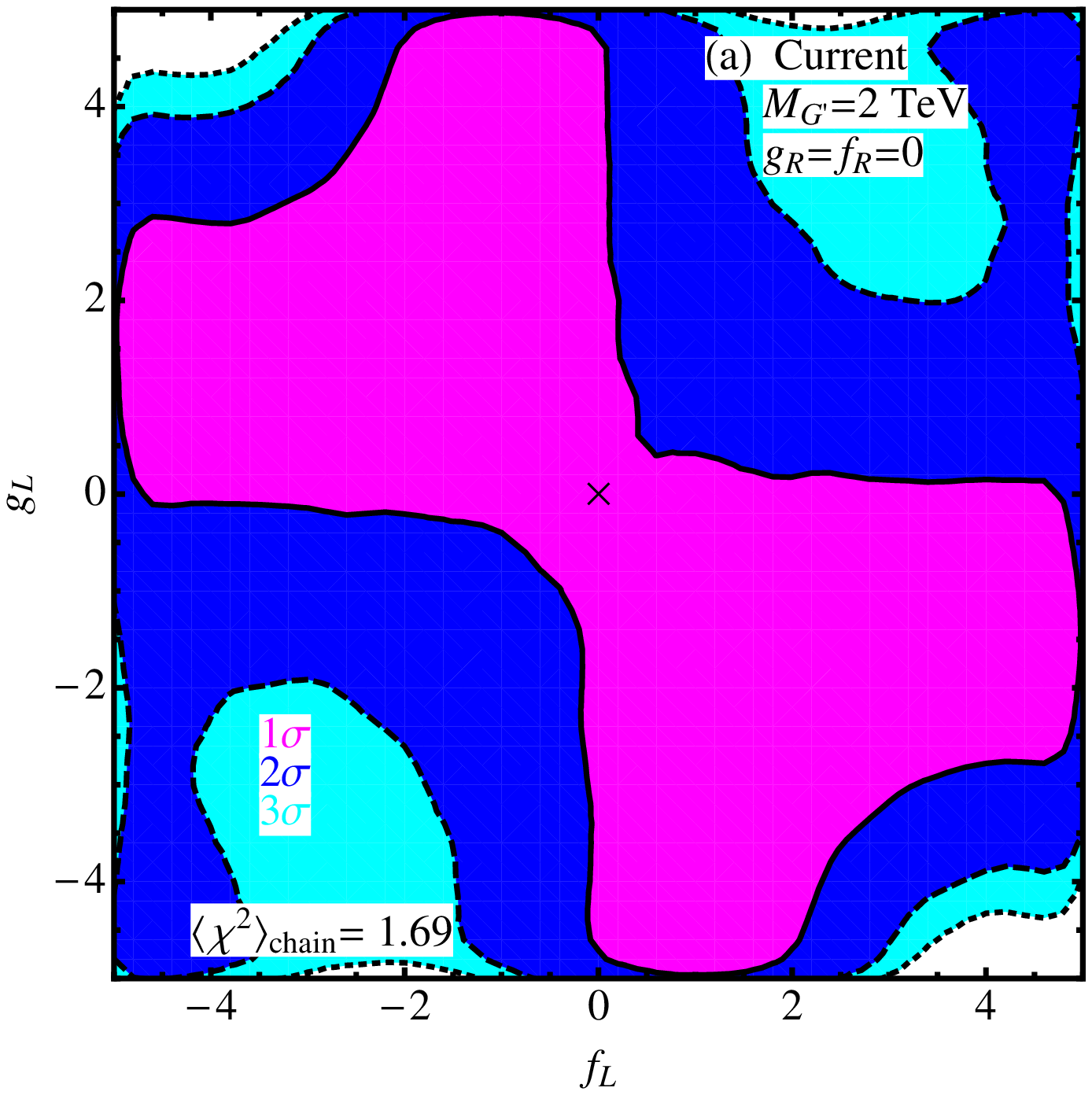}%
~~~~\includegraphics[scale=0.4]{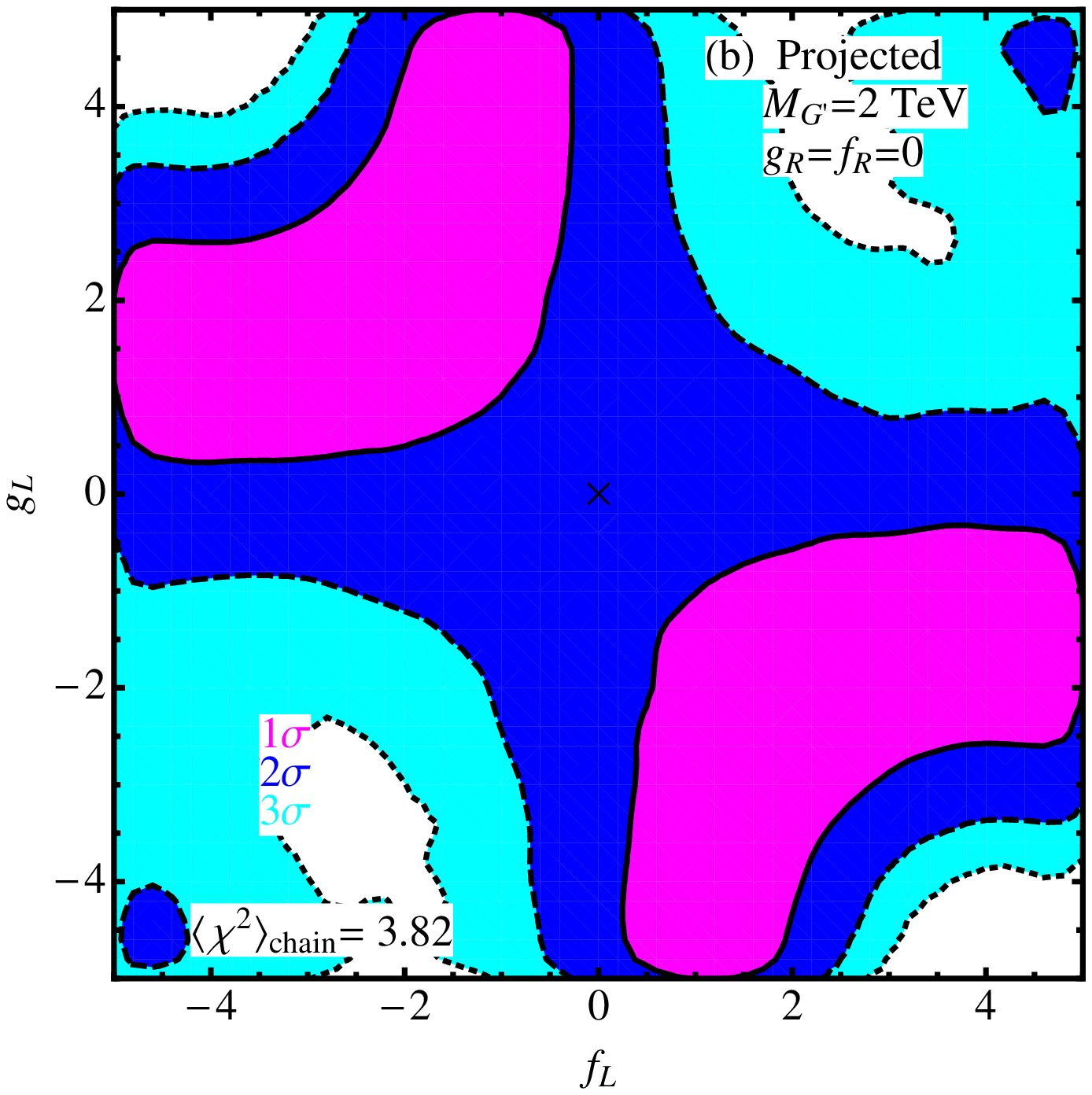} 
\caption{Same as Fig.~\ref{fig:gp1000} but for $m_{G^{\prime}}=2000\,{\rm GeV}$.  The region bounded by the solid curve defines the $1\sigma$ region, the areas bounded by the dashed lines are within $2\sigma$, and the areas bounded by the dotted lines are within $3\sigma$.
\label{fig:gp2000}}
\end{figure}

When the $G^{\prime}$ boson is very heavy, only the interference
term in Eq.~(\ref{eq:dsdz_int}) contributes to $A_{FB}$, leading to
\begin{equation}
A_{FB}^{INT}\propto\frac{\left(f_{L}-f_{R}\right)\left(g_{L}-g_{R}\right)}
{\left(f_{L}+f_{R}\right)\left(g_{L}+g_{R}\right)}.
\label{eq:AFB_INT}
\end{equation}
This dependence is illustrated in Fig.~\ref{fig:theo-gp2000}. In order to
get positive corrections to $A_{FB}^{obs}$ and the top pair production cross section, the product
$f_{L} g_{L}$ needs to be negative to compensate the negative sign from the
denominator of the propagator $1/(\hat{s}-m_{G^{\prime}}^{2})$ which is what we see in the upper left and lower right quadrants of Figs.~\ref{fig:theo-gp2000} (a) and (b).  The results of the MCMC scan are shown in Fig.~\ref{fig:gp2000} for $m_{G^{\prime}}= 2000\,{\rm GeV}$. The contours look quite different from those in Fig.~\ref{fig:gp1000}.   We note that values of $f_L$ and $g_L$ in the upper-left and lower-right quadrants are preferred which is where a large positive asymmetry is generated as seen in Fig.~\ref{fig:theo-gp2000}.  This shows that the $M_{t\bar{t}}$ distribution is less constraining than in the $m_{G^{\prime}}= 1\,{\rm TeV}$ case as one would expect. The fit to the three experiments gives $\left< \chi^2 \right>_{\rm chain} = 1.69$ for the current integrated luminosity which is slightly better than in the SM where $\left< \chi^2 \right>_{\rm chain} = 1.75$.  The fit is improved relative to the SM at the upgraded luminosity with $\left< \chi^2 \right>_{\rm chain} = 3.82$ if the central values do not change as compared to the SM value of $\left< \chi^2 \right>_{\rm chain} = 4.22$.

In Fig.~\ref{fig:gp2000}, we observe that the upper-right and lower-left quadrants are not as tightly constrained as in the $m_{G^{\prime}}= 1\,{\rm TeV}$ case in Fig.~\ref{fig:gp1000}.  Here, the $M_{t\bar{t}}$ distribution is improved.\footnote{Naively, one might expect that the $M_{t\bar{t}}$ distribution would not be very important for a $G^\prime$ with $m_{G^\prime}=2~{\rm TeV}$.  However, for couplings $f_L,~g_L\sim 4$ the width of the $G^\prime$ can be comparable to $m_{G^\prime}$ (for $f_L=4,~g_L=0$, $\Gamma_{G^\prime}\simeq 0.5 m_{G^\prime}$) and the $2~{\rm TeV}$ $G^\prime$ can contribute to the $800~{\rm GeV}<M_{t\bar{t}}<1400~{\rm GeV}$ bin.}  In the case of the $1~{\rm TeV}$ $G^\prime$ large couplings in these quadrants decrease the total cross section and asymmetry too much.  However, in the $2~{\rm TeV}$ case for large couplings in the upper-right and lower-left quadrants the NPS term becomes important due to the large width effects and this can mitigate the negative contribution to $\sigma_{t\bar{t}}$ and $A_{FB}$ from the INT term allowing for a better fit.  This is why the $2$ and $3\sigma$ regions are not tightly constrained in the upper-right and lower-left quadrants for a $2~{\rm TeV}$ $G^\prime$ in Fig.~\ref{fig:gp2000}.  Again, we note that the bands along the $f_L$ and $g_L$ axes are not symmetric due to their asymmetric contributions to $\Gamma_{G^\prime}$.

% This is Figure 8

%  sub-sub-section : other combination of couplings

\subsection{Axi-gluon: $f_R = -f_L$ and $g_R = - g_L$}\label{sec:axigluon}
Now we study the axi-gluon case, in which $G^\prime$ only has axial couplings to the quark sector.  This type of model has been explicitly proposed as an explanation of the $A_{FB}$ measurement without significantly affecting the total cross section~\cite{Frampton:2009rk}.  There, the SM prediction for the $t\bar t$ cross section was taken to be larger than our value due to differences in $m_t$ and including incomplete NNLO calculations.  Therefore, they did not need a significant correction to the cross section.

% This is Figure 9
\begin{figure}
\includegraphics[scale=0.5]{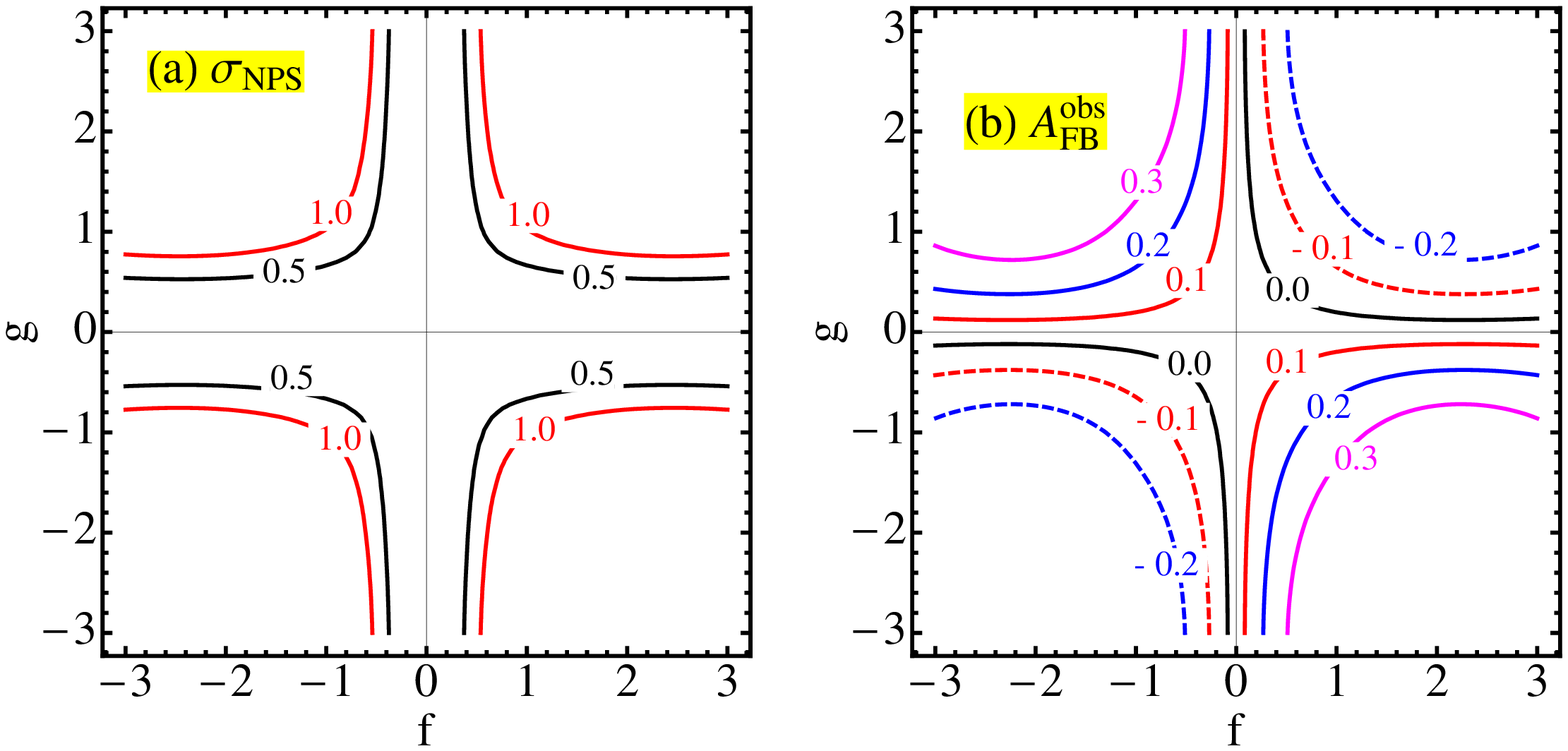}\\ 
\includegraphics[scale=0.5]{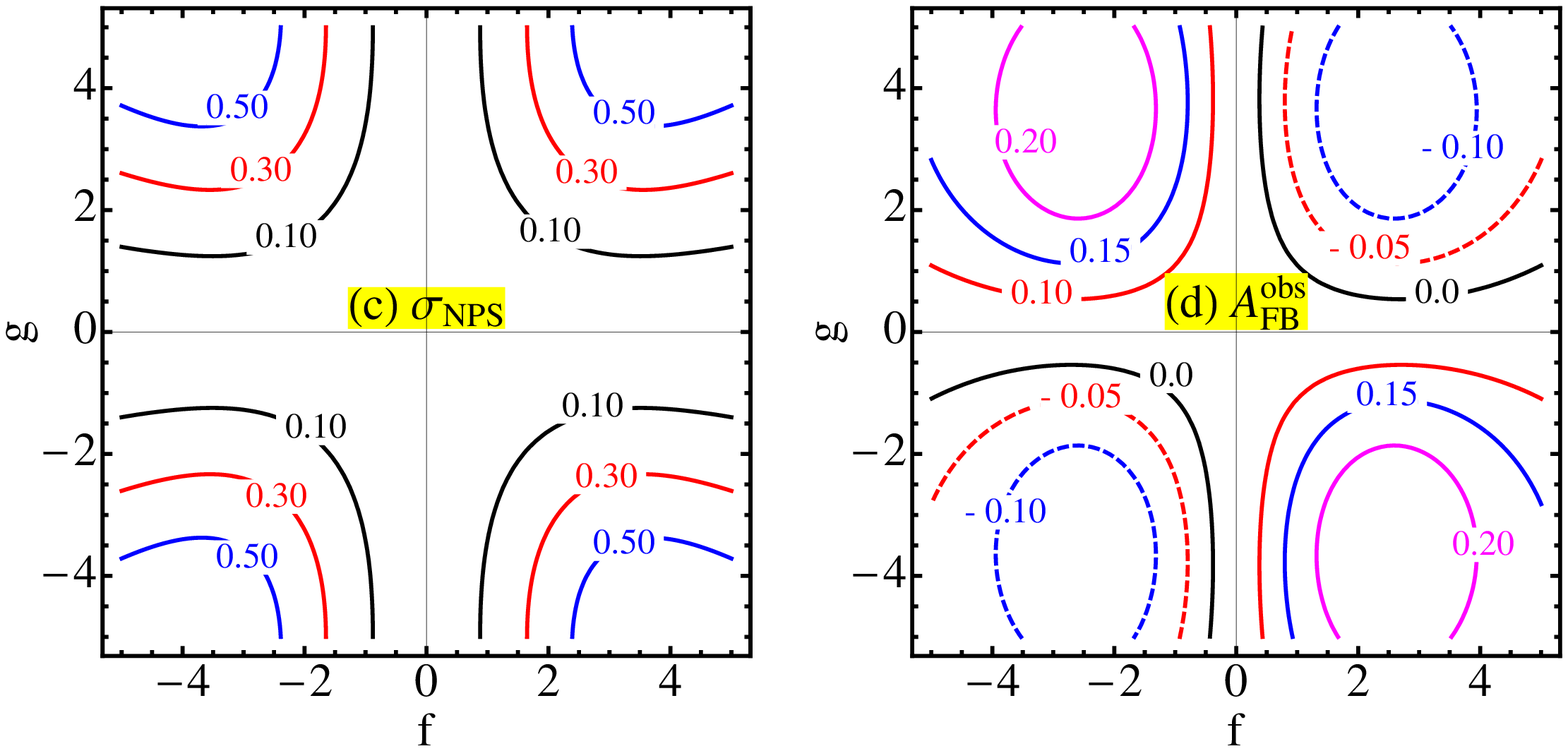} 
\caption{Theoretical prediction contours for an axi-gluon of $\sigma_{NPS}=\sigma_{NP}$ (since $\sigma_{INT}=0$) and $A_{FB}^{obs}$ for (a,b) $m_{G^\prime} = 1000~\rm{GeV}$ and  (c,d) $m_{G^\prime} = 2000~\rm{GeV}$. 
\label{fig:theo-axigluon1000}}
\end{figure}

In the axial limit, only the asymmetry-generating term of the INT in Eq.~(\ref{eq:dsdz_np}) remains.  In general, all terms in the NPS remain.  Therefore, at large $m_{G^\prime}$, the INT term dominates and a rather large asymmetry can arise without a sizable contribution to $\sigma_{t\bar{t}}$ or to $\frac{d\sigma}{dM_{t\bar{t}}}$.  At lower values of $m_{G^\prime}$, the NPS term is increasingly relevant.  

\subsubsection{$m_{G^{\prime}}= 1000\,{\rm GeV}$}\label{sec:axigluon-1TeV}
 
\begin{figure}
\includegraphics[scale=0.43]{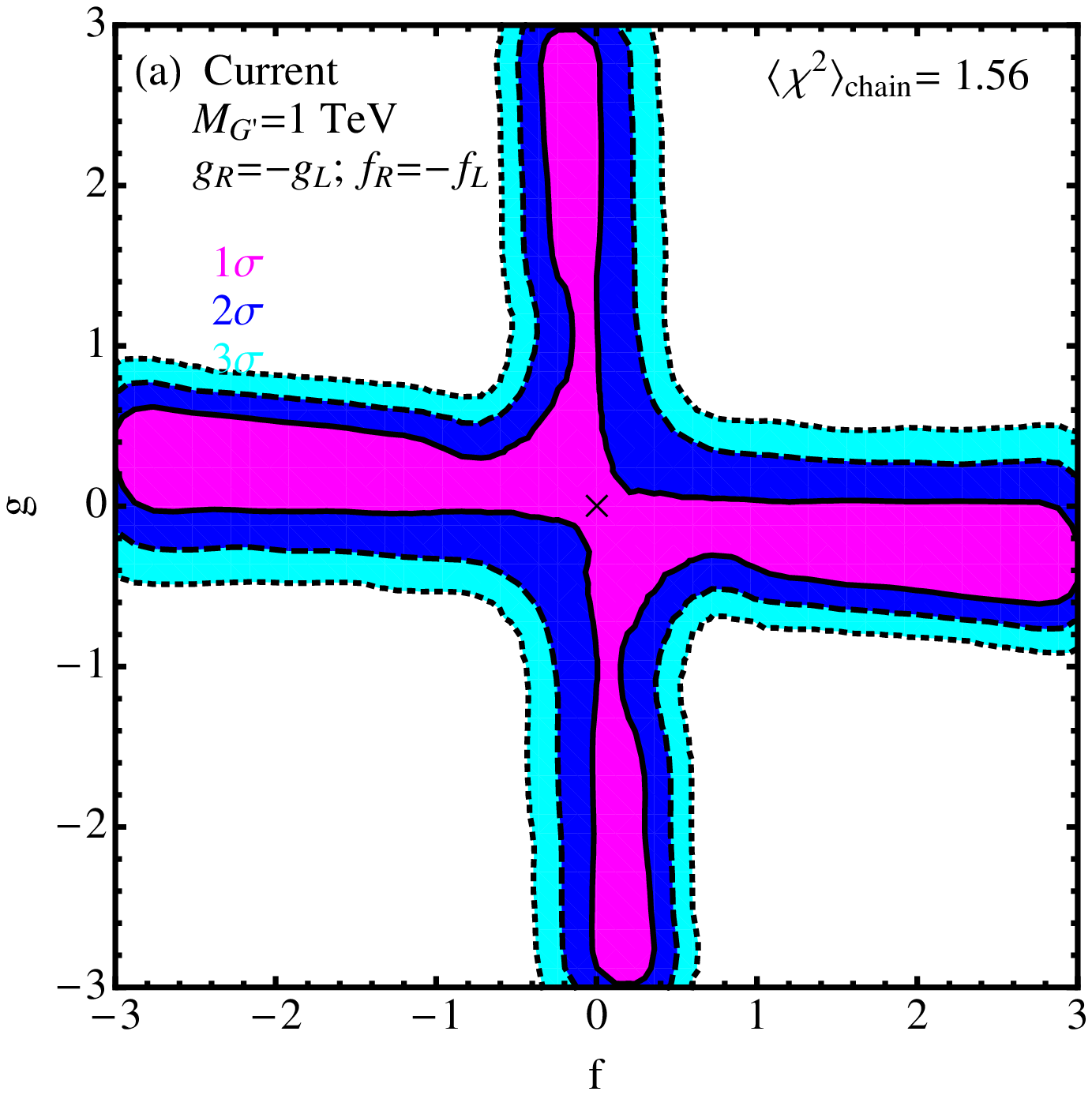}
\includegraphics[scale=0.43]{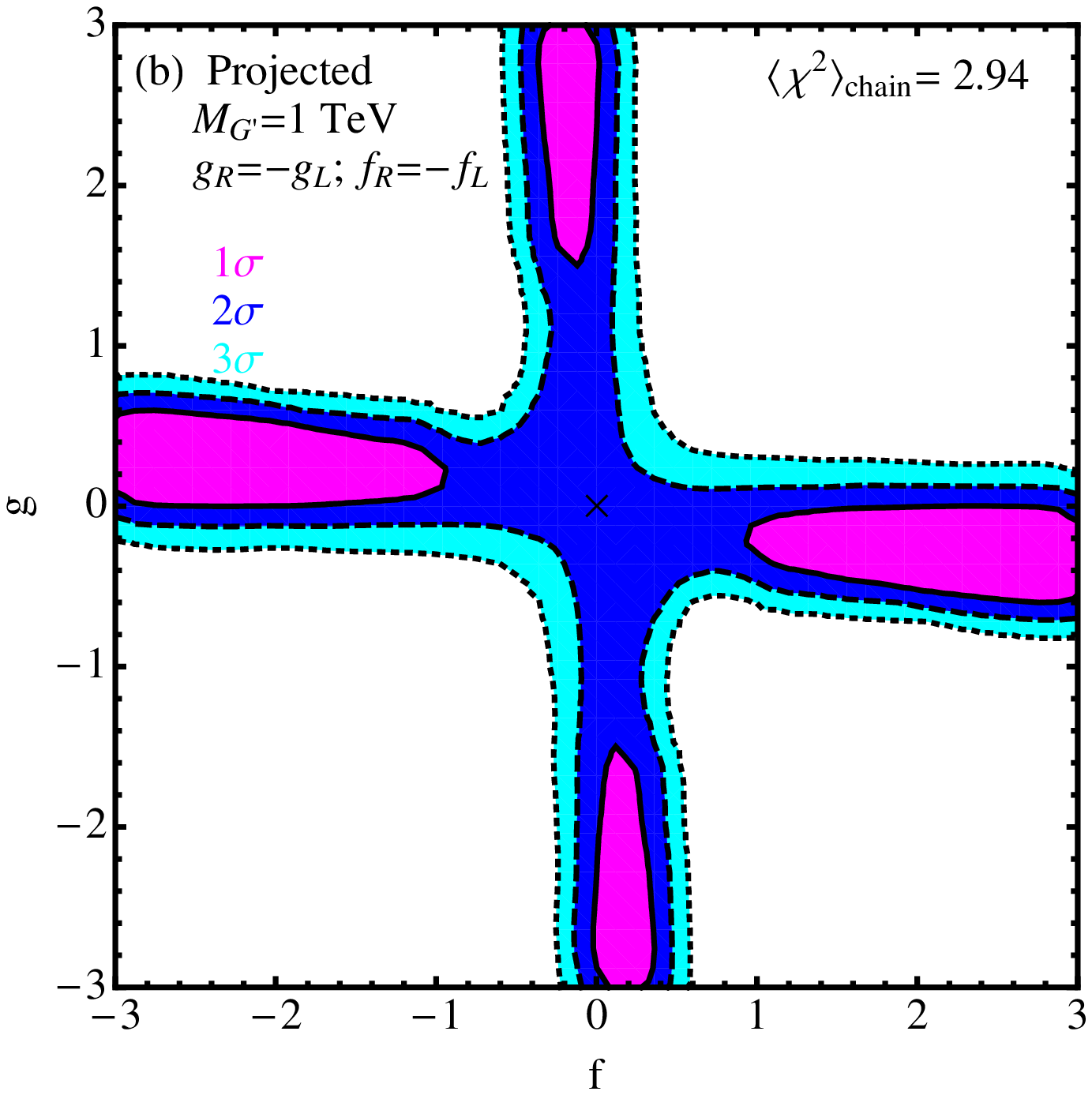}\\
\includegraphics[scale=0.43]{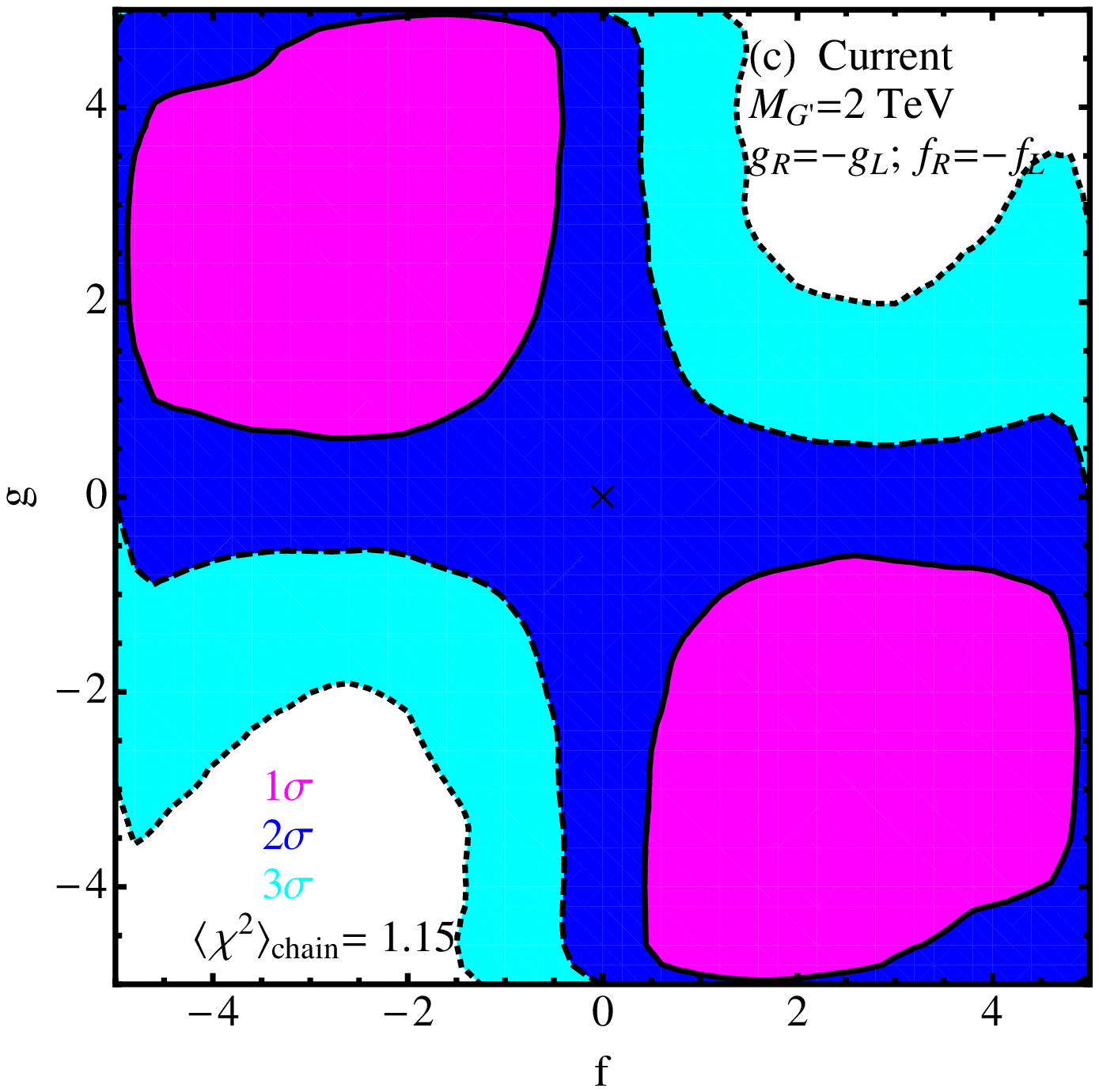}
\includegraphics[scale=0.43]{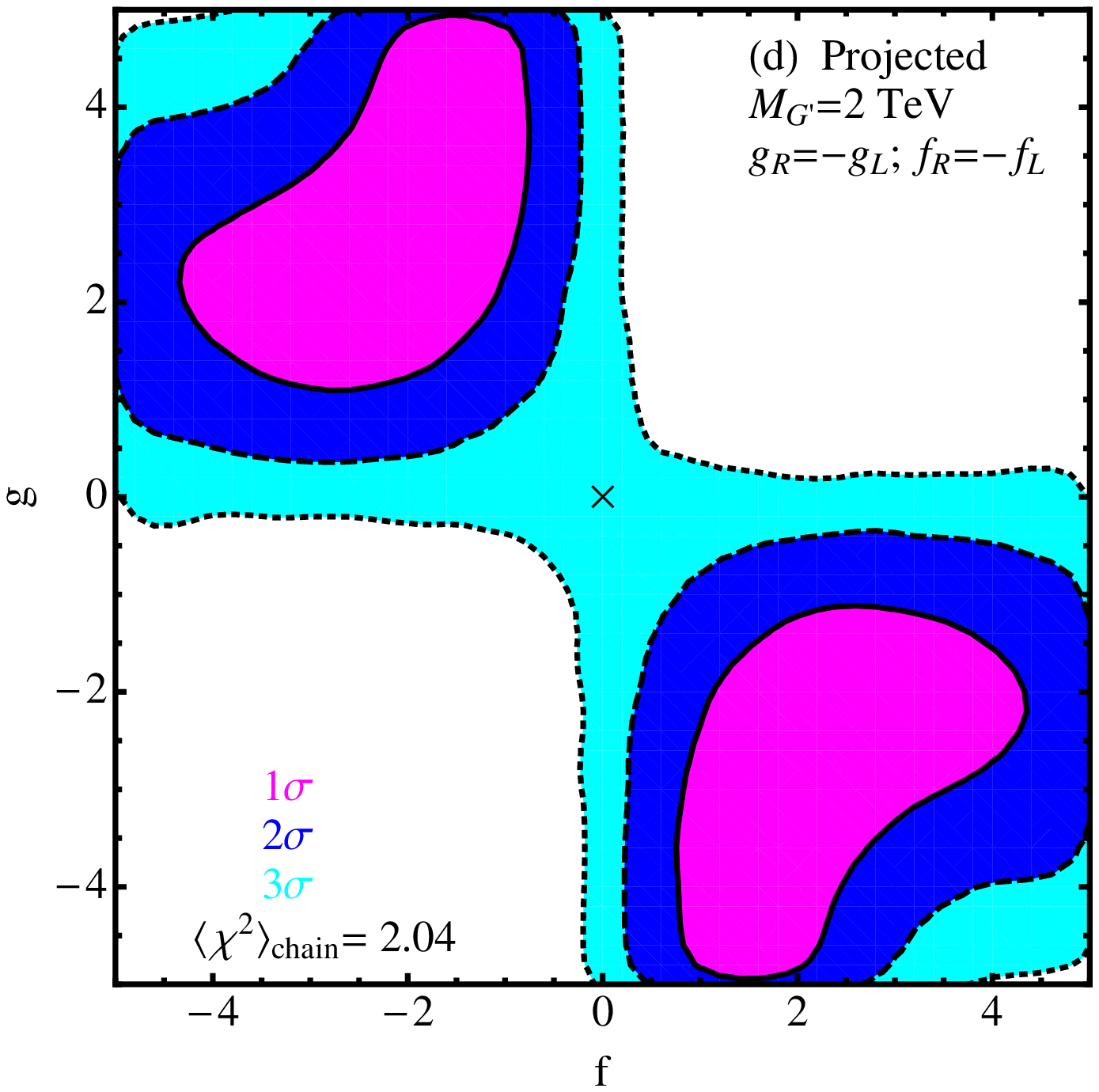}\\
\caption{Correlation of couplings for $m_{G^{\prime}}= 1000\,{\rm GeV}$ (top panels) and $m_{G^{\prime}}=2000\,{\rm GeV}$ (bottom panels) in the axi-gluon limit
with the current integrated luminosity (left panels) and $\int \mathcal{L} dt =10\,{\rm fb}^{-1}$ (right panels) assuming the central values of the measurements do not change.
\label{fig:axigluon}}
\end{figure}

We show theoretical contours of $\sigma_{NPS}$ (since $\sigma_{INT}$ vanishes when integrated over $\cos\theta$ in the axi-gluon case) and $A_{FB}^{obs}$ in Fig.~\ref{fig:theo-axigluon1000}(a) and (b) for an axi-gluon of mass $1~{\rm TeV}$.  We observe that a positive asymmetry is generated when the product $fg$, with $f=f_L=-f_R$ and $g=g_L=-g_R$, is negative as we expect from Eq.~\ref{eq:dsdz_int}.  In Fig.~\ref{fig:axigluon}(a) and (b), we perform MCMC scans and find a fit with $\langle \chi^2\rangle_{chain}= 1.56$ for the current luminosity and $\langle \chi^2\rangle_{chain}= 2.94$ for 10 fb$^{-1}$ assuming the central values of the measurements do not change.  Small values of either $f$ or $g$ are preferred due to the constraint on the $M_{t\bar{t}}$ distribution.  These values of $\langle \chi^2\rangle_{chain}$ for the $1~{\rm TeV}$ axi-gluon are better than the corresponding values in the SM fit.  This indicates that a $1~{\rm TeV}$ axi-gluon has less tension with the current data than the SM and would also offer an improvement over the SM if the central values of the data remain the same with an upgraded luminosity of $10~{\rm fb}^{-1}$.

% This is Figure 10

\subsubsection{$m_{G^{\prime}}= 2000\,{\rm GeV}$}\label{sec:axigluon-2TeV}

For a $2~{\rm TeV}$ axi-gluon, we plot theoretical contours of $\sigma_{NPS}$ and $A_{FB}$ in Fig.~\ref{fig:theo-axigluon1000} (c) and (d).  In Figs.~\ref{fig:axigluon}(c) and (d), we show the results of a scan in the case of an axi-gluon with $m_{G^{\prime}}=2000\,{\rm GeV}$.  The fit shows better agreement with the data in this case than in the SM, with $\langle \chi^2\rangle_{chain}= 1.15$ for the current luminosity and $\langle \chi^2\rangle_{chain}= 2.04$ for 10 fb$^{-1}$ if the central values do not change.  Due to the lessening of the $M_{t\bar{t}}$ constraint for a heavier axi-gluon, the scans show somewhat different structure than in the $1~{\rm TeV}$ case.  The $1\sigma$ allowed regions are located in the quadrants where $fg<0$ which is where a positive $A_{FB}$ is generated as seen in Fig.~\ref{fig:theo-axigluon1000} (d).  The regions of large coupling are constrained by the $M_{t\bar{t}}$ distribution.  There is again a slight asymmetry in the width of the allowed regions near the $f$ and $g$ axes due to the asymmetry in the width.

Our results suggest that a heavy axi-gluon can offer a good explanation of the large $A_{FB}$ observed without increasing the disagreement in the $M_{t\bar{t}}$ distribution too much, as proposed in Ref.~\cite{Frampton:2009rk}.

\subsection{Other combinations of couplings}\label{sec:other_couplings}

% This is Figure 11
\begin{figure}[b]
\includegraphics[scale=0.37]{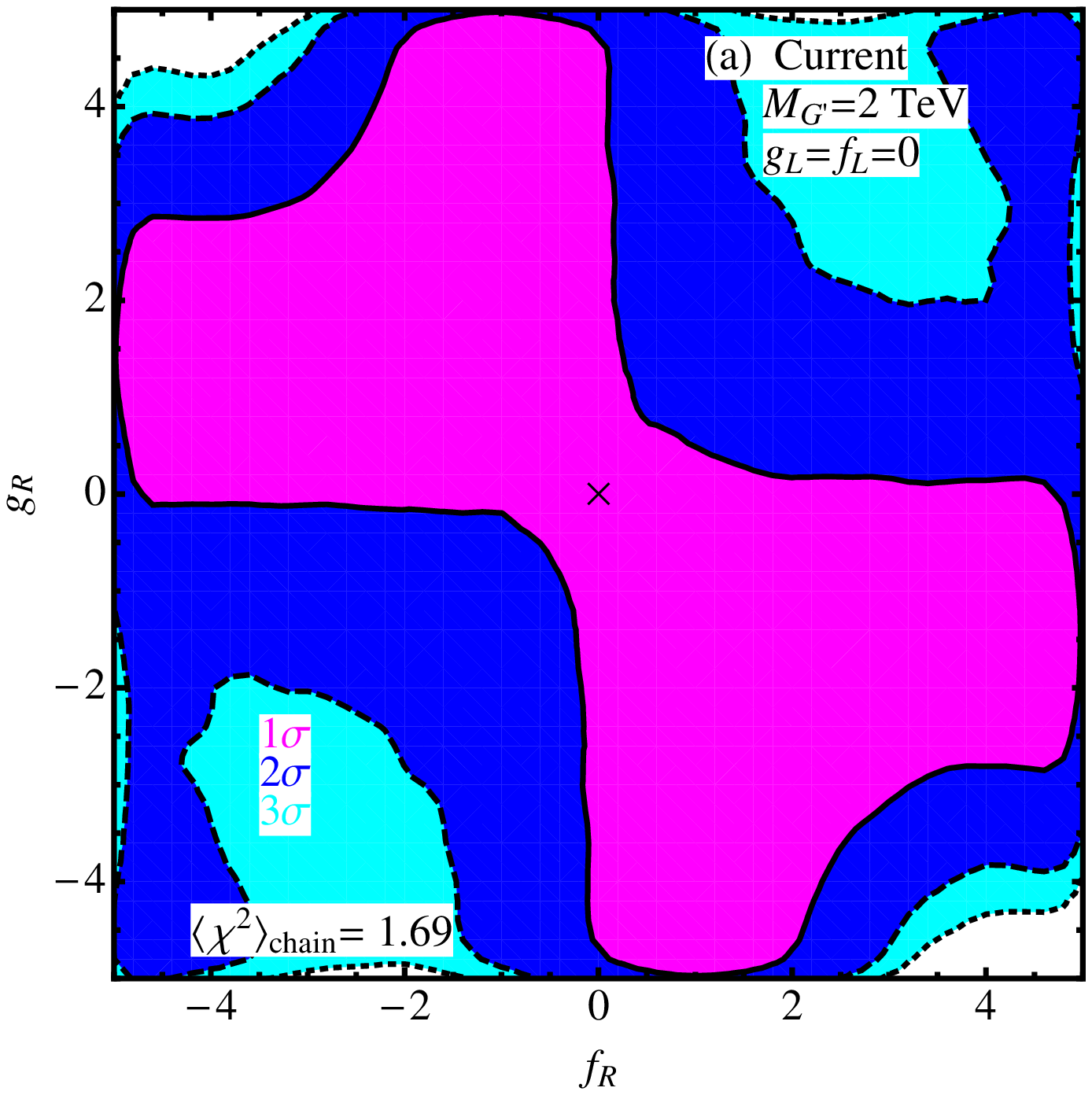}
\includegraphics[scale=0.37]{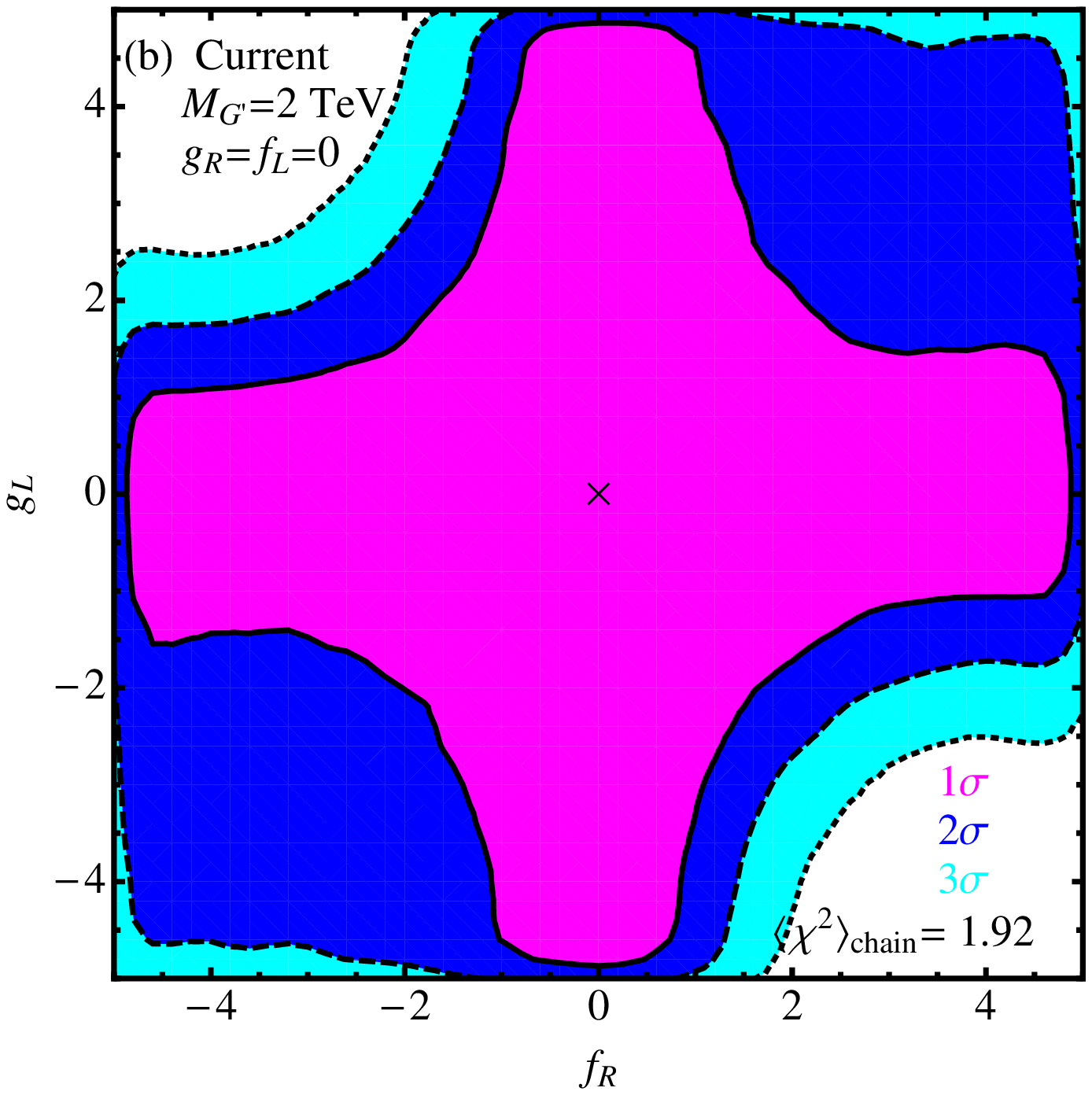}
\includegraphics[scale=0.37]{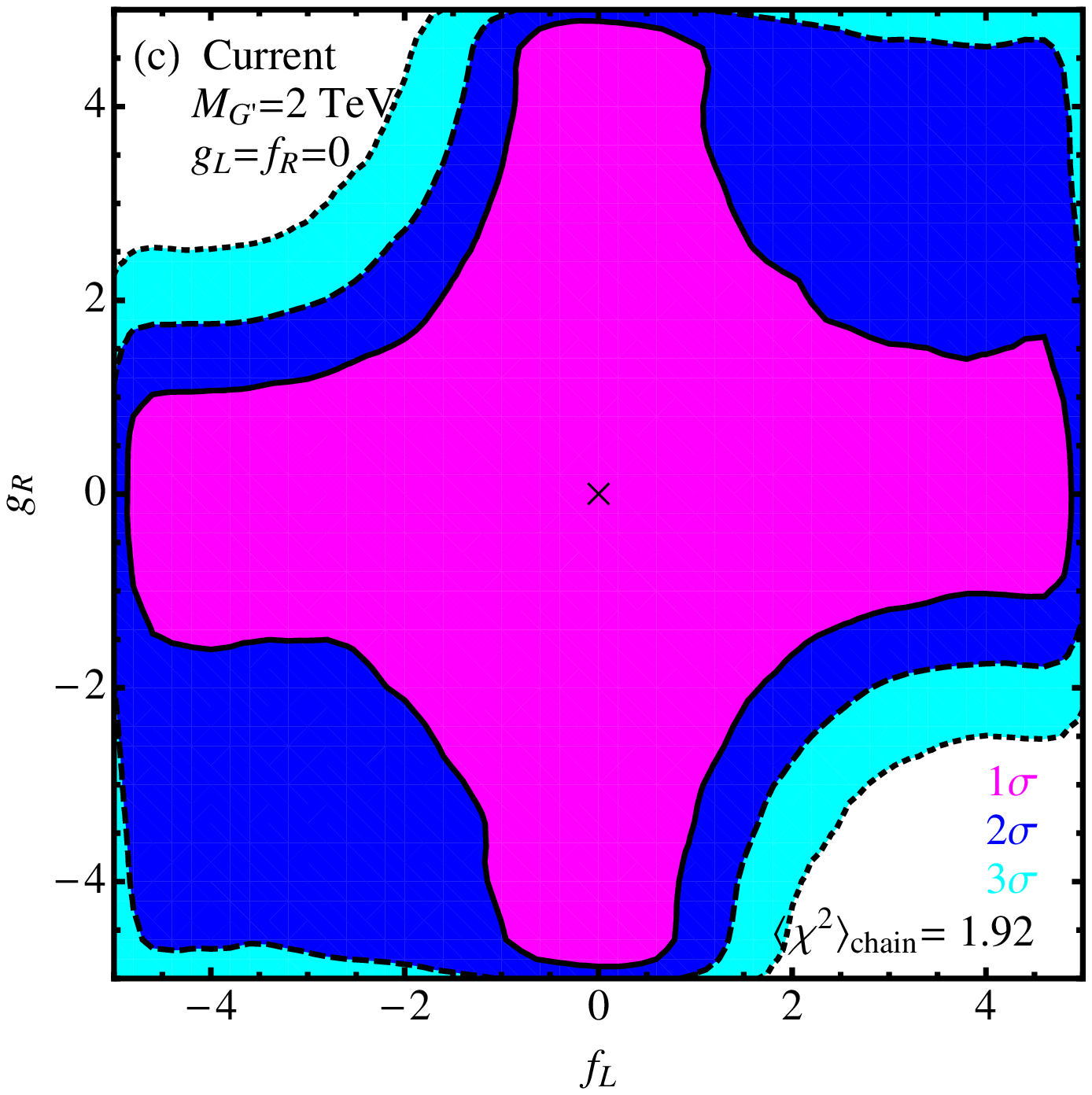}
\caption{Correlation of couplings for $m_{G^{\prime}}= 2\,{\rm TeV}$
with the current integrated luminosity for (a) $g_L=f_L=0$, (b) $g_R=f_L=0$, (c) $g_L=f_R=0$.
\label{fig:coupl_combo}}
\end{figure}  

Now let us study different combinations of couplings, e.g., $f_L=g_L=0$,
$f_L=g_R=0$ and $f_R=g_L=0$. Fig.~\ref{fig:coupl_combo} shows the MCMC scan
results of various combinations of couplings with $m_{G^{\prime}}=2000\,{\rm GeV}$ and with the current luminosity.  Purely right-handed couplings
in the $q$-$\bar{q}$-$G^\prime$ interaction give rise to the exactly same
result as purely left-handed couplings, cf.\ Fig.~\ref{fig:gp2000}(a). But mixed
combinations of left-handed and right-handed couplings, e.g., $g_L=f_R=0$ and
$g_R=f_L=0$, result in a worse fit, with $\left<\chi^2\right>_{chain}=1.92$, which is worse than the SM.
This is mainly due to
the INT effects which are sensitive to the signs of couplings; 
see Eqs.~(\ref{eq:dsdz_int}) and (\ref{eq:AFBINT}). Choosing $f_L=g_R=0$
or $f_R=g_L=0$ causes the INT effects to generate a negative $A_{FB}^{INT}$,
leading to the bad fit.

%--------------------------------------------------------------------%
%  Section IV                                                        %
%                                                                    %
%   S-channel diagram for effective field theory                     %
%--------------------------------------------------------------------%  
\section{Effective field theory\label{sec:eft}}

For a $2~{\rm TeV}$ $G^\prime$ boson, due to the broad decay width, the NPS contribution is still sizable for large couplings in the above MCMC scans.  It is interesting to ask what the effects are if only the INT term contributes.  To that end, in this section we consider dim-6 effective operators that can interfere with the SM top quark pair production channel $q\bar{q}\to g\to t\bar{t}$.  We further assume that the scale of the new physics is large enough that the NPS contributions (i.e. $\propto 1/\Lambda^4$ with $\Lambda$ the new physics scale) can be neglected.  For illustration we focus only on the operators which couple left-(right-)handed light quarks to left-(right-)handed top quarks so that contact with Sec.~\ref{sec:lefthanded} can be made.
They are listed as follows: 
\begin{eqnarray}
\mathcal{O}_{qq}^{(8,1)} & = &
\left(\bar{q}\gamma_{\mu}t^{A}q\right)\left(\bar{Q}\gamma^{\mu}t^{A}Q\right),\\
\mathcal{O}_{qq}^{(8,3)} & = &
\left(\bar{q}\gamma_{\mu}t^{A}\tau^{I}q\right)
\left(\bar{Q}\gamma^{\mu}t^{A}\tau^{I}Q\right),\\
\mathcal{O}_{ut}^{(8,1)} & = &
\left(\bar{u}\gamma_{\mu}t^{A}u\right)\left(\bar{t}\gamma^{\mu}t^{A}t\right),\\
\mathcal{O}_{dt}^{(8,1)} & = & 
\left(\bar{d}\gamma_{\mu}t^{A}d\right)\left(\bar{t}\gamma^{\mu}t^{A}t\right),
\end{eqnarray}
where $q$ and $Q$ denote the $SU(2)_{L}$ doublets of the light
(first two generation) quarks and heavy (third generation) quark,
respectively, and $u(d,\, t)$ are the right-handed gauge singlets. Here,
$t^{A}$ and $\tau^{I}$ are the $SU(3)$ and $SU(2)$ matrices; appropriate
contractions are understood. The first index in the superscripts of
operators labels the color octet and the second index denotes the
weak isospin. Other color and weak singlet operators are omitted as
they cannot interfere with the SM channel.

The effective Lagrangian of the four fermion interaction $q\bar{q}t\bar{t}$
is thus given by
\begin{eqnarray}
\mathcal{L}^{(4f)} & = &
 g_{s}^2\frac{\kappa_{L}^{q}}{\Lambda^{2}}\left(\bar{q}\gamma_{\mu}P_{L}q\right)
 \left(\bar{t}\gamma_{\mu}P_{L}t\right)+g_{s}\frac{\kappa_{R}^{q}}{\Lambda^{2}} 
 \left(\bar{q}\gamma_{\mu}P_{R}q\right)\left(\bar{t}\gamma_{\mu}P_{R}t\right),
\end{eqnarray}
where we explicitly factor out a strong coupling strength $g_{s}^2$,
and the reduced coefficients are given as follows: 
\[
\kappa_{L}^{u}=C_{qq}^{(8,1)}+C_{qq}^{(8,3)},\quad\kappa_{L}^{d}=
C_{qq}^{(8,1)}-C_{qq}^{(8,3)},\quad\kappa_{R}^{u}=C_{ut}^{(8)},
\quad\kappa_{R}^{d}=C_{dt}^{(8)}.
\]
Here the $SU(3)$ generators are omitted and $\Lambda$ denotes the
new physics scale.

The differential cross section of the EFT can be easily derived from
Eq.~(\ref{eq:dsdz_int}) by taking the limit of $m_{G^\prime}=\Lambda\gg\sqrt{\hat{s}}$,
\begin{eqnarray}
\mathcal{A}_{INT}^{EFT} & = &
 -\frac{\pi\beta\alpha_{s}^{2}}{18}
\left[ \frac{\kappa_{L}^{q}+\kappa_{R}^{q}}{\Lambda^{2}}\right]
 \left\{ \left(2-\beta^{2}\right)+2\beta\cos\theta+
 \left(\beta\cos\theta\right)^{2}
 \right\} .
\label{eq:dsdz_EFT}
\end{eqnarray}
Obviously, the coefficients $\kappa_{L/R}^{q}$ only affect $R$ but not
$A_{FB}^{NP}$. We extract the cutoff scale and coefficients as follows: 
\begin{equation}
\mathcal{A}_{INT}^{EFT} = 
 \frac{\kappa_{L}^{q}+\kappa_{R}^{q}}
 {\left(\frac{\Lambda}{{\rm TeV}}\right)^{2}}
 \times
 \left\{ -\frac{\pi\beta\alpha_{s}^{2}}{18\left({\rm TeV}\right)^{2}}
 \left[\left(2-\beta^{2}\right)+2\beta\cos\theta+
 \left(\beta\cos\theta\right)^{2}\right]\right\}. 
 \label{eq:dsdz_EFT_TeV}
\end{equation}
which yields, after integration over $\hat{s}$ and convolution with
PDFs,
\[
\frac{d\sigma_{EFT}^{INT}}{d\cos\theta}=
\frac{\kappa_{L}^{q}+\kappa_{R}^{q}}{\left(\frac{\Lambda}{{\rm TeV}}\right)^{2}}
\left(A_{EFT}+B_{EFT}\cos\theta+C_{EFT}\cos^{2}\theta\right).
\]
The parameters ($A_{EFT},\, B_{EFT},\, C_{EFT}$) are listed in
Table~\ref{tab:eft} for various choices of factorization scale, where
$A_{FB}^{EFT}$ and $\sigma_{EFT}$ are evaluated using Eq.~(\ref{eq:AFB_para}).
It is clear that one needs positive $\kappa_{L}^{q}$ or $\kappa_{R}^{q}$
to get positive $A_{FB}$ but this inevitably gives rise to a negative
contribution to $\sigma(t\bar{t})$. Hence, it is difficult to fit 
both the asymmetry and the top pair production cross section simultaneously
at the $1\sigma$ level.

% This is Figure 11
\begin{figure}
\includegraphics[scale=0.43]{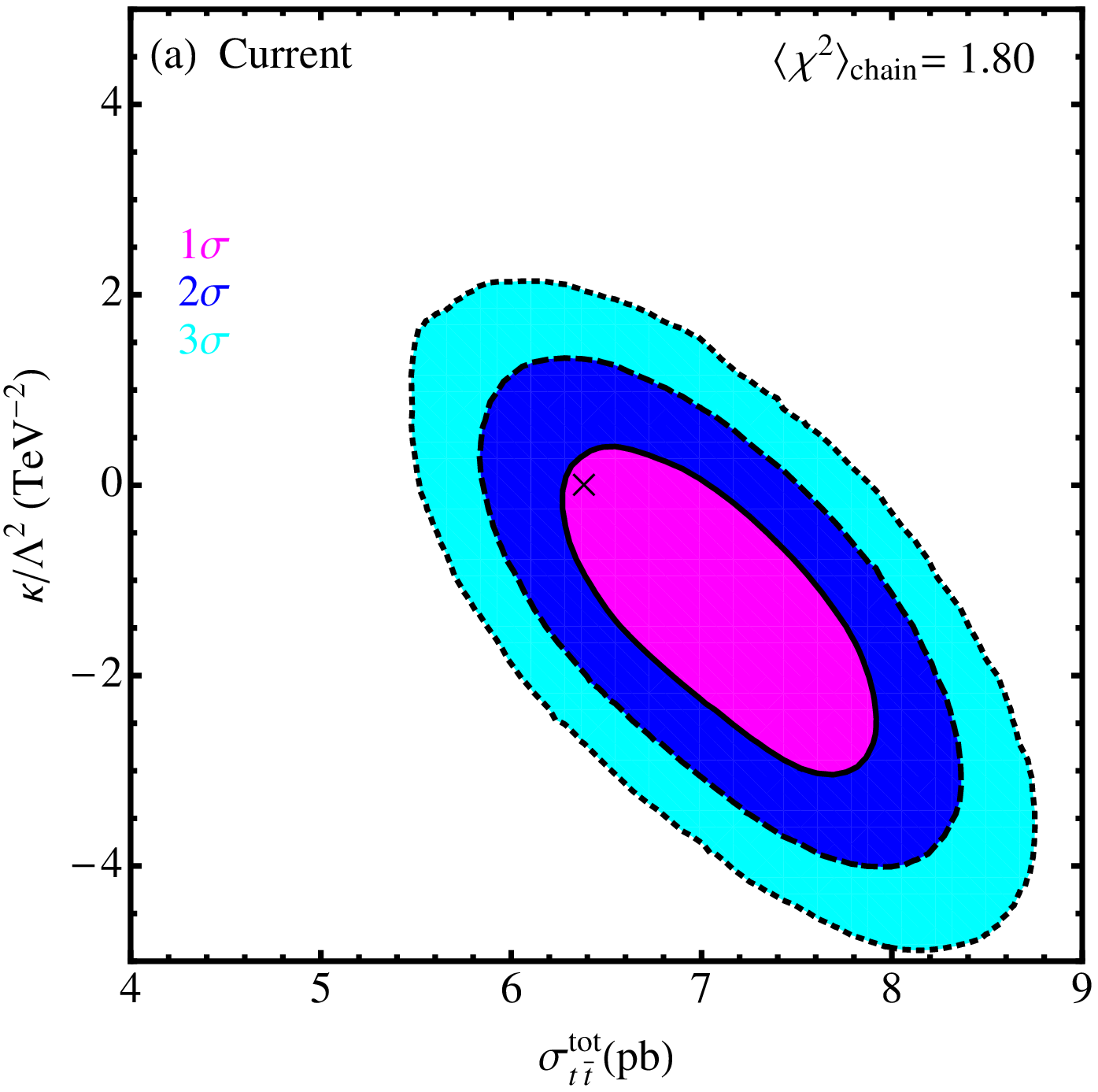}
\includegraphics[scale=0.43]{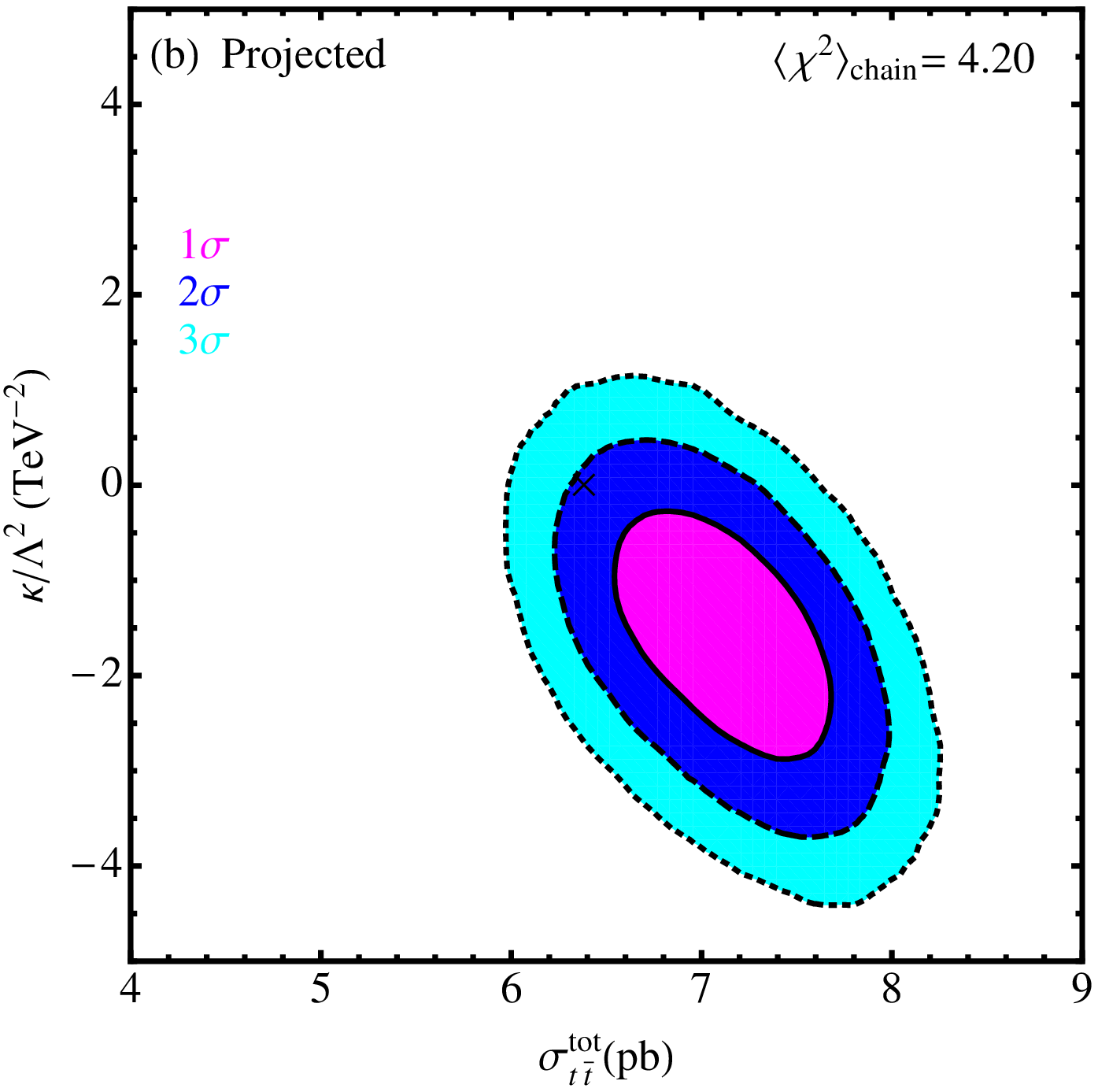}\\
\includegraphics[scale=0.43]{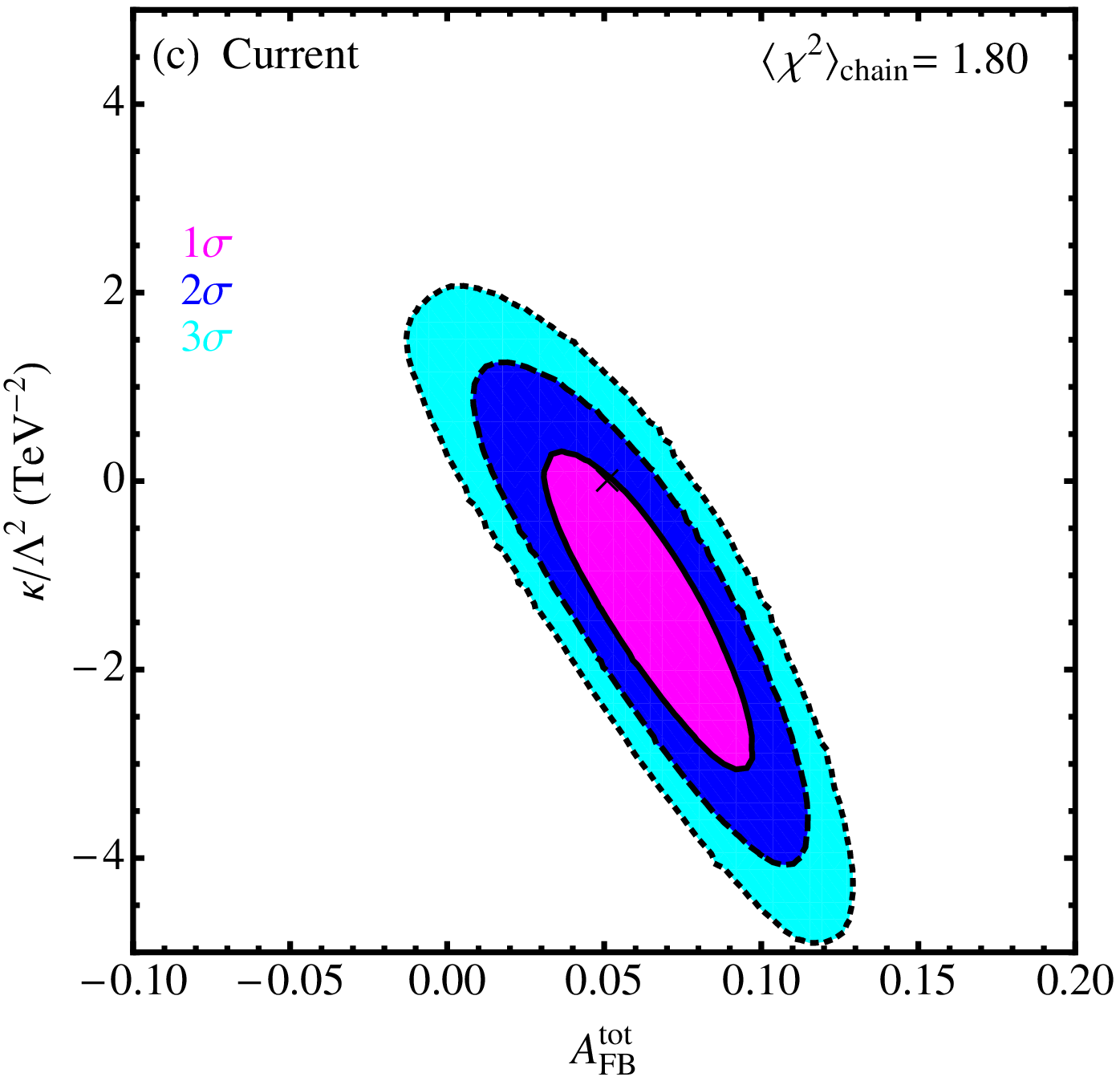}
\includegraphics[scale=0.43]{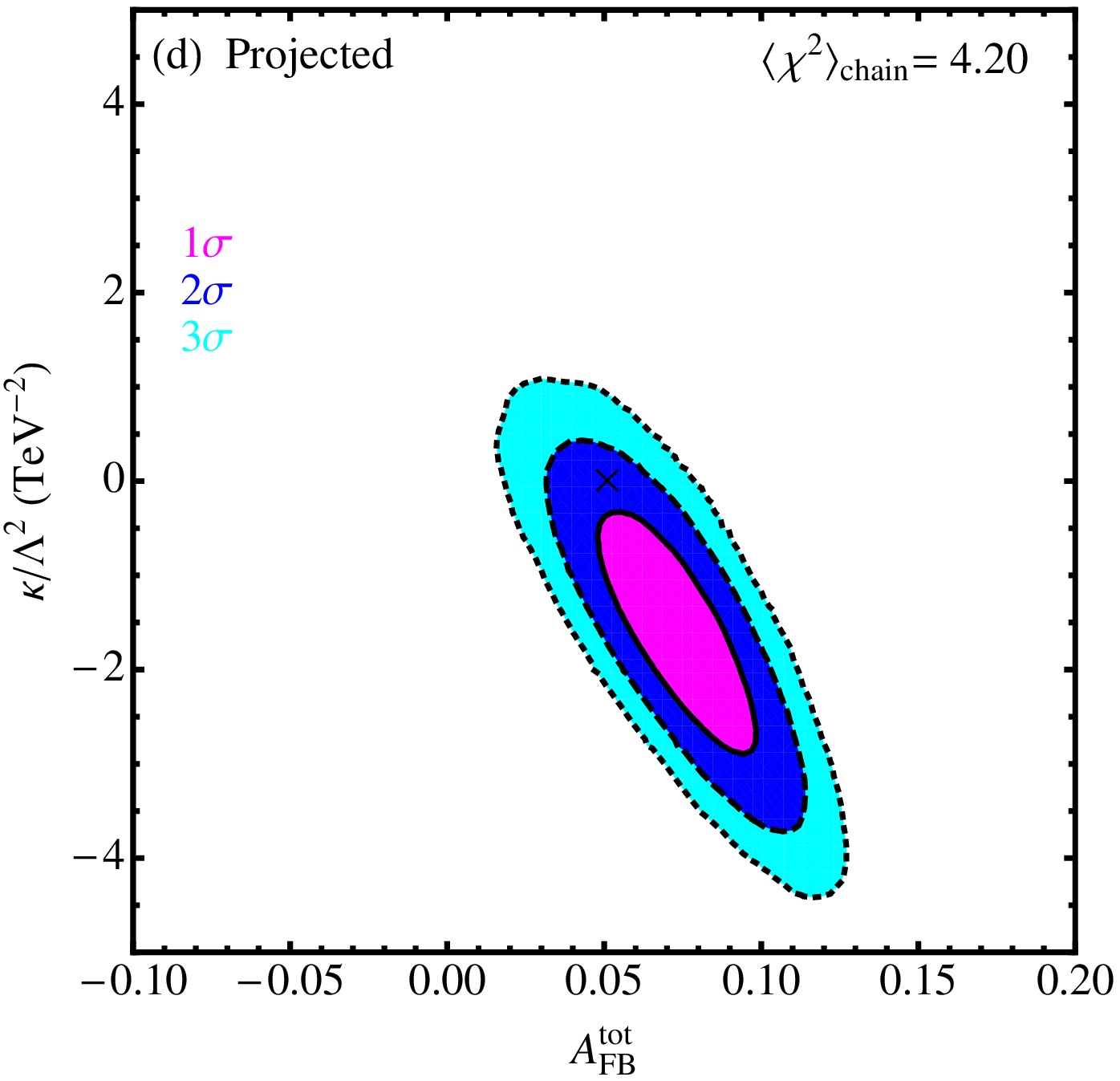}
\caption{(Left): Fitting contours in the EFT with the current integrated luminosity:
(a) in the plane of $A_{FB}^{obs}$ and $\kappa/(\Lambda/{\rm TeV})^2$,
(b) in the plane of $\sigma(t\bar{t})$ and $\kappa/(\Lambda/{\rm TeV})^2$.
(Right): same as the left column but for $\int
\mathcal{L} dt =10\,{\rm fb}^{-1}$.
\label{fig:eft}}
\end{figure} 

% This is Table III
\begin{table}
\caption{Parameter ($A_{EFT}$, $B_{EFT}$, $C_{EFT}$) for EFT where 
$A_{FB}^{INT}$ and $\sigma_{EFT}^{INT}$ are calculated after 
integrating over the angle $\theta$. \label{tab:eft}}
\begin{tabular}{cccccc|ccccc}
\hline 
\multicolumn{6}{c|}{$u\bar{u}\to t\bar{t}$} & 
\multicolumn{5}{c}{$d\bar{d}\to t\bar{t}$}\tabularnewline
\hline 
$\mu_{R}$ & $A_{EFT}$ & $B_{EFT}$ & $C_{EFT}$ & $A_{FB}^{EFT}$ & 
$\sigma_{EFT}$ & $A_{EFT}$ & $B_{EFT}$ & $C_{EFT}$ & $A_{FB}^{EFT}$
 & $\sigma_{EFT}$\tabularnewline
\hline
$m_{t}/2$ & -0.294 & -0.256 & -0.092 & 0.395 
& -0.648 & -0.052 & -0.040 & -0.040 & 0.355 & -0.113\tabularnewline
$m_{t}$ & -0.215 & -0.185 & -0.066 & 0.392 & -0.473 
& -0.037 & -0.0288 & -0.10 & 0.353 & -0.062\tabularnewline
$2m_{t}$ & -0.165 & -0.141 & -0.050 & 0.389 & -0.363 
& -0.028 & -0.022 & -0.007 & 0.350 & -0.082\tabularnewline
\hline
\end{tabular}
\end{table} 

Since one cannot separate the coefficient $\kappa$ from the cutoff $\Lambda$,
we scan over the combination $\kappa/(\Lambda/{\rm TeV})^2$ and limit
ourselves to the region of $|\kappa/(\Lambda/{\rm TeV})^2| < 10$ in the MCMC
scan. In Fig.~\ref{fig:eft} we plot the correlations between
$\kappa/(\Lambda/{\rm TeV})^2$ and $\sigma_{tot}$ (top row) and between
$\kappa/(\Lambda/{\rm TeV})^2$ and $A_{FB}^{obs}$ (bottom row). For the current luminosity, the fit quality of the EFT is worse than the SM,
$\left<\chi^2\right> _{\rm chain}=1.80$.  The fit is marginally better than the SM, $\left<\chi^2\right> _{\rm chain}=4.20$, for an integrated luminosity of 10 fb$^{-1}$ if the central values of the measurements remain the same.  The fit is worse than that of the $2~{\rm TeV}$ left-handed $G^\prime$ due to the lack of a NPS term to balance the contributions of the INT term.  This indicates the importance of resonance effects in the fit.

%--------------------------------------------------------------------%
%  Section V                                                         %
%                                                                    %
%   S-channel diagram for Z-prime boson                              %
%--------------------------------------------------------------------%  
\section{Flavor-conserving $Z^{\prime}$ boson\label{sec:zprime-fd}}

An additional $Z^{\prime}$ can generate a nonzero $A_{FB}$ if its coupling
to quarks does not respect parity, 
\begin{eqnarray}
Z^{\prime}q\bar{q} & : & ie\gamma^{\mu}\left(f_{L}P_{L}+f_{R}P_{R}\right),\\
Z^{\prime}t\bar{t} & : & ie\gamma^{\mu}\left(g_{L}P_{L}+g_{R}P_{R}\right).
\end{eqnarray}
where $e$ denotes the electromagnetic coupling strength.  In contrast to $G^{\prime}$, there is no interference between the
$Z^{\prime}$-mediated top quark pair production and the SM process.
Even though the $Z^{\prime}$ amplitude interferes with the SM process
$q\bar{q}\to\gamma^{*}/Z^{*}\to t\bar{t}$, the latter contribution is negligible
at the Tevatron. Only the NP resonance itself 
contributes to $A_{FB}$ when the collider energy is large enough
to see the resonance effects.  
We consider the case where both the up and down quarks are gauged, 
but it is also possible to gauge the up and down quarks 
differently~ \cite{Rosner:1996eb}.

Since the interference is absent for the color singlet $Z^{\prime}$, 
only $\mathcal{A}_{NPS}$ contributes to NPS. 
For the $s$-channel diagram, the differential cross section of 
$Z^{\prime}$ can be easily derived from that of $G^{\prime}$ by 
omitting the color factor $2/9$ in Eq.~(\ref{eq:dsdz_np}) and replacing $\alpha_{s}$ by $\alpha_{em}$,
yielding 
\begin{eqnarray}
\frac{d\sigma}{d\cos\theta}\Biggr|_{Z^{\prime}} & = &
 \frac{\pi\beta\alpha_{s}^{2}}{9\hat{s}}
 \frac{9}{2}\mathcal{A}_{NPS}^{G^{\prime}}
 \biggr|_{m_{G}\to m_{Z^{\prime}},\alpha_{s}\to\alpha_{W}}
 \nonumber \\
 & = &
 \frac{\pi\beta\alpha_{em}^{2}}{8\hat{s}}
 \frac{\hat{s}^{2}}{\left(\hat{s}-m_{G}^{2}\right)^{2}+m_{G}^{2}\Gamma_{G}^{2}}
 \left(f_{L}^{2}+f_{R}^{2}\right)\left(g_{L}^{2}+g_{R}^{2}\right)
 \nonumber \\
 & \times & 
 \left\{ 1+\frac{2g_{L}g_{R}}{g_{L}^{2}+g_{R}^{2}}
 \left(1-\beta^{2}\right)+2\frac{\left(f_{L}^{2}-f_{R}^{2}\right)
 \left(g_{L}^{2}-g_{R}^{2}\right)}{\left(f_{L}^{2}+f_{R}^{2}\right)
 \left(g_{L}^{2}+g_{R}^{2}\right)}\beta\cos\theta
 +\left(\beta\cos\theta\right)^{2}
 \right\}.
 \label{eq:dsdz_zprime}
\end{eqnarray}
Negative searches for the $Z^{\prime}$ boson at the Tevatron impose
several lower bounds on the $Z^{\prime}$ mass, roughly above 1~TeV for couplings of order electroweak size.
For a leptophobic $Z^{\prime}$ boson, the bound is slightly looser,
$m_{Z^{\prime}}>700\,{\rm GeV}$. Owing to the rapid drop of the PDFs,
the $Z^{\prime}$ boson contributes significantly only in the resonance
region, where $\beta\to1$. Further noting that the
coefficient of the $(1-\beta^{2})$ term in Eq.~(\ref{eq:dsdz_zprime})
is always less than one, we can drop this term and obtain 
\begin{equation}
A_{FB}^{Z^{\prime}}\propto\frac{\left(f_{L}^{2}-f_{R}^{2}\right)
\left(g_{L}^{2}-g_{R}^{2}\right)}{\left(f_{L}^{2}+f_{R}^{2}\right)
\left(g_{L}^{2}+g_{R}^{2}\right)}.
\label{eq:AFB_NP}
\end{equation}

% This is Figure 12
\begin{figure}
\includegraphics[scale=0.43]{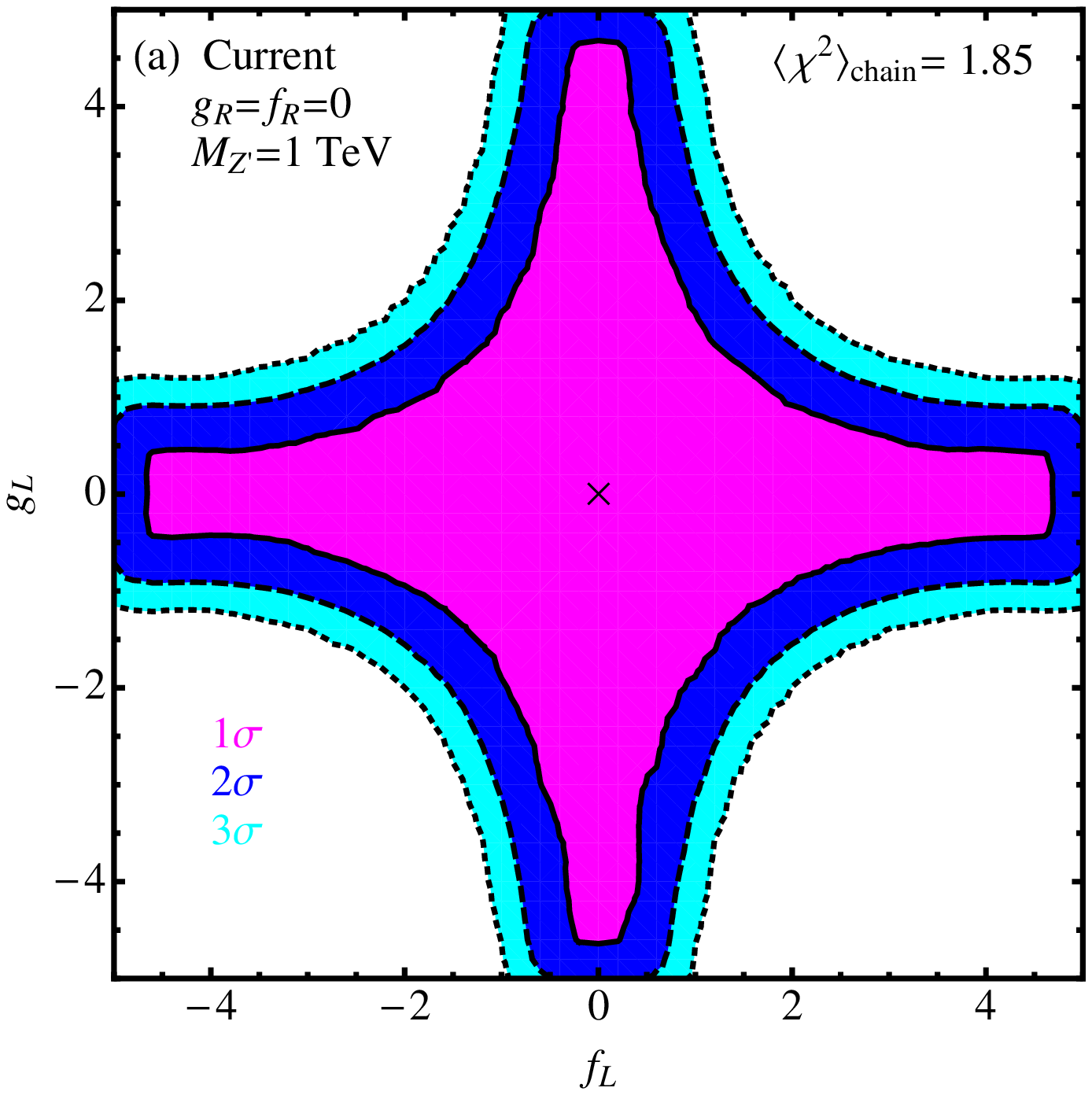}
\includegraphics[scale=0.43]{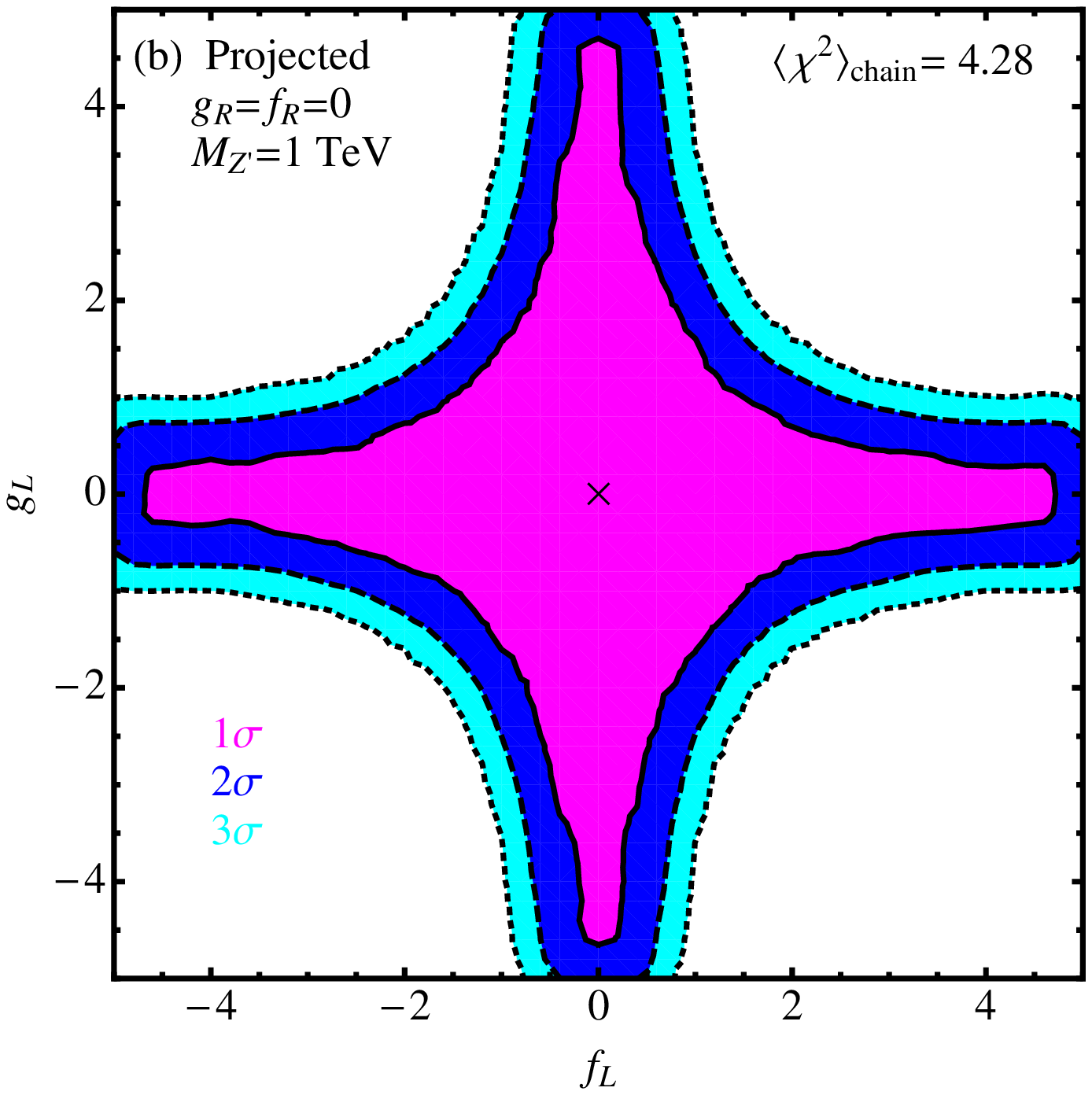}\\
\caption{Same as Fig.~\ref{fig:gp1000} but for 1000~GeV $Z^{\prime}$.
\label{fig:zp1000}}
\end{figure}

\begin{figure}
\includegraphics[scale=0.43]{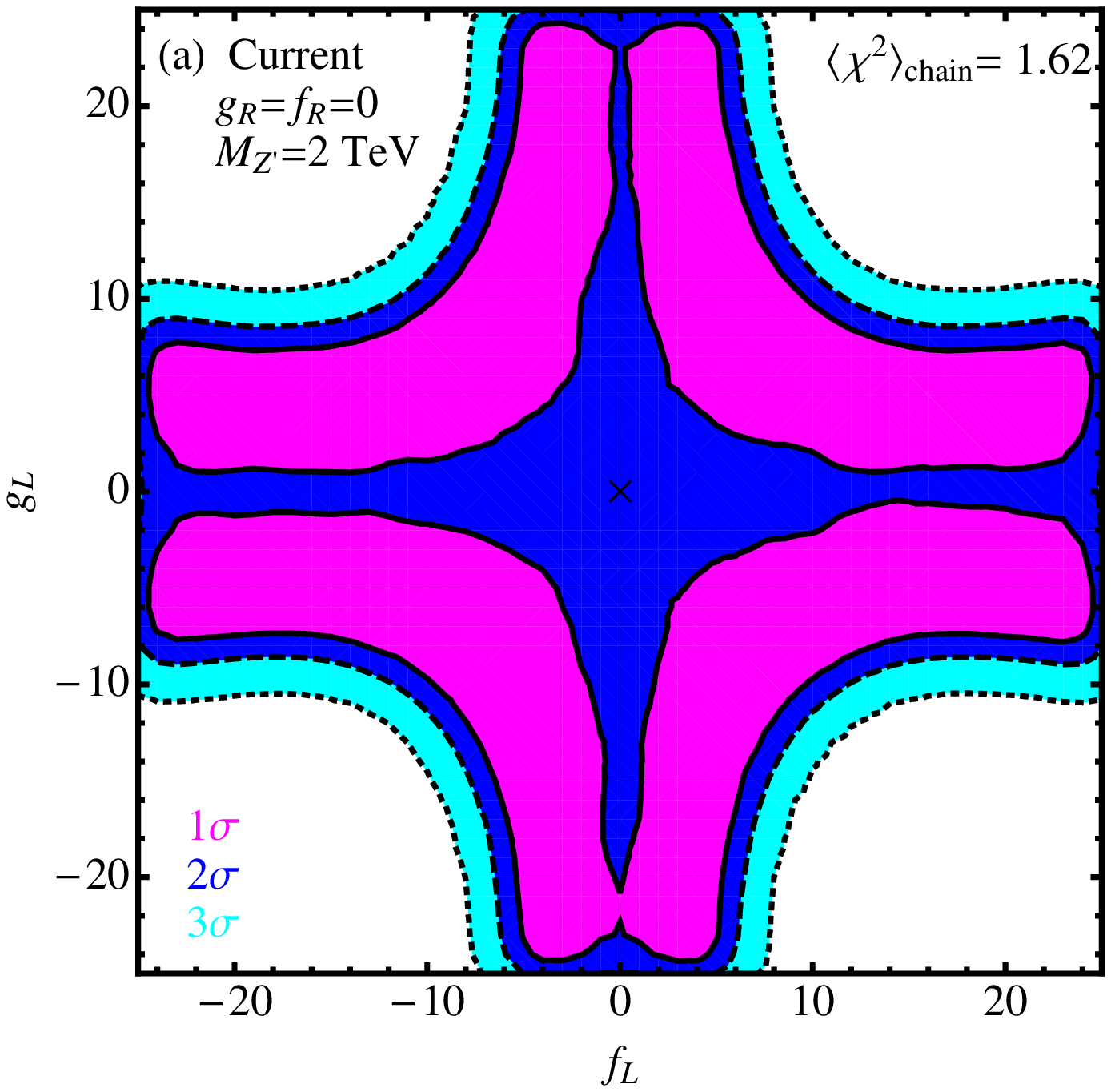}
\includegraphics[scale=0.43]{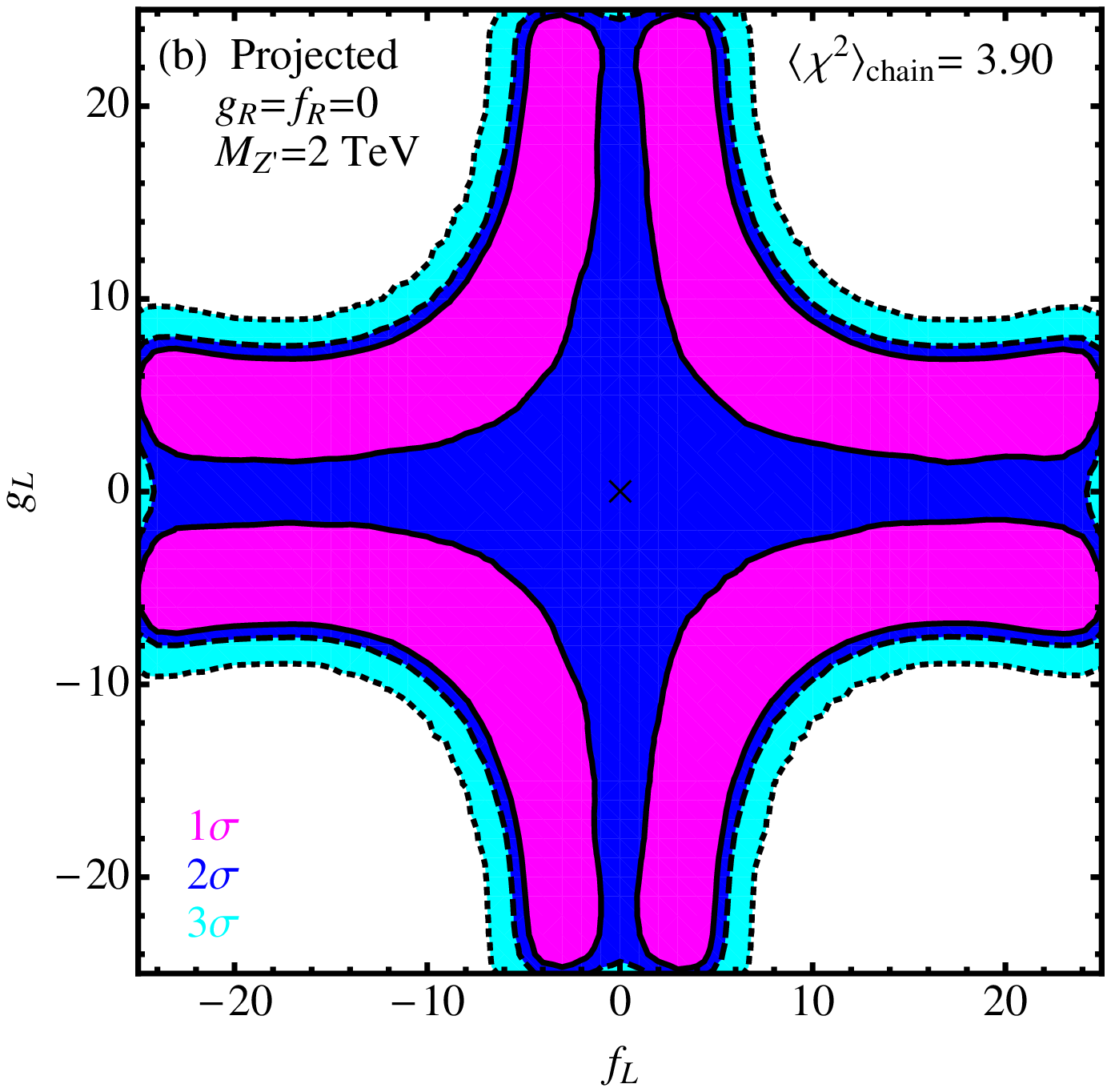} \\
\caption{Same as Fig.~\ref{fig:gp1000} but for 2~TeV $Z^{\prime}$.
\label{fig:zp2000}}
\end{figure}

As in the $G^\prime$ study, we turn off the right-handed couplings first. 
Fig.~\ref{fig:zp1000} displays the correlation between $f_L$ and $g_L$
couplings for a 1000~GeV $Z^\prime$ boson with $f_R=g_R=0$.  
Like the 1~TeV $G^\prime$, the $M_{t\bar{t}}$ distribution favors smaller $f_L$ or $g_L$.  The fit is worse than in the SM with $\left<\chi^2\right>_{chain}=1.85$ for the current luminosity and $\left<\chi^2\right>_{chain}=4.28$ for $10~{\rm fb}^{-1}$ if the central values remain the same.
For a 2~TeV $Z^\prime$ boson with $f_R=g_R=0$, we see the results of the MCMC scan in Fig.~\ref{fig:zp2000}.   The fit is somewhat better than the SM: $\left<\chi^2\right>_{chain}=1.62$ for the current luminosity and $\left<\chi^2\right>_{chain}=3.90$ for $10~{\rm fb}^{-1}$ due to the lessening of the importance of $M_{t\bar{t}}$ for the higher mass $Z^\prime$.  Furthermore, large couplings are allowed at the $1\sigma$ level.  However, note that the unitarity 
constraint derived for the process 
$u\bar{u}\to Z^\prime \to t\bar{t} $ requires $|f_R|\lesssim28$; see 
Appendix~\ref{sec:unitarity} for further details.  Now, the heavy $Z^\prime$ contributions are very sensitive to width effects.  For $f_L=g_L=10$ we obtain $\Gamma_{Z^\prime}\simeq 0.55 m_{Z^\prime}$.  The positive contributions to the $M_{t\bar{t}}$ distribution, particularly in its last bin, are somewhat constrained.  We conclude that this model can offer a small improvement over the SM in describing $A_{FB}^{tot}$, $\sigma_{t\bar t}^{tot}$, and the $t\bar{t}$ invariant mass distribution simultaneously.

% This is Figure 13

%--------------------------------------------------------------------%
%  Section VI                                                        %
%                                                                    %
%   t-channel diagram for Z-prime and W-prime bosons                 %
%                                                                    %
%--------------------------------------------------------------------%  
\section{Flavor-violating $Z^{\prime}$ and $W^{\prime\pm}$ models
\label{sec:Flavor-violating}}

In this section we consider a flavor-violating $Z^\prime$ model, which
includes a $u$-$t$-$Z^{\prime}$ interaction. Such a FCNC could appear at tree-level or loop-level. 
Rather than focus on a specific model, we consider the following effective
coupling of $u$-$t$-$Z^{\prime}$~\cite{Jung:2009jz}:
\begin{equation}
\mathcal{L}=e\bar{u}\gamma^{\mu}(f_{L}P_{L}+f_{R}P_{R})tZ_{\mu}^{\prime},
\label{eq:lang-zprime}
\end{equation}
where $e$ denotes the electromagnetic coupling strength. In addition to the SM QCD 
production channel, $u\bar{u} \to g \to t\bar{t}$, the top quark pair can also
be produced via the process $u\bar{u}\to t\bar{t}$ with a $t$-channel
$Z^{\prime}$ boson propagator.  The top quark asymmetry is naturally generated
by this new process which also interferes with the SM production mode.  
Therefore, the differential cross section versus the cosine of the top
production angle is $\theta$ given as follows~\cite{Cheung:2009ch}:
\begin{equation}
\frac{d\hat{\sigma}}{d\cos\theta}=\mathcal{A}_{SM}+
\mathcal{A}_{INT}^{Z^{\prime}}+\mathcal{A}_{NP}^{Z^{\prime}},
\label{eq:tch-zprime}
\end{equation}
where 
\begin{eqnarray}
\mathcal{A}_{INT}^{Z^{\prime}} & = &
 \frac{\beta}{72\pi\hat{s}}
 \frac{e^{2}g_{s}^{2}(g_{L}^{2}+g_{R}^{2})}{\shat(t-M_{Z^{\prime}}^{2})}
 \left[2(\hat{u}-m_{t}^{2})^{2}+2\shat m_{t}^{2}+
 \frac{m_{t}^{2}}{M_{Z^{\prime}}^{2}}((t-m_{t}^{2})^{2}
 +\shat m_{t}^{2})\right],\\
\mathcal{A}_{NP}^{Z^{\prime}} & = &
 \frac{\beta}{128\pi\shat}
 \frac{e^{4}}{(t-M_{Z^{\prime}}^{2})^{2}}
 \biggl\{4\left[(g_{L}^{4}+g_{R}^{4})(\hat{u}-m_{t}^{2})^{2}+
 2g_{L}^{2}g_{R}^{2}\shat(\shat-2m_{t}^{2})
 \right]\nonumber \\
 &  &
 \qquad\qquad\qquad\qquad+\frac{m_{t}^{4}}{M_{Z^{\prime}}^{4}}
 (g_{L}^{2}+g_{R}^{2})^{2}
 \left(4\shat M_{Z^{\prime}}^{2}+(t-M_{Z^{\prime}}^{2})^{2}\right)
 \biggr\}, 
\end{eqnarray}
and $\mathcal{A}_{SM}$ is given in Eq.~(\ref{eq:dsdz_sm}).
The interference between the QCD and EW processes can be easily understood
as follows. The $SU(N)$ gluon propagator can be split into a $U(N)$ gluon 
propagator and a $U(1)$ gluon propagator~\cite{Maltoni:2002mq}, 
\begin{equation}
   \sum_a t^a_{ij} t^a_{kl} = \frac{1}{2}\left(\delta_{il}\delta_{kj}
   -\frac{1}{N}\delta_{ij}\delta_{kl}\right),   
\end{equation}
where the $U(1)$ gluon, carrying a factor $1/N$, is unphysical.  Color flow of
the SM QCD channel (i.e., $u\to t$ and $\bar{t} \to \bar{u}$) is then exactly
the same as the $Z^\prime$ induced $t$-channel diagram, resulting in
interference between both processes.  
            
Within the SM, the FCNC coupling $u$-$t$-$Z$ vanishes at tree-level, but can 
be generated at one loop.  However, the one-loop generated coupling is strongly
suppressed by the GIM mechanism, making the FCNC top interactions very small.
In models beyond the SM this GIM suppression can be relaxed, and one-loop
diagrams mediated by new particles may also contribute, yielding effective
couplings orders of magnitude larger than those of the SM. Since the coupling
strength of this FCNC interaction is typically at the order of the SM
weak interaction, the coefficients $f_{L}$ and $f_{R}$ are expected
to be much smaller than 1. Therefore, it is not easy to generate a large asymmetry from a loop-induced $u$-$t$-$Z$ interaction.  However, the couplings $f_{L}$ and $f_{R}$
could be larger if they are generated at tree-level. 

While the value of $f_R$ is not well constrained by direct or indirect search
experiments, the value of $f_L$ is tightly bounded by the $B$-sector.  
The left-handed coupling $f_{L}$ in Eq.~(\ref{eq:lang-zprime}) originates
from the gauge interaction of the $Z^{\prime}$ boson to the first and
third generation quark doublets, 
\begin{equation}
\mathcal{L}=\bar{q}_{L}i \gamma^{\mu} D_{\mu} Q_{L}+h.c.\,,
\label{eq:zprime-1-3-coupling}
\end{equation}
where $q_{L}(Q_{L})$ denotes the first (third) generation quark doublet
and the covariant derivative is
$D_{\mu}=i\partial_{\mu}+iB^{\prime}Z_{\mu}^{\prime}$, where $B^\prime$ is the charge.
The flavor violating interaction $d_{L}$-$b_{L}$-$Z^{\prime}$ then follows 
directly from the gauge invariance, which can contribute
to the $B_{d}^{0}$-$\bar{B}_{d}^{0}$ mixing at the tree level~\footnote{
A similar correlation among the gauge boson and the third
generation quarks in the SM has been studied in Ref.~\cite{Berger:2009hi}.}.
A coupling of the form
\begin{equation}
\mathcal{L}=e\bar{d}\gamma^{\mu}(f_{L}V_{ud}^*V_{tb}P_{L})bZ_{\mu}^{\prime}~~,
\label{eq:d-b-zprime-coupling}
\end{equation}
follows from Eq.~\ref{eq:zprime-1-3-coupling} after rotating to the mass eigenstate basis with $f_L$ as in Eq.~\ref{eq:lang-zprime} and $V_{ud}$ and $V_{tb}$ elements of the CKM matrix.
Assuming no additional NP effects arise, this gives a contribution to the mass 
difference between  $B_{d}^{0}$-$\bar{B}_{d}^{0}$ of
\begin{equation}
\Delta m=\frac{4e^2 f_{L}^2}{3}\left|V_{ud}^*V_{tb}\right|^2\frac{f_{B_d}^2\hat{B}
M_{B_d}}{M_{Z^{\prime}}^{2}}
\label{eq:delta-m-zprime}
\end{equation}
where $f_{B_d}$ is the $B_d$ decay constant and $\hat{B}$ is the 
``bag parameter" that characterizes the deviation from the vacuum saturation 
approximation.  If we conservatively require that this contribution does not exceed 
the experimental value of $3.34\times10^{-10}~{\rm MeV}$~\cite{Amsler:2008zzb}, 
we can set a limit on $f_L$ of
\begin{equation}
f_{L}<3.5\times 10^{-4}\left(\frac{M_{Z^\prime}}{100~{\rm GeV}}\right),
\label{eq:fL-limit-zprime}
\end{equation}
where we use $f_{B_d}\sqrt{\hat{B}}=216\pm 15~{\rm MeV}$~\cite{Gamiz:2009ku} and $\left|V_{ud}^*V_{tb}\right|\simeq1$~\cite{Amsler:2008zzb}.
As a result, we choose $f_L=0$ hereafter. Furthermore, the unitarity 
constraint derived for the process 
$u\bar{t}\to Z^\prime \to \bar{u} t$ requires only $|f_R|\lesssim28$; see 
Appendix~\ref{sec:unitarity} for further details.  

\begin{figure}
\includegraphics[scale=0.45]{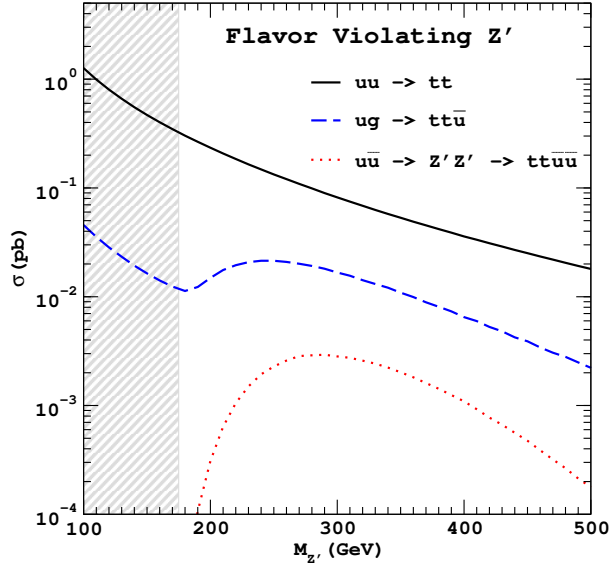}
\caption{Value of the same-sign top quark production cross section at the Tevatron for a flavor violating $Z^\prime$ with $f_L=0$, $f_R=1$ (Eq.~(\ref{eq:lang-zprime})). 
\label{fig:zprime-tevatron}}
\end{figure}

The most striking signal of the FCNC $Z^{\prime}$ model is same-sign top pair
production through the processes $uu \to tt$ via a $t$-channel diagram mediated
by the $Z^\prime$ boson.  Recently, the CDF collaboration searched for the
same-sign top pair signature induced by the maximally flavor-violating scalars
at the Tevatron and found no evidence of new physics beyond the
SM~\cite{Aaltonen:2008hx}. In their analysis, the upper limit to the production cross section of
same-sign top pairs is of the order of $0.7~\rm{pb}$.   We show, in Fig.~\ref{fig:zprime-tevatron}, the same-sign top production cross section at the Tevatron for couplings $f_R=1$ and $f_L=0$.  The cross section scales with the right coupling as $\sigma(tt+\bar t\bar t) \sim f_R^4$ if $f_L=0$.  Direct production via $t$-channel $Z'$ exchange dominates and is severely constrained as the couplings increase.  Note that we do not consider the possible effects that same-sign top production could have on a measurement of $A_{FB}$ which requires a delicate analysis and will be presented elsewhere~\cite{future}.

% This is Figure 14
% This is Figure 15
\begin{figure}[h!]
\includegraphics[scale=0.45]{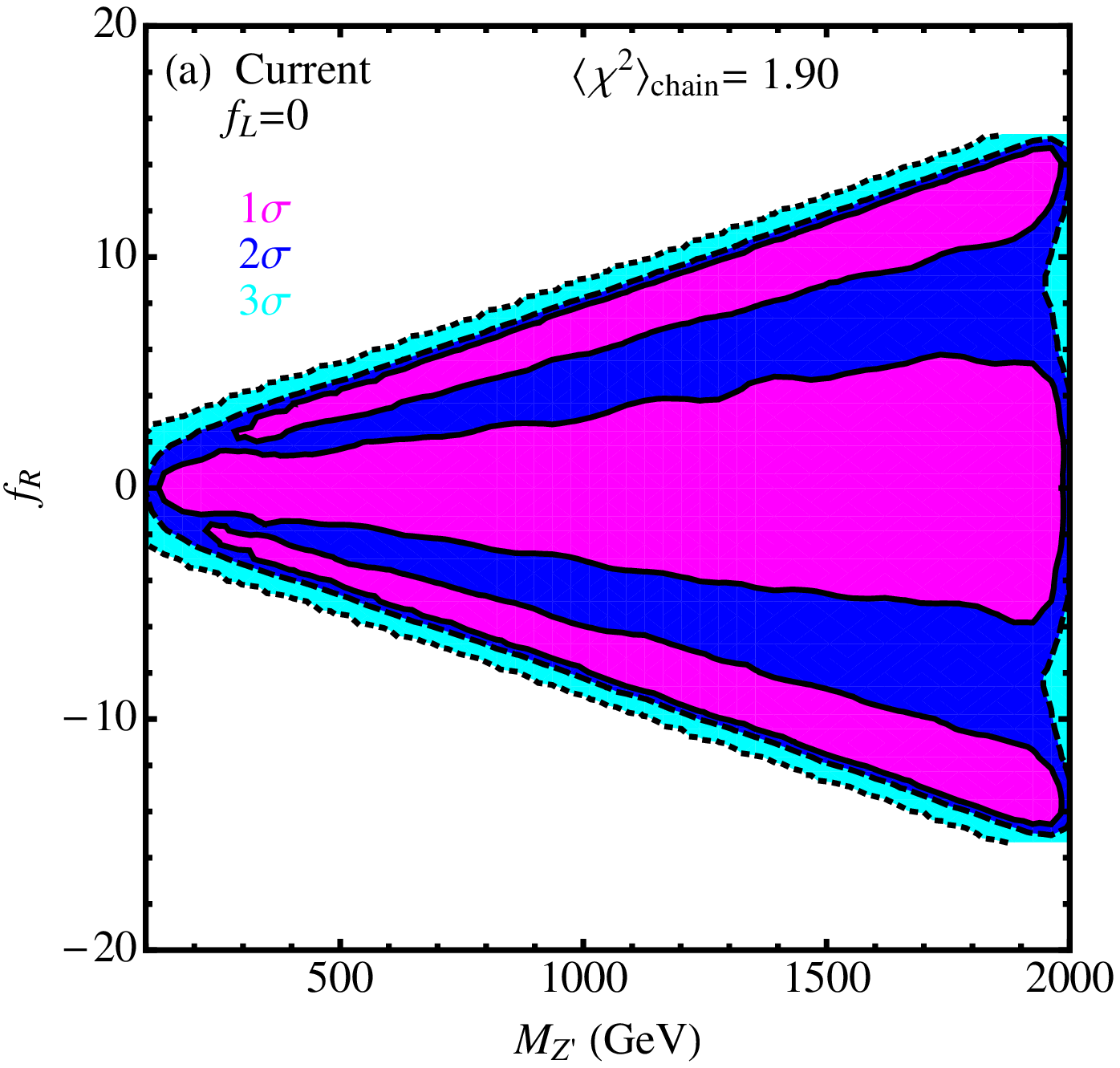}
\includegraphics[scale=0.45]{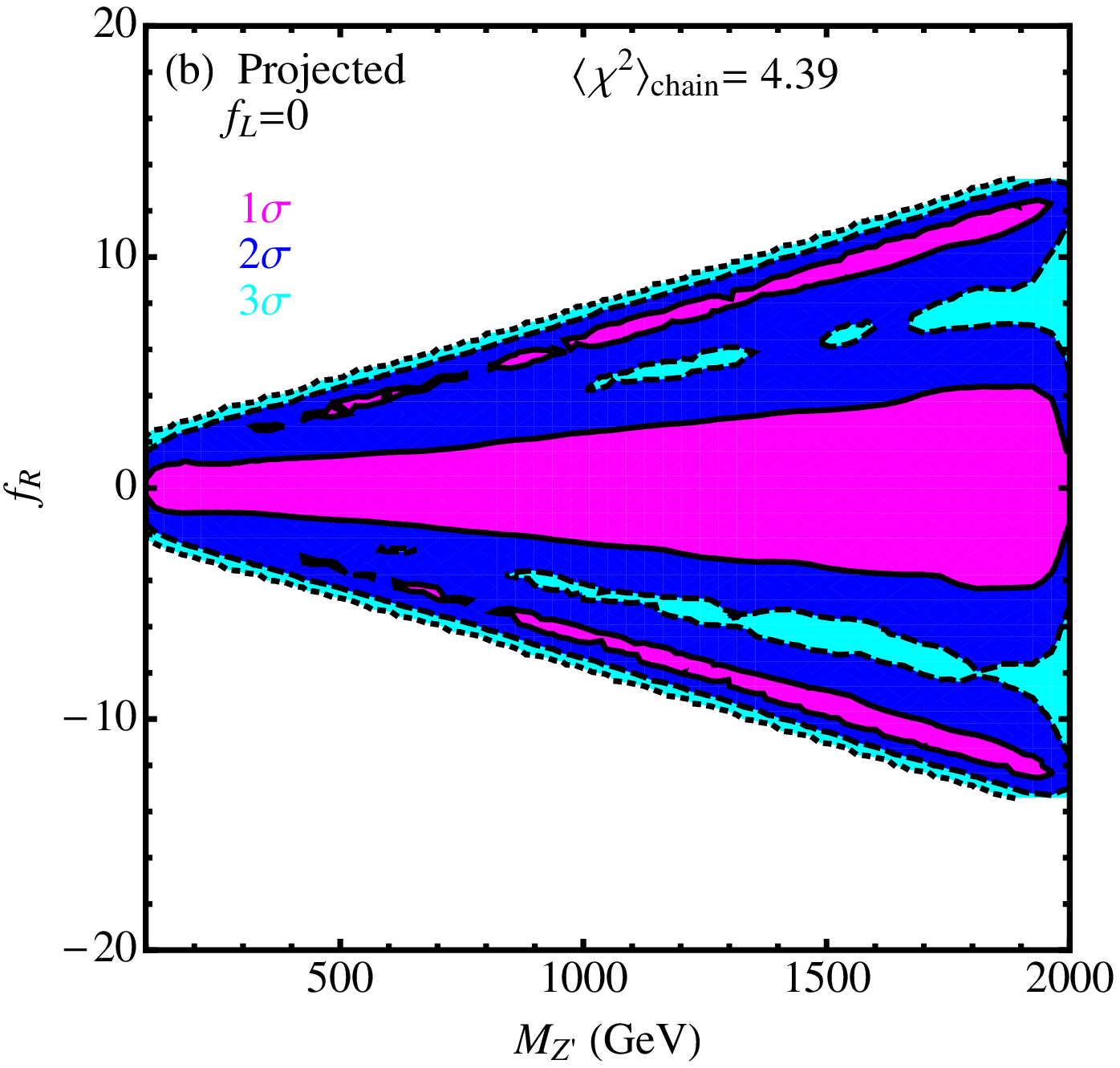}\\
\caption{Correlation of $f_R$ and $m_{Z^\prime}$
with $f_{L}=0$. The left panel is for the current integrated luminosity while the right panel is for an upgraded
luminosity of $\int \mathcal{L} dt =10\,{\rm fb}^{-1}$.  
Tension with the same-sign top production constraint from the Tevatron prevents this model from becoming a better fit than the SM.
\label{fig:tchan-zp-1000-right}}
\end{figure}

In Fig.~\ref{fig:tchan-zp-1000-right} we show the result of the  MCMC scan 
in the plane of $m_{Z^\prime}$ and $f_R$: (a) for the current integrated luminosity and 
(b) for expected $10~{\rm fb}^{-1}$.  For the current integrated luminosity, we impose the constraint of
$\sigma(tt+\bar t\bar t)<0.7~\rm{pb}$ \cite{Aaltonen:2008hx}, whereas for 10$~{\rm fb}^{-1}$, we assume the cross section limit scales with $1/\sqrt{\cal L}$, giving $\sigma(tt+\bar t\bar t)<0.4~\rm{pb}$.
The value of $\left<\chi^2\right>_{chain}= 1.90$ for the current integrated luminosity
indicates that the FCNC $Z^\prime$ model fits $\sigma(t\bar{t})$ 
$A_{FB}$, and the $M_{t\bar{t}}$ distribution worse than the SM.  Note the quality of fit is maintained even if we fix the $Z'$ mass to be specific values as in the flavor-conserving $G'$ and $Z'$ cases.  Since the INT effects lead to a negative asymmetry, 
one needs a large NPS contribution to overcome the negative INT contributions 
to generate the positive asymmetry. 
That requires a very large $f_R$ coupling, as seen in the upper and lower $1\sigma$ contours of Fig.~\ref{fig:tchan-zp-1000-right}(a) and (b), which is near 
the constraint of $\sigma(tt+\bar t\bar t)<0.7~\rm{pb}$. In this model, the predicted value for the same-sign top pair production cross section is pushed to just below the limit taken.  Overall, while there is tension between the positive asymmetry and small $\sigma(tt+\bar t\bar t)$, we find a fit not much worse than the SM.  With $10~{\rm fb}^{-1}$, the fit remains worse than in the SM with $\left<\chi^2\right>_{chain}= 4.39$.

The observed top asymmetry may also be induced by a flavor-changing interaction
via a charged $W^{\prime}$ boson ~\cite{Cheung:2009ch, Barger:2010mw} . The top
quark pair can be produced in the channel $d\bar{d}\to t\bar{t}$
via a $t$-channel $W^{\prime}$ boson propagator.
As in the flavor-violating $Z'$ case, we consider the following effective
$d$-$t$-$W^{\prime}$ coupling:
\begin{equation}
\mathcal{L}=e\bar{d}\gamma^{\mu}(f_{L}P_{L}+f_{R}P_{R})tW_{\mu}^{\prime},
\end{equation}
where $e$ denotes the electromagnetic coupling strength. The differential cross section
of $d\bar{d}\to t\bar{t}$ is the same as Eq.~(\ref{eq:tch-zprime}) with the 
substitution $u(\bar{u})\to d(\bar{d})$.

% This is Figure 16
\begin{figure}
\includegraphics[scale=0.5]{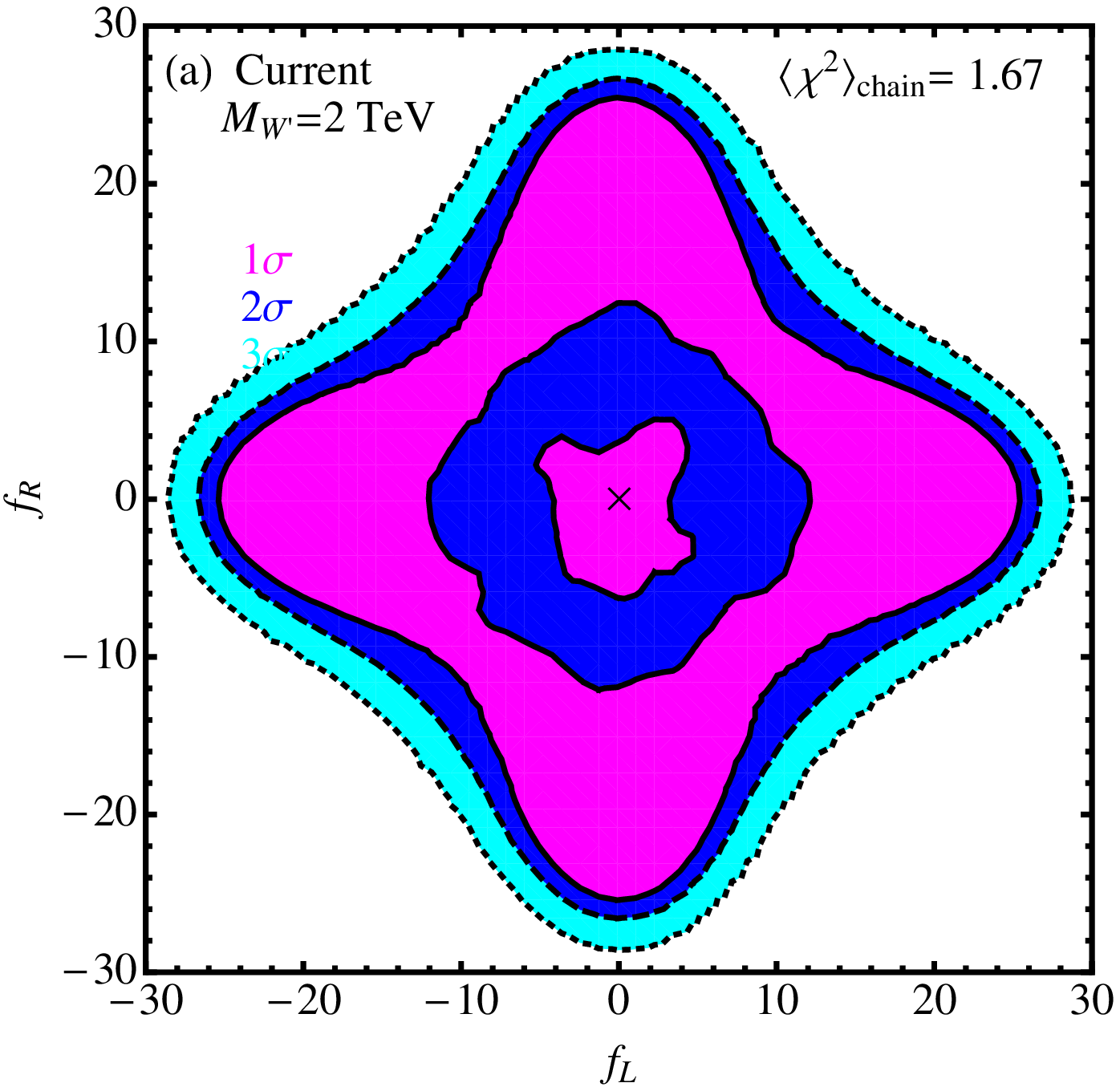}
\includegraphics[scale=0.5]{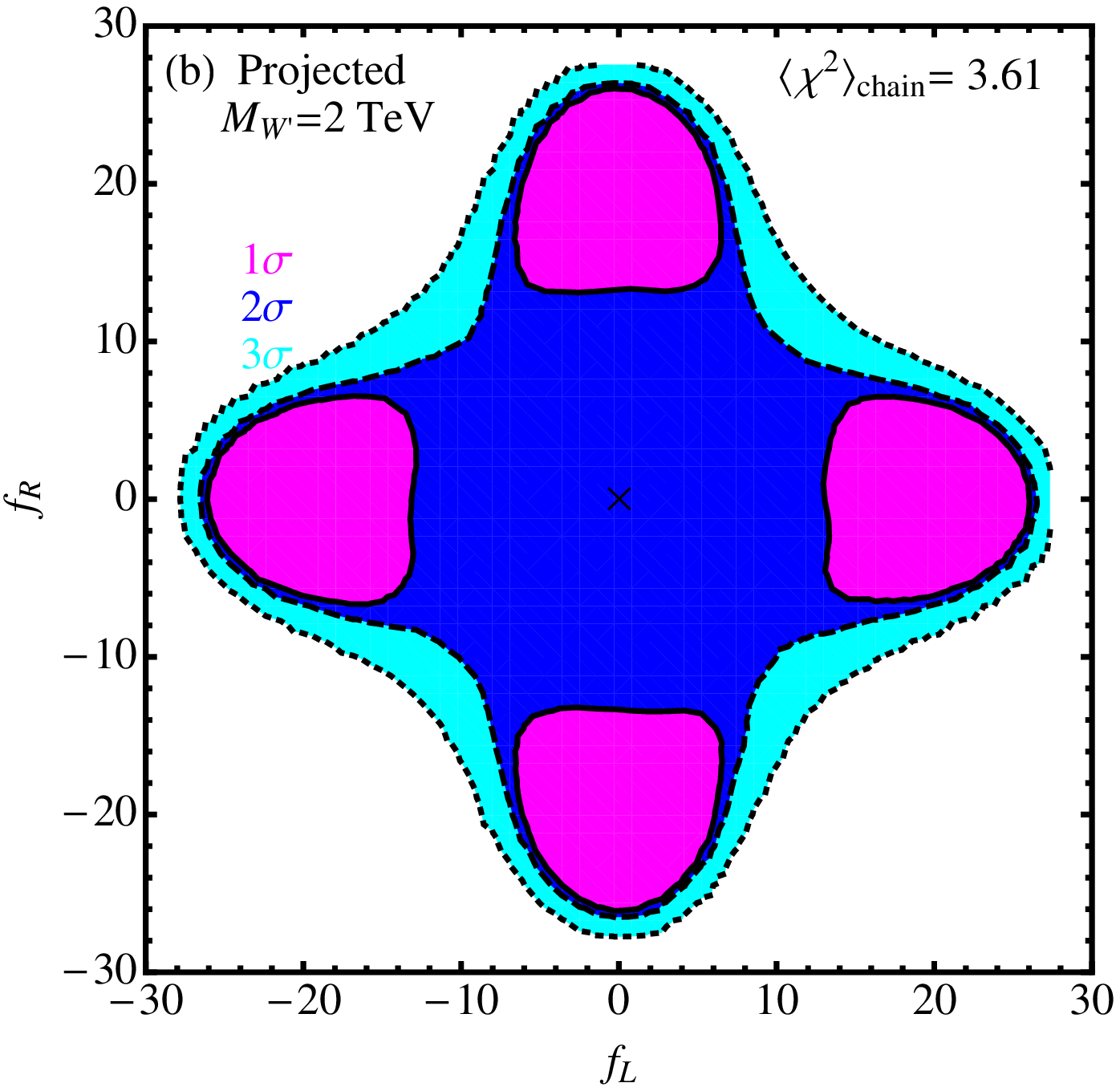}\\
\caption{
Correlation of couplings for $m_{W^{\prime}}=2000\,{\rm GeV}$.  The left panel
is for the current integrated luminosity
while the right panel is for an upgraded luminosity of $\int \mathcal{L} dt
=10\,{\rm fb}^{-1}$.  The $1\sigma$ region is bounded by the solid curves, the $2\sigma$ by the dashed curves, and the $3\sigma$ by the dotted curves.
Couplings of either $f_L\sim0$ and $f_R\sim \pm10$ or $f_R\sim0$ and $f_L\sim
\pm10$ are consistent with the $A_{FB}$ and $\sigma_{t\bar t}$ measurements.
\label{fig:tchan-wp-2000-fl-fr}}
\end{figure}  

One advantage of the flavor-violating $W^\prime$ model is that it does not 
suffer from the constraint of same-sign top pair production at the Tevatron. 
Fig.~\ref{fig:tchan-wp-2000-fl-fr} displays the correlation between 
$f_L$ and $f_R$ couplings for a 2000~GeV $W^\prime$ boson. For the current luminosity the $1\sigma$
contours are symmetric for $f_L$ and $f_R$ and the innermost one includes $f_L=f_R=0$.  
In order to generate positive asymmetry, the couplings $f_L$ and $f_R$ need 
to be large enough to overcome the negative INT contributions. The typical
values of couplings $f_L$ and $f_R$ are in the range of $\pm10$ to $\pm20$.
For the current luminosity, $\langle\chi^2\rangle_{chain}=1.67$ which indicates a better overall fit than
the FCNC $Z^\prime$ boson due to the lack of the same-sign top constraint, and a better fit than the SM.  Due to the PDF dependence, the $d$-quark initiated $t\bar t$ production via the $W^\prime$ is smaller than the $u$-quark initiated production through a $Z^\prime$.  Therefore, larger couplings are required to maintain the production cross section than in the $Z^\prime$ case.  With upgraded luminosity, $\left<\chi^2\right>_{chain}=3.61$ provided that the central values of $A_{FB}^{tot}$ and $\sigma_{t\bar t}^{tot}$ are maintained which offers more improvement over the SM.

We focus on a heavy $W^\prime$ due to general constraints from electroweak precision and flavor measurements.  In general, a $W^\prime$ is associated with a broken non-abelian gauge group and one must also consider a neutral gauge boson, $Z^\prime$, whose mass is typically degenerate or nearly so with that of the $W^\prime$.  If this $Z^\prime$ has predominantly flavor-changing couplings to top quarks, then it falls into the previous case we analyzed.  If its coupling to top quarks is flavor-conserving then one would expect to produce top quark pairs through s-channel $Z^\prime$ exchange.  However, such a process suffers from PDF suppression and is negligible in comparison to the $W^\prime$ contribution considered above and therefore we ignore it here.  \footnote{A model with a light $W^\prime$ and $Z^\prime$ has been proposed in Ref.~\cite{Barger:2010mw} and may lead to a naturally good fit.  This model has potential implications for precision electroweak observables which have not yet been fully explored.}

%--------------------------------------------------------------------%
%  Section VII                                                       %
%                                                                    %
%   t-channel diagram for S-prime and S+/- prime scalar              %
%                                                                    %
%--------------------------------------------------------------------% 
\section{Flavor-violating scalar $S/S^{\pm}$\label{sec:scalar}}

In addition to spin-1 exchange, we also consider the FCNC top interaction with
a new color singlet scalar $S^\prime$: 
\begin{equation}
\mathcal{L}\supset e S^\prime
\left(f_{L}\bar{u}_{R}t_{L}+f_{R}\bar{u}_{L}t_{R}\right),
\label{eq:sprime}
\end{equation}
where $S^\prime$ is an $SU(2)$ doublet and we parameterize the overall 
coupling strength with respect to the weak coupling $e$. 
If we assume 
$S^\prime$ to be the SM Higgs boson, then the FCNC top interaction originates 
from the dimension-6 operator
$$\mathcal{L}=\frac{v^{2}}{\Lambda^{2}}h\left(f_{L}\bar{u}_{R}t_{L}+
f_{R}\bar{u}_{L}t_{R}\right).$$
Results for this operator can be obtained from those for Eq.~(\ref{eq:sprime}) with the substitution 
$e \to {v^2\over \Lambda^2}$. 
As will be shown below, such an effective coupling is too small to 
generate a sizable asymmetry however. Hence, it is difficult to explain
the asymmetry with the SM Higgs boson effective coupling without
introducing additional heavy scalars. 

The differential cross section is written as
\begin{equation}
\frac{d\hat{\sigma}}{d\cos\theta}=
\mathcal{A}_{SM}+\mathcal{A}_{INT}^{S^{\prime}}+\mathcal{A}_{NPS}^{S^{\prime}}
\end{equation}
where 
\begin{eqnarray}
\mathcal{A}_{INT}^{S^{\prime}} & = & 
\frac{2\pi\alpha_{s}\alpha_{em}\beta \left(f_{L}^{2}+f_{R}^{2}\right)
\left[\hat{s}m_{t}^{2}+
\left(\hat{t}-m_{t}^{2}\right)^{2}\right]}
{9\shat^{2}\left(\hat{t}-m_{S^{\prime}}^{2}\right)},\\
\mathcal{A}_{NPS}^{S^{\prime}} & = &
 \frac{\pi\alpha_{em}^{2}\beta\left(f_{L}^{2}+f_{R}^{2}\right)^{2}\left(\hat{t}
 -m_{t}^{2}\right)^{2}}
 {8\shat\left(\hat{t}-m_{S^{\prime}}^{2}\right)^{2}}, 
\end{eqnarray}
with $\alpha_{em}\equiv e^2/(4\pi)$. 
Due to the repulsive scalar interaction, the NPS contributions generate a
negative $A_{FB}^{NP}$.  In order to generate a positive $A_{FB}^{NP}$, the 
scalar $S^\prime$
needs to be very light, generally $m_{S^\prime}< m_{t}$, and to have large
couplings with the top quark.  However, such a light scalar leads to a new top
quark decay channel $t\to S^{\prime} u$, which is tightly constrained
\cite{Amsler:2008zzb}.  Therefore, we consider scalar masses that are larger
than top quark mass to forbid this new decay channel.  Furthermore, the 
flavor-violating coupling will be highly constrained by $D^0$-$\bar{D}^0$ 
mixing (a $\Delta C=2$ process) if one assumes a universal flavor-violating 
coupling among the three families of quarks.  However, from a purely 
phenomenological perspective, we assume that the second generation quarks are 
not involved in the flavor-violating Yukawa interaction which leads to no 
constraint on the Yukawa couplings $f_L$ and $f_R$.  In other words, $f_{ut}$ 
($f_L$,$f_R$) is taken as a free parameter and is only constrained by 
considerations of unitarity (see Appendix~\ref{sec:unitarity} for details).

% This is Figure 17
\begin{figure}
\includegraphics[scale=0.4]{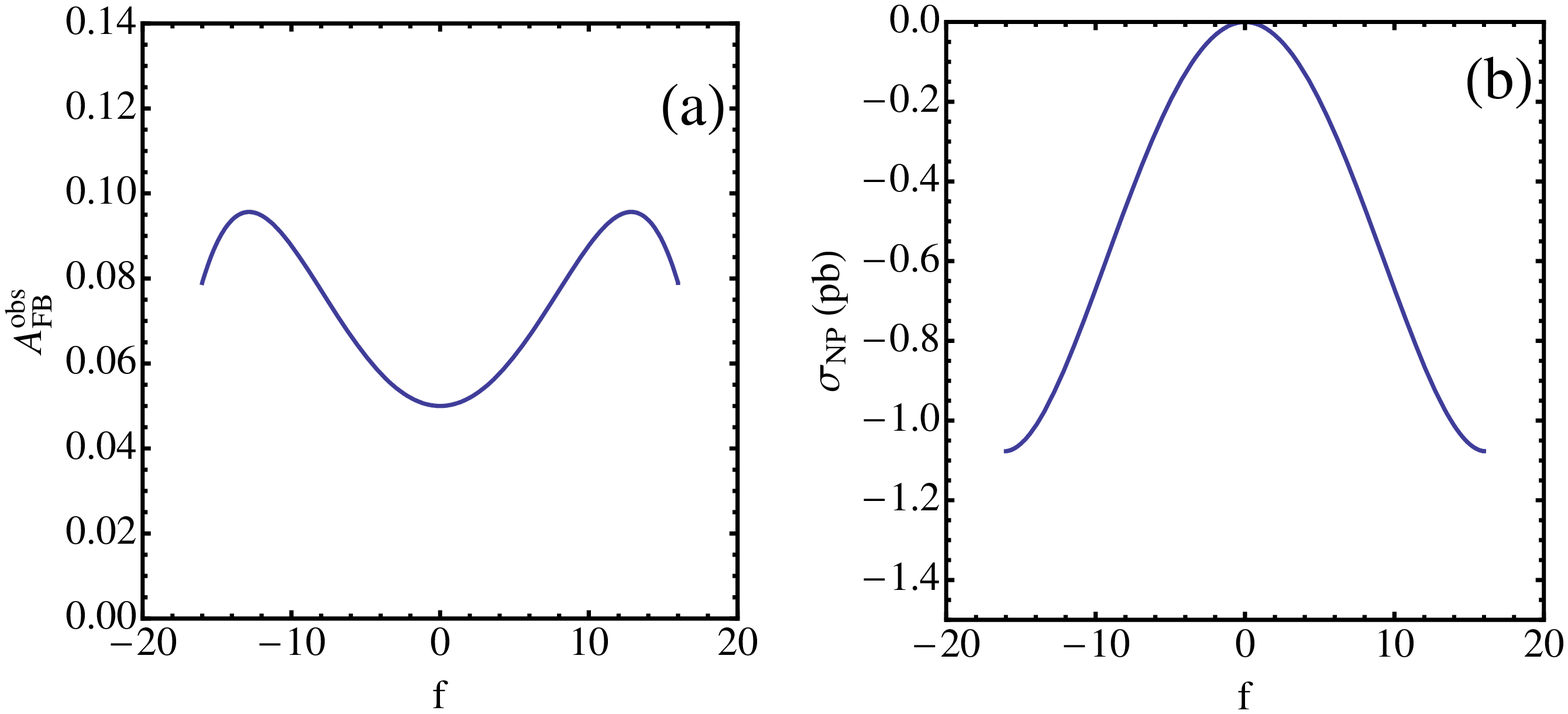}\\
~~~~\includegraphics[scale=0.38]{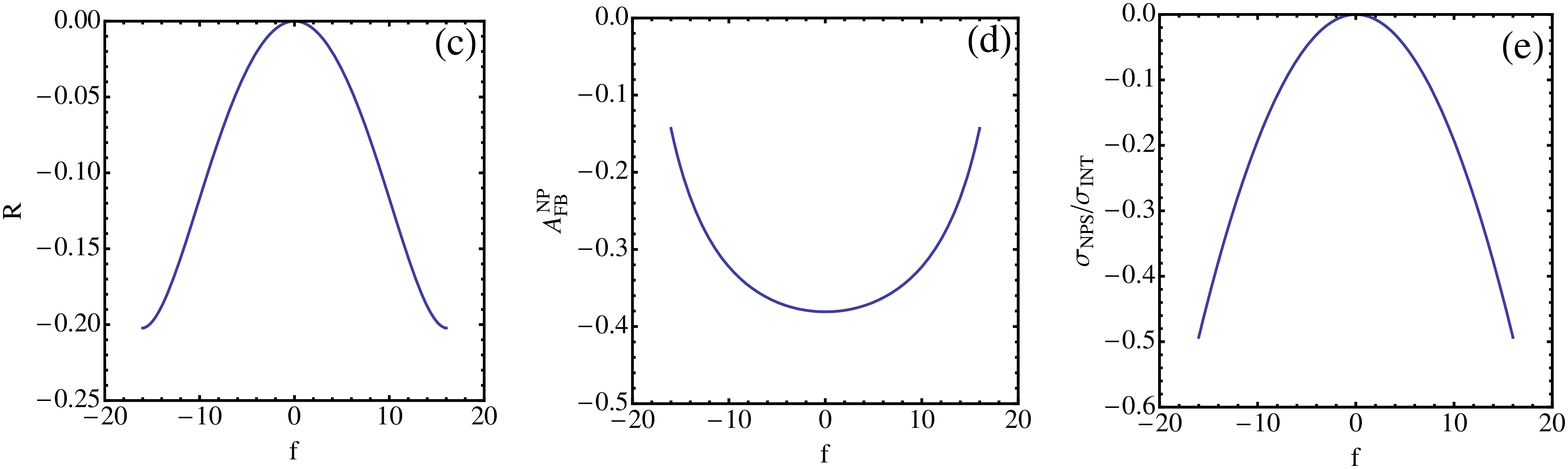}
\caption{Theoretical expressions for a flavor-violating neutral scalar $S$.  
(a) $A_{FB}^{tot}$ (cf. Eqs.~(\ref{eq:AFB1})-(\ref{eq:AFB})) as a function of 
$f=\sqrt{f_L^2+f_R^2}$ for a flavor violating scalar with $m_S=2~{\rm TeV}$.  
(b) $\sigma_{NP}=\sigma_{INT}+\sigma_{NPS}$ vs. $f$.  Note that the 
interference is destructive.  (c) $R$, as defined in 
Eq.~(\ref{eq:def-afbnp-R}).  Note that it is negative for all $f$.  (d)  
$A_{FB}^{NP}$ vs. $f$ which is negative, as mentioned in the text.  (e) The 
ratio of $\sigma_{NPS}$ to $\sigma_{INT}$.  The INT term dominates over the 
NPS term for a scalar of mass $2~{\rm TeV}$.
\label{fig:tchan-scalar}}
\end{figure}

% This is Figure 18
\begin{figure}
\includegraphics[scale=0.45]{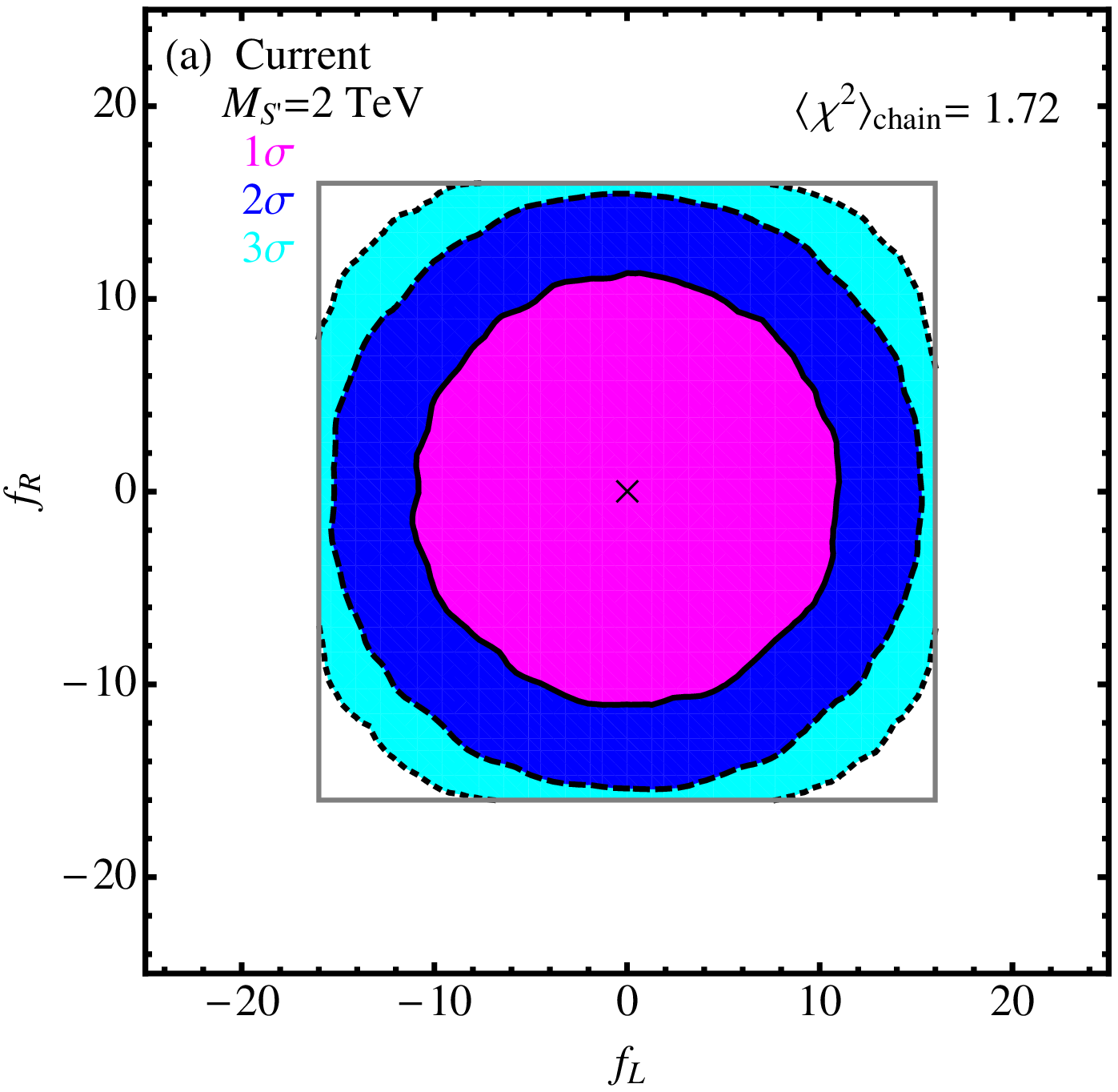}%
~~~~\includegraphics[scale=0.45]{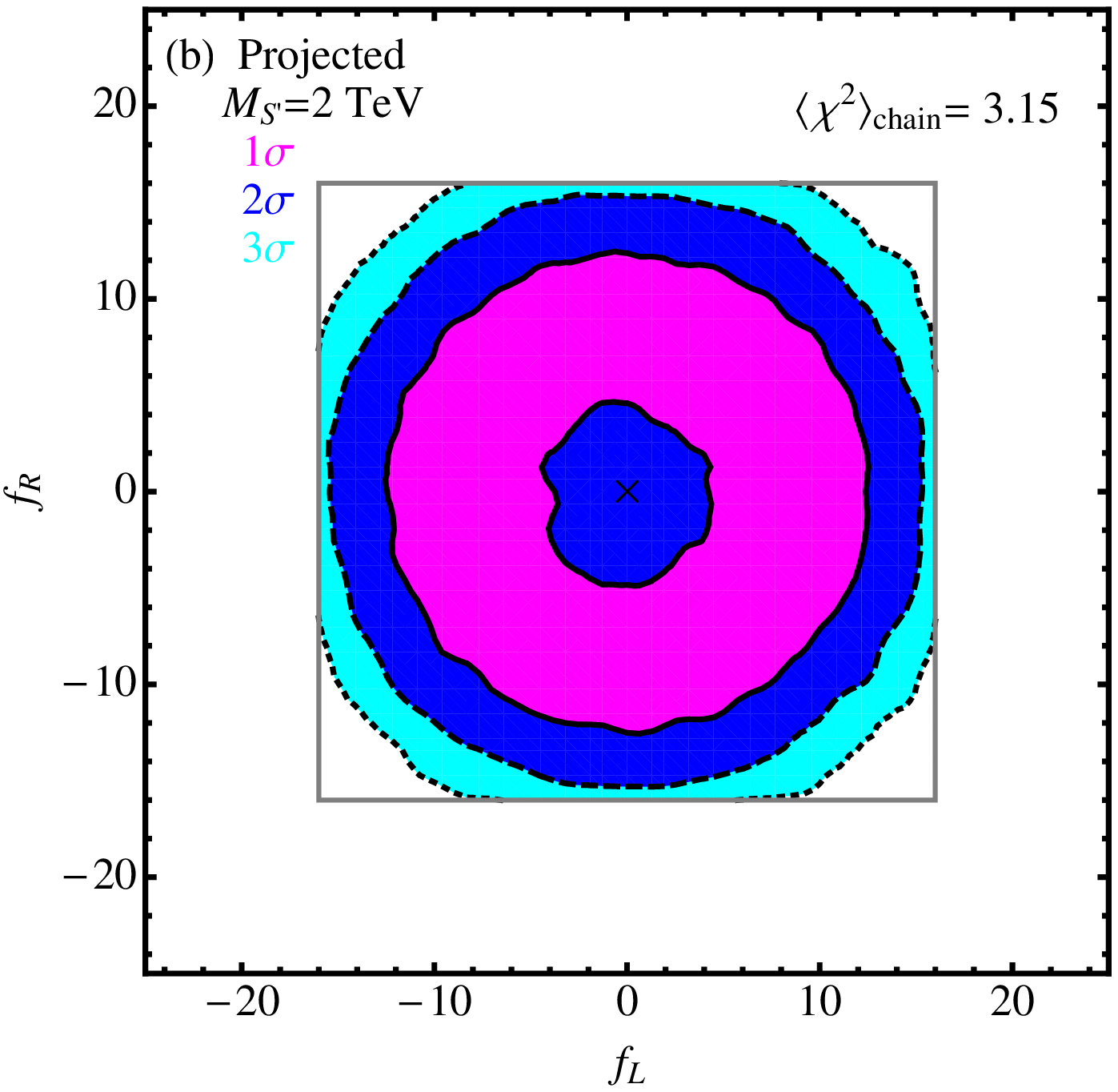}
\includegraphics[scale=0.45]{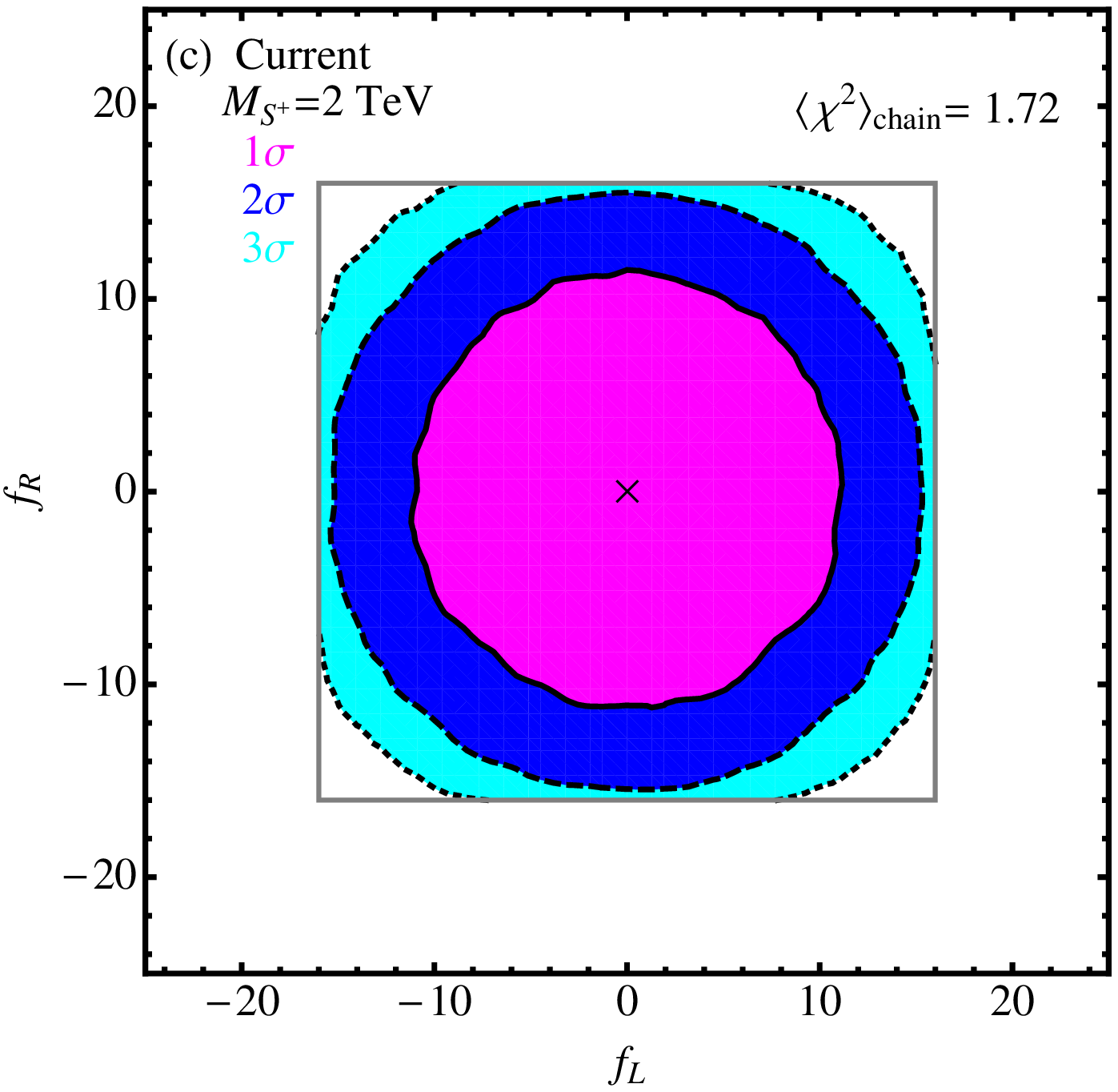}%
~~~~\includegraphics[scale=0.45]{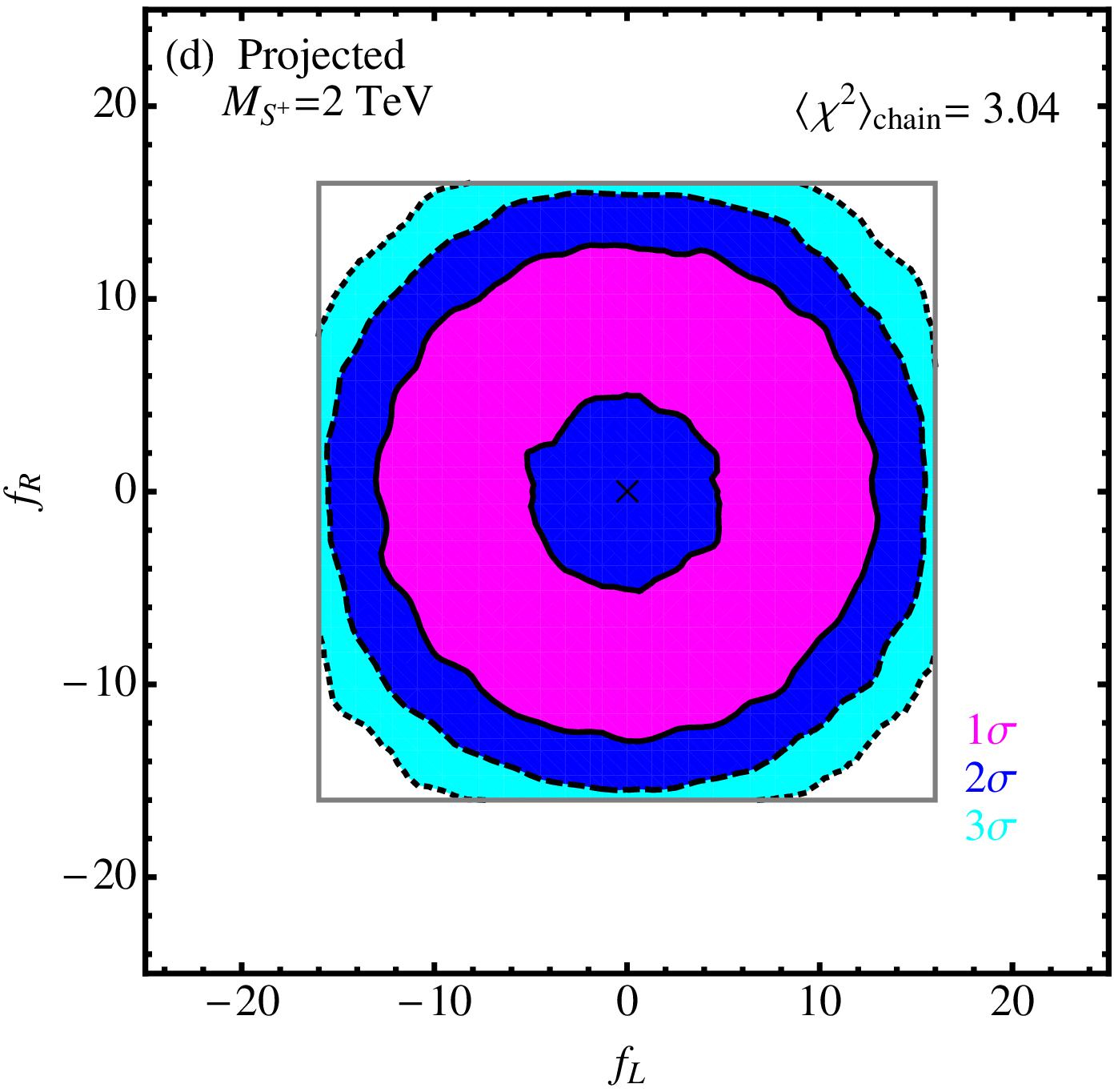}
\caption{Correlation of couplings for (a,b) $m_{S^{\prime}}=2000\,{\rm GeV}$ and (c,d) $m_{S^{\pm}}=2000\,{\rm GeV}$.
The left panels are for the current integrated luminosity while the right panels are for an upgraded
luminosity of $\int \mathcal{L} dt =10\,{\rm fb}^{-1}$.
 The regions of each figure, 
from innermost to outermost, are (a) within $1\sigma$, within 
$2\sigma$, within $3\sigma$, and greater than $3\sigma$;  (b) within $2\sigma$, within $1\sigma$, within $2\sigma$, within $3\sigma$, and 
greater than $3\sigma$;  (c) within $1\sigma$, within $2\sigma$, within 
$3\sigma$, and greater than $3\sigma$; (d) within $2\sigma$, within $1\sigma$, 
within $2\sigma$, within $3\sigma$, and greater than $3\sigma$.
The couplings are varied only within their allowed values from unitarity 
considerations (see Appendix~\ref{sec:unitarity}). 
\label{fig:tchan-sp-1000-fl-fr}}
\end{figure}

% This is Figure 19
\begin{figure}
\includegraphics[scale=0.5]{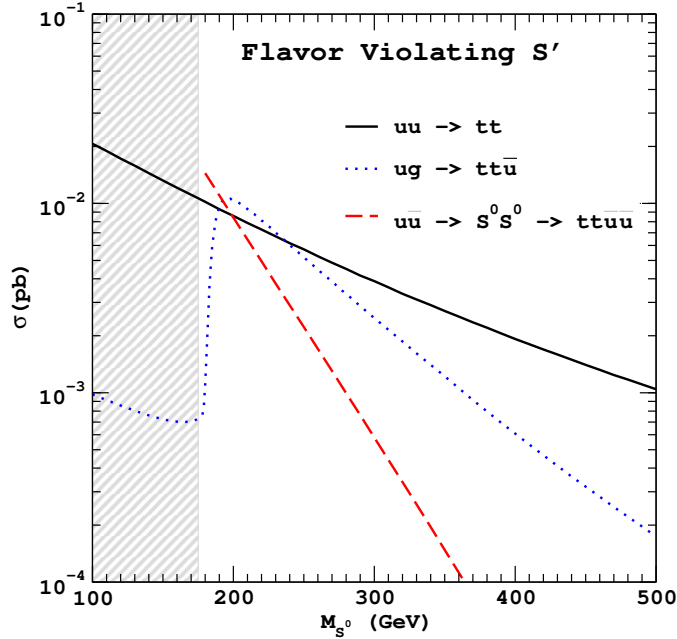}
\caption{Value of the same-sign top quark production cross section at the 
Tevatron for a flavor violating $S^\prime$ with $f_L=f_R=1$ 
(Eq.~(\ref{eq:sprime})).  The gray region indicates $m_{S^{\prime}}<m_t$.
\label{fig:sprime-tevatron}}
\end{figure}

We plot $A_{FB}^{NP}$ as a function of $f=\sqrt{f_L^2+f_R^2}$ for a scalar of 
mass $2~{\rm TeV}$ in Fig.~\ref{fig:tchan-scalar} (d) and see that it is 
indeed negative.  This does not pose a problem with respect to the measurement 
of a positive asymmetry since the scalar interferes destructively with the SM 
which gives a negative $R$, defined in Eq.~(\ref{eq:def-afbnp-R}) and shown in 
Fig.~\ref{fig:tchan-scalar} (c).  Thus, the total asymmetry, which is related 
to $A_{FB}^{NP}$ and $R$ in Eq.~(\ref{eq:AFB}), is positive.  This is plotted 
in Fig.~\ref{fig:tchan-scalar} (a).

We also consider a charged scalar, $S^{\pm}$, which couples to top quarks as 
in Eq.~(\ref{eq:sprime}), but with the replacement $u\to d$.  In 
Fig.~\ref{fig:tchan-sp-1000-fl-fr}, we see the result of the MCMC scan 
in the plane of $f_L$ and $f_R$ for a 1000 GeV neutral ${S^\prime}$ and 
charged $S^{\pm}$ scalar: 
(a,c) for the current luminosity and (b,d) for expected $10~{\rm fb}^{-1}$, again
assuming the central value of experimental data is not changed.
For the neutral scalar, we impose the same-sign top pair production constraint 
discussed in Sect.~\ref{sec:Flavor-violating}.  We plot 
the same-sign top pair production cross section for a neutral scalar with 
$f_L=f_R=1$ at the Tevatron in Fig.~\ref{fig:sprime-tevatron}.
In the case of the neutral scalar, with the current luminosity we find a marginally better fit than in the SM, 
$\left<\chi^2\right>_{chain}=1.72$.  The fit improves relative to the SM with 
$\left<\chi^2\right>_{chain}=3.15$ for $10~{\rm fb}^{-1}$ if the central values 
remain the same.  For the charged scalar we obtain a similar fit at the current luminosity with $\left<\chi^2\right>_{chain}=1.72$.  At $10~{\rm fb}^{-1}$ with the same central values we obtain a slightly better fit, $\left<\chi^2\right>_{chain}=3.04$, than in the case of the neutral scalar due to differences between the $u$ and $d$ PDFs.   Although the negative $\sigma_{INT}$ decreases the 
total cross section 
$\sigma_{t\bar{t}}$, it allows for good agreement with the 
$\frac{d\sigma}{dM_{t\bar{t}}}$ distribution and gives a positive contribution 
to $A_{FB}^{tot}$ which results in a better overall fit.

%--------------------------------------------------------------------%
%  Section IX                                                        %
%                                                                    %
%   Conclusion                                                       %
%                                                                    %
%--------------------------------------------------------------------% 
\section{Conclusion\label{sec:conclusion}}

We have examined a number of models for new physics in top quark pair
production which could account for the larger-than-expected forward-backward
asymmetry observed at the Fermilab Tevatron, while not significantly
disturbing the approximate agreement of the cross section $\sigma_{t \bar t}$ and its $M_{t\bar{t}}$ distribution
with the Standard Model predictions.  Our results are summarized in Table
\ref{tab:sum}.

% This is Table IV
\begin{table}[ht]
\caption{Models for top quark pair production and their ability to account
simultaneously for the cross section and forward-backward asymmetry in top
quark pair production at the Tevatron.  Flavor Conserving (FC) and Flavor Violating (FV) models are considered.
\label{tab:sum}}
\begin{center}
\begin{tabular}{l c l} \hline \hline
\multicolumn{2}{l}{Model} & Result \\ \hline
FC $G^\prime$ & 1 TeV & Poor fit for $f_R=g_R=0$ considered due to $\frac{d\sigma}{dM_{t\bar{t}}}$ constraint; \\
             & & Good fit for axial couplings\\ 
             & 2 TeV & Fit not significantly improved with respect to the SM for $f_R=g_R=0$;\\ 
             &  & Excellent fit for axial couplings\\ 
\multicolumn{2}{l}{EFT} & Poor fit; $A_{FB}$ consistently smaller than measured value.\\
FC $Z'$ & 1 TeV & Poor fit due to $\frac{d\sigma}{dM_{t\bar{t}}}$ constraint\\
             & 2 TeV & Fit not significantly improved with respect to the SM for $f_R=g_R=0$\\
\multicolumn{2}{l}{FV $Z'$} & Poor fit due to $\frac{d\sigma}{dM_{t\bar{t}}}$ constraint and same-sign top constraint\\
FV $W'$ & 2 TeV & Good fit although large couplings necessary with a large amount of fine-tuning\\
FV $S'$, $S^\pm$ & 2 TeV & Tension with small predicted $\sigma_{t\bar t}$ leads to poor fit with current data; \\
	  &  & however, a good fit would be obtained if central values were unchanged after $10~{\rm fb}^{-1}$ \\
\hline \hline
\end{tabular}
\end{center}
\end{table}

% This is Table V
\begin{table}
\caption{
Values of $\chi^2$ for selected models. 
Assuming the current integrated luminosities, the $\langle\chi^2\rangle_{point}$ contributions from the production cross section, forward-backward asymmetry and invariant mass distribution in top quark pair production at the Tevatron are given for the chain only in the small neighborhood of the points that provide the best fit to the data in each model. 
The value of $\langle\chi^2\rangle_{point}$ is the Total/$N_{dof}$, where $N_{dof}=3$, and provides a measure of how well the best fit point in each model fits the data.  The value $\delta\langle\chi^2\rangle_{point}$ is the $\langle\chi^2\rangle_{point}$ value of a box that is $\pm10\%$ wide in the couplings for the best fit point and is a rough measure of fine tuning.  Note the SM value for $\langle \chi^2\rangle_{point}$ is equivalent to that given in Eq.~\ref{eq:chisqsm}.  The $\langle\chi^2\rangle_{chain}$ values for each model are also listed.
\label{tab:bestmodel}}
\begin{center}
\begin{tabular}{|c|ccc|ccc|c|} \hline \hline
Model& $\sigma_{t\bar t} $ & $A_{FB}$  & ${d \sigma\over d M_{t\bar t}}$ & Total &$\langle\chi^2\rangle_{point}$ &$\delta\langle\chi^2\rangle_{point}$ &$\langle\chi^2\rangle_{chain}$ \\
\hline
FV $W^\prime$ ${(f_L=17.0, f_R=0.0,m_{W^\prime}=2~\rm{TeV})}$:  &0.63 &0.06& 0.27 & 0.96& 0.32  & 41.4\% & 1.67\\
axial $G^\prime$ ${(f=2.5, g=-2.5,m_{G^\prime}=2~\rm{TeV})}$:  &0.75 &0.47& 0.75 & 1.97& 0.66  & 1.6\% & 1.15\\
chiral $G^\prime$ ${(f_L=2.0, g_L=-2.0,m_{G^\prime}=2~\rm{TeV})}$:  &0.59 &2.91& 0.62 & 4.12& 1.37 & 0.4\% & 1.69 \\
FC $Z^\prime$ ${(f_L=8.0, g_L=8.0,m_{Z^\prime}=2~\rm{TeV})}$:  &0.67 &2.66& 0.91 & 4.24& 1.41   & 1.0\% & 1.62\\
FV $S^\prime$ ${(f_L=7.0, f_R=0.0,m_{S^\prime}=2~\rm{TeV})}$:  &2.01 &2.85& 0.09 & 4.96& 1.65  & 0.2\% & 1.72\\
FV $S^\pm$ ${(f_L=7.5, f_R=0.0,m_{S^\prime}=2~\rm{TeV})}$:  &2.01 &2.85& 0.12 & 4.98& 1.66& 0.2\% & 1.72 \\
\hline
SM : &1.12 &4.07& 0.06& 5.25& 1.75& -- & 1.75\\
\hline \hline
\end{tabular}
\end{center}
\end{table}

The results summarized in Table \ref{tab:sum} show that it is not easy to
account for the larger-than-expected value of $A_{FB}(t \bar t)$ observed at
the Fermilab Tevatron while maintaining the good agreement between theory
and experiment for the production cross section $\sigma_{t \bar t}$ and differential rate ${d \sigma \over d M_{t \bar t}}$.  

Of the models considered, those that provide a fit better than the SM for the applicable data are a $1~{\rm TeV}$ or $2~{\rm TeV}$ flavor-conserving $G^\prime$ with axial couplings, a $2~{\rm TeV}$ $W^\prime$ (or a $2~{\rm TeV}$ flavor-conserving $G^\prime$ or $Z^\prime$) with chiral couplings. 
Other models we considered provide at most a mild
improvement with respect to the SM case.
The 1 TeV cases often give large corrections to the $M_{t\bar t}$ distribution since the additional signal is well inside the data region.  Finally, in Table~\ref{tab:bestmodel}, we examine in detail the contribution to the $\langle\chi^2\rangle_{chain}$ in the small neighborhood around the best points in parameter space of the axial $G^\prime$,  chiral $G^\prime$, FV chiral $W^\prime$, and  FC chiral $Z^\prime$ models as well as the SM.  Assuming the current integrated luminosity for each measurement and a mass of the new states responsible for the $A_{FB}$ deviation of 2 TeV, we find:
\begin{itemize}
\item The $2~{\rm TeV}$ axial $G^\prime$ model provides the best overall fit to the experimental data
considered in this article.  It improves the agreement with experiment of  both the total top quark production cross section and the forward-backward asymmetry. The fit to the 
$M_{t\bar t}$ distribution is slightly worse than in the SM case but still in good agreement with data.  
\item The $2~{\rm TeV}$ $W^\prime$ can also lead to a good fit to the data. It is able to generate a large asymmetry and to improve the agreement of total cross section with data without disturbing the differential cross section sizably for some regions of parameter space.  However, large couplings are needed.  Note that in Table~\ref{tab:bestmodel}, the best-fit point in the $W^\prime$ case has a lower $\chi^2$ than the axial $G^\prime$ although the $\langle\chi^2\rangle_{chain}$ is lower for the axial $G^\prime$ indicating that the axial $G^\prime$ gives a better overall fit.  Stated differently, the $W^\prime$ requires a greater amount of fine-tuning of its parameters to fit the data than the axial $G^\prime$.  This is seen in the large value of $\delta\langle\chi^2\rangle_{point}$; a slight perturbation of the best fit points greatly decreases the quality of the fit.  The $W^\prime$ model provides such a large value since the $\chi^2$ contributions from $\sigma_{t\bar t}$ and $A_{FB}$ are aligned and increase together with couplings that deviate from the minimum $\chi^2$ couplings.  This is to be contrasted with the other models in which the increasing $\chi^2$ contribution from, say, $A_{FB}$ is compensated by a smaller $\chi^2$ contribution from $\sigma_{t\bar t}$, resulting in a total $\chi^2$ value that remains relatively flat.
\item The $2~{\rm TeV}$ chiral $G^\prime$ and $Z^\prime$ do not lead to significant improvement over the SM. They reduce the discrepancy with the asymmetry measurement although they are unable to reduce it below $2\sigma$ without disturbing the $M_{t\bar{t}}$ distribution due to their large widths.
\item The $2~{\rm TeV}$ FV scalars $S^\prime$ and $S^\pm$ have fits that are not significantly improved with respect to the SM.  They lead to a significant discrepancy in $\sigma_{t\bar{t}}$ and only slightly improve the fit to $A_{FB}$ and $d\sigma/dM_{t\bar{t}}$ with respect to the SM.
\end{itemize}

In this work, we have used the full NLO QCD $t\bar{t}$ production cross
section. 
It is worth noting that partial NNLO QCD corrections to the $t\bar{t}$ cross section have been calculated 
in Ref.~\cite{Kidonakis:2008mu} and give rise to an enhancement of the total cross section of about 0.3~pb. This indicates that higher order QCD corrections
might improve the agreement between the measured total cross section and its value in the SM, and therefore areas of NP parameter space which give negative contributions to the total cross section will be less constrained.  Such negative contributions, however, may be in
tension with the observed $M_{t \bar{t}}$ invariant mass distribution. 
A detailed collider simulation, including the complete NNLO corrections,  
would be therefore highly desirable in order to make a more reliable comparison of
the predictions of these models with data. 

Crucial to the test of any model is also the accumulation of more integrated
luminosity at the Tevatron, in order to demonstrate deviations from the
Standard Model exceeding $3 \sigma$.  Until then, the observed $A_{FB}$ values
cannot be regarded as anything more than a hint of new physics.

\begin{acknowledgments}
Q.-H.~C. is supported in part by the Argonne National Laboratory and
University of Chicago Joint Theory Institute (JTI) Grant 03921-07-137,
and by the U.S.~Department of Energy under Grants No.~DE-AC02-06CH11357
and DE-FG02-90ER40560.  
D.~M. and J.~L.~R. are supported by the U.~S.~Department of Energy under 
Grant No.~DE-FG02-90ER40560. 
G.~S. is supported in part by the U.~S.~Department of Energy under Grants 
No.~DE-AC02-06CH11357 and DE-FG02-91ER40684.
C.~E.~M.~W. is supported in part by U.~S.~Department of Energy
under Grants No.~DE-AC02-06CH11357 and DE-FG02-90ER40560.
J.~L.~R., G.~S., and C.~E.~M.~W. thank the Aspen Center for Physics for hospitality.
The authors thank M.~Neubert for useful discussions.
\end{acknowledgments} 

\appendix
%--------------------------------------------------------------------%
%  Section VI                                                        %
%                                                                    %
%   Asymmetry calculation and useful formula                         %
%                                                                    %
%--------------------------------------------------------------------% 
\section{Unitarity constraints\label{sec:unitarity}}
In this appendix, we explore the unitarity constraints on new physics
models considered in this work. The weak isospin amplitude $\mathcal{M}^I$
($I$ being isospin index) can be decomposed with respect to orbital angular
momentum according to
\begin{equation}
\mathcal{M}^I=16\pi\sum_{0}^{\infty}(2\ell +1) P_{\ell}(\cos\theta) a_\ell^I.    
\end{equation}
With the normalization $\Im a_\ell^I=|a_\ell^I|^2$,
the unitarity constraint requires
\begin{equation}
 |\Re a_\ell^I|<\frac{1}{2},
\label{eq:unitarity_limit}     
\end{equation}
where $a_\ell^I$ could be projected via: 
\begin{equation}
a_\ell^I=\frac{1}{32\pi}\int_{-1}^{1} d\cos\theta P_\ell(\cos\theta)
\mathcal{M}^I.
\label{eq:partialwave}      
\end{equation}

First consider the flavor-conserving $Z^\prime$ models, which involve the 
$s$-channel diagram only. Note that the constraints of the $G^\prime$ model
can be easily derived from flavor-conserving $Z^\prime$ model. 
The helicity amplitudes for $q\bar{q} \to t \bar{t}$ are represented by  
$A(\lambda_q,\lambda_{\bar{q}},\lambda_t,\lambda_{\bar{t}]})$, where 
$\lambda_t=-,+$, respectively, indicates a left-handed and a right-handed
top quark. Apart from the common factor 
$$ \frac{2 e^2 E}{s-m_{Z^\prime}^2},$$ the non-vanishing helicity amplitudes
from the diagram mediated by the $Z^\prime$ boson are
\begin{eqnarray}
A(-,+,-,-) & = &  ~~ f_L \sin\theta     
                  ~m_t \left[  g_L + g_R  \right],
                  \label{mpmm}\\
A(-,+,-,+) & = & -f_L (1+\cos\theta) 
                  E   \left[ (1+\beta_t) g_L + (1-\beta_t) g_R \right],
                  \label{mpmp}\\
A(-,+,+,-) & = &  ~~ f_L (1-\cos\theta) 
                  E   \left[ (1-\beta_t) g_L + (1+\beta_t) g_R \right],
                  \label{mppm}\\
A(-,+,+,+) & = & -f_L \sin\theta     
                  ~m_t \left[ g_L + g_R  \right],
                  \label{mppp}\\
A(+,-,-,-) & = &  ~~ f_R \sin\theta     
                  ~m_t \left[ g_L + g_R  \right],
                  \label{pmmm}\\
A(+,-,-,+) & = &  ~~ f_R (1-\cos\theta) 
                  E   \left[ (1+\beta_t) g_L  + (1-\beta_t) g_R \right], 
                  \label{pmmp}\\
A(+,-,+,-) & = & -f_R (1+\cos\theta) 
                  E   \left[ (1-\beta_t) g_L  + (1+\beta_t) g_R \right],
                  \label{pmpm}\\
A(+,-,+,+) & = & -f_R \sin\theta     
                  ~m_t \left[ g_L + g_R  \right],\label{pmpp}
\label{eq:hel-schan}      
\end{eqnarray}
where $\beta_t=\sqrt{1-m_t^2/E^2}$. In the c.m.\ frame of the $t\bar{t}$ pair
the 4-momenta of the particles are chosen to be 
\begin{eqnarray}
  p_{q}       & = & E~(1,~0,~0,~1)\\
  p_{\bar{q}} & = & E~(1,~0,~0,~-1)\\
  p_{t}       & = & E~(1,~\beta_t\sin\theta,0,~\beta_t\cos\theta)\\
  p_{\bar{t}} & = & E~(1,-\beta_t\sin\theta,0,-\beta_t\cos\theta).
\label{eq:hel-momenta}      
\end{eqnarray}
In the high energy limit $E \gg m_t$, only $(1+\beta_t)$ terms contribute. 
The $J=1$ partial-wave of $A(-,+,-,+)$ amplitude is
\begin{equation}
a_{s}^{J=1}(q\bar{q}\to Z^\prime \to t\bar{t})= \frac{\alpha_{em}}{12} f_L g_L,   
\end{equation} 
yielding the following limit
$ |f_L g_L| \leq 6/\alpha_{em}$. 
Similarly, one can derive the following constraints
$$|f_L|\lesssim 28\qquad {\rm and} \qquad |g_L| \lesssim 28 $$ 
from the $q\bar{q}\to Z^\prime \to q\bar{q}$ and 
$t\bar{t} \to Z^\prime \to t\bar{t}$ processes. 

Now consider the flavor-violating $Z^\prime$ model. We consider the scattering $u\bar{t}\to Z^\prime \to t\bar{u}$, the calculation of which is identical
to the flavor-conserving $Z^\prime$ model. In the high energy limit 
$\sqrt{s}\gg m_t$, we obtain the following unitarity bound on $f_R$ from 
the helicity amplitude $A(+,~-,~+,~-)$ in Eq.~\ref{pmpm},
\begin{equation}
   \left|f_R\right| \lesssim 28~.   
\end{equation}

Finally, we consider the scattering $u\bar{t} \to S \to \bar{u} t$ to derive
the unitarity constraint for the flavor-violating $S$ model. 
The helicity amplitudes are represented by 
$A(\lambda_q,\lambda_{\bar{t}},\lambda_{\bar{u}},\lambda_t)$. 
In the high energy limit $s \gg m_t$, the non-vanishing helicity amplitudes
are
\begin{eqnarray}
A(+,+,+,+) & = & e^2 f_{R}^2,\\
A(+,+,-,-) & = & e^2 f_{R}f_{L},\\
A(-,-,+,+) & = & e^2 f_{L}f_{R},\\
A(-,-,-,-) & = & e^2 f_{L}^2,
\label{eq:hel-schan-s}      
\end{eqnarray}
where $f_L$ and $f_R$ are given in Eq.~\ref{eq:sprime}.
The $J=0$ partial-wave of $A(+,+,+,+)$ is
\begin{equation}
a_{s}^{J=0}(u\bar{t}\to S \to \bar{u}t) = \frac{\alpha_{em}}{4} f_R^2,   
\end{equation} 
yielding the unitarity limit $f_L^2 \leq 2/\alpha_{em}$. 
Hence, $|f_L| \lesssim 16$, $|f_R| \lesssim 16$, and $|f_L f_R|< 256$.

\bibliographystyle{apsrev}
\bibliography{reference}

\end{document}